\documentclass[onecolumn,nofootinbib,notitlepage]{revtex4-1}
\usepackage{color}
\usepackage{subfigure}
\usepackage{bm}
\usepackage[toc,page]{appendix}
\usepackage[usenames,dvipsnames,svgnames,table]{xcolor}
\usepackage{mathrsfs}
\usepackage{graphicx}
\usepackage{setspace}
\usepackage{latexsym,lipsum}
\usepackage[colorlinks=true, linkcolor=red, urlcolor=blue]{hyperref}
\usepackage{footnotebackref}
\definecolor{red}{rgb}{0.75,0,0}
\definecolor{blue}{rgb}{0,0,0.75}
\definecolor{green}{rgb}{0,0.5,0}
\definecolor{yellow}{rgb}{0,0.5,0}
\usepackage{amsmath}
\usepackage{amssymb}

\def\be{\begin{equation}}
\def\ee{\end{equation}}
\def\bea{\begin{eqnarray}}
\def\eea{\end{eqnarray}}
\def\ito{It$\hat{\mbox{o}}$}

\def\besub{\begin{subequations}}
\def\eesub{\end{subequations}}

\def\v{{\bf v}}

\def\bwd{\begin{widetext}}
\def\ewd{\end{widetext}}

\newcommand{\bsf}[1]{\textsf{\textbf{#1}}}

\begin{document}
\title{Origins and diagnostics of the nonequilibrium character of active systems}
\author{Lokrshi Prawar Dadhichi}
\email{lpdadhichi@gmail.com}
\affiliation{Tata Institute of Fundamental Research, Gopanpally, Hyderabad 500 107}
\author{Ananyo Maitra}
\email{nyomaitra07@gmail.com}
\affiliation{LPTMS, CNRS, Univ. Paris-Sud, Universit\'e Paris-Saclay, 91405 Orsay, France}
\author{Sriram Ramaswamy}
\email{sriram@iisc.ac.in}
\affiliation{Centre for Condensed Matter Theory, Department of Physics, Indian Institute of Science, Bangalore 560012, India}
\begin{abstract}
We present in detail a Langevin formalism for constructing stochastic dynamical equations for active-matter systems coupled to a thermal bath. We apply the formalism to clarify issues of principle regarding the {sources} and signatures of nonequilibrium behaviour in a variety of polar and apolar single-particle systems and polar flocks. We show that distance from thermal equilibrium depends on how time-reversal is implemented and hence on the reference equilibrium state. We predict {characteristic forms} for the frequency-resolved entropy production for an active polar particle in a harmonic potential, which should be testable in experiments. 
\end{abstract}
\maketitle
\section{Introduction}\label{intro} 
Active systems are {held away from thermal} equilibrium by free energy supplied directly to the constituent particles, which transduce it into systematic movement\cite{ramaswamy2010mechanics,marchetti2013hydrodynamics}. This breaking of time-reversal symmetry alone at the scale of the microscopic units, while retaining spatial homogeneity and isotropy, is the defining feature of active matter and sets it apart from familiar driven systems forced by an imposed spatial gradient of temperature, potential or velocity. In this article we construct active-particle dynamics from the Langevin equation for {thermal Brownian} particles coupled to the chemical kinetics of fuel and offer a fresh perspective on their nonequilibrium character. {One aim of this work is to point out that the answer to the question ``How far from equilibrium is active matter?'' \cite{fodor2016far} is not unique. It depends on the reference equilibrium state which, in turn, depends on the time-reversal signature assigned to the dynamical variables in question. In addressing this question we show precisely what observed behaviours, as defined by a suitable notion of entropy production, distinguish an active system from a passive counterpart with the same spatial symmetries. A contrast with the treatment of \cite{fodor2016far} is that in the present work we consider systems in which all degrees of freedom are coupled to a common heat bath at temperature $T$.} 

The first part of our discussion highlights the source of nonequilibrium behaviour. We begin by writing down \emph{equilibrium} Langevin equations for a set of coordinates and momenta coupled to an explicit chemical degree of freedom. We show that the stochastic equations for the dynamics of active particles emerge naturally if the chemical driving force, which is the difference $\Delta\mu$ between the chemical potentials of reactants and products, is then held fixed. This amounts to an extension of the approach of refs. \cite{julicher1997modeling,kruse2005generic,joanny2007hydrodynamic,juelicher2007active,julicher2018hydrodynamic} to include noise, inherited in the simplest case from equilibrium Langevin dynamics, {but allowing the possibility of strong modification in the presence of driving}. In the second part of our treatment we examine the resulting behaviour, that is, how the imposition of a maintained chemical-potential difference leads to entropy production. Note that we define the latter through probability ratios of forward and backward processes with assigned time-reversal signatures \cite{pietzonka2017entropy,neri2017statistics,shankar2018hidden}, 
with no necessary connection to heat generation \cite{pigolotti2017generic}.  
{This is operationally useful because in experiments one tracks only a few of the many possible degrees of freedom -- e.g., the position and vectorial orientation of a single self-propelled particle \cite{kumar2015anisotropic}.} We shall see that such partial entropy-production functions do estimate {how far the observed dynamics is from thermal equilibrium}. Of course, the calculation of entropy production requires the definition of a time-reversal operation, and such a definition should lead to vanishing entropy production rate in the limit of vanishing $\Delta\mu$. However, two different equilibrium limits are possible in standard models for active polar particles -- one in which the self-propulsion velocity remains a velocity and another in which it lapses back to being simply the orientation vector. We discuss this both for single polar active particles {-- Active Brownian Particles (ABPs) or Active Ornstein-Uhlenbeck Particles (AOUPs) \cite{schimansky1995structure,ebeling1999active,romanczuk2012active,maggi2014generalized,koumakis2014directed,fodor2016far} --} and the Toner-Tu field theory \cite{toner1995long,toner1998flocks,toner2005hydrodynamics} for a collection of such particles. The latter is then seen as the active variant of either a polar liquid crystal or a Navier-Stokes fluid, with different entropy production rates. 

\begin{figure}
 \includegraphics[width=\columnwidth]{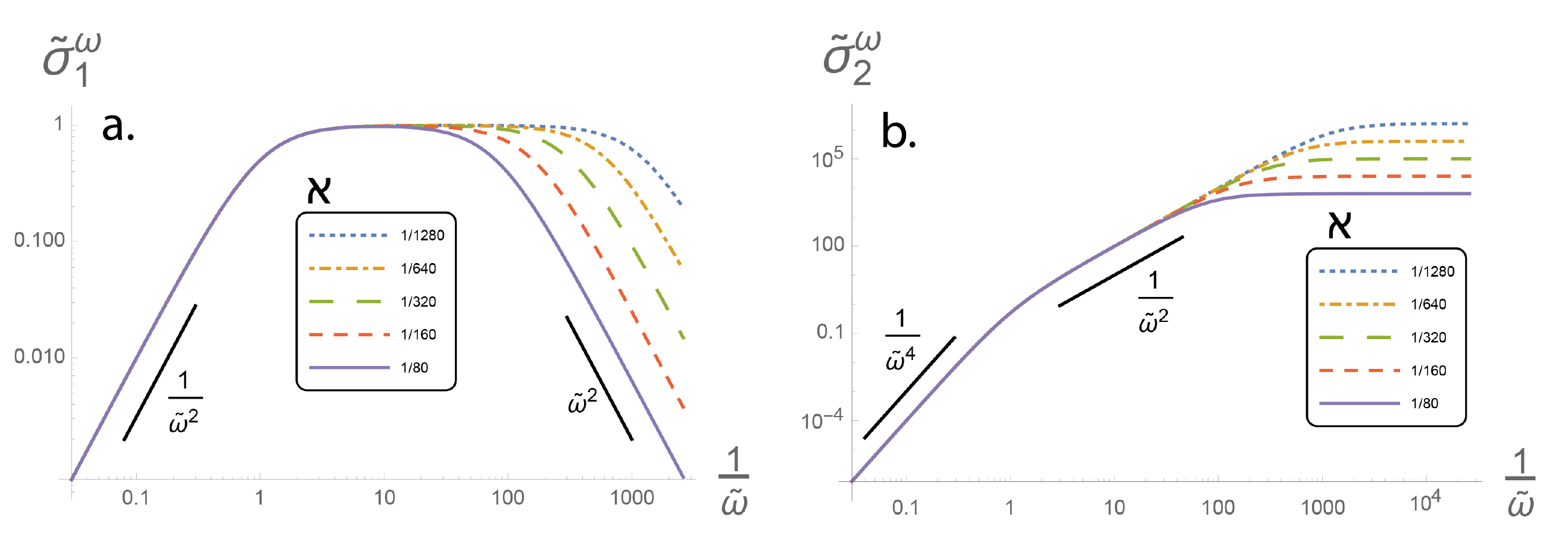}
\caption{(a) {Frequency-dependent entropy production rate for a harmonically confined active polar dimer with translational diffusion, when time-reversal does not include a polarity-flip: As discussed, and consistent with \cite{wang2016entropy}, the rate vanishes both at small and large frequencies with a frequency-independent plateau in {between} which increases with decreasing $\aleph$. (b) When time-reversal includes a polarity-flip, the rate} differs from (a) only by a factor of $1/\tilde{\omega}^2$.}
\label{EntFig1}
\end{figure}

{Here are our main results. 
	\begin{itemize}  
		\item All standard stochastic dynamical models for active matter, including nominally two-temperature cases such as those for apolar active particles, as well as those in which activity enters as a negative linear-order damping, emerge naturally from our single-bath framework. 
		\item A non-unit coefficient for the ``advective nonlinearity'' in the Toner-Tu model arises only if both Galilean invariance and detailed balance are broken. 
		\item Translational Brownian motion in the dynamics of polar active particles leads to a statistical steady state qualitatively different from that obtained in \cite{fodor2016far, sandford2017pressure}, with nonzero entropy production even for a particle in a harmonic potential. The value and form of the entropy production rate depends on whether the time-reversal involution includes an inversion of the polarity. The characteristic profiles we predict for frequency-resolved entropy production rates (see Fig. \ref{EntFig1}), from a Harada-Sasa approach \cite{harada2005equality}, should be testable in experiments on granular or colloidal active matter. {The predicted profile for the case where the polarisation is treated as even under time-reversal is consistent with that discussed in \cite{wang2016entropy} for Markovian systems with timescale separation.
		\item{An active polar particle with translational diffusion in a harmonic trap is distinguishable from a passive one only when the dynamics of the auxiliary variable is also recorded. 
		}	
			}
\end{itemize}
}

{This paper is organised as follows. In section \ref{sec:gendev} we summarise the approach to the construction of stochastic equations for active systems. In section \ref{sec:applic} we apply the formalism to generate the dynamics of a variety of single-particle and spatially extended systems, staring from polar active particles (Active Brownian or Active Ornstein-Uhlenbeck) as motile dimers. In section \ref{sec:entpo} we evaluate entropy production rates via time-reversal of the stochastic action, and construct Harada-Sasa relations connecting entropy production to correlation-response differences. 
We close in section \ref{conclusion} with a summary. {A more detailed technical exposition of many relevant points can be found in a series of \hyperref[Appendicesmark]{Appendices}.}} 
\section{General derivation of active equations}
\label{sec:gendev}
In this section 
\footnote{The discussion in this and the next sections completely disposes of all issues raised in \cite{Brand2014}}
we review the general derivation of stochastic active dynamics as presented in \cite{ramaswamy2017active}. In line with \cite{julicher1997modeling}, we describe active systems as those in which one or more chemical degrees of freedom drive the observed mechanical degrees of freedom. We construct equations of motion for dynamical variables $\bm{\mathcal{C}}$ with position-like and momentum-like components $\bm{\mathcal{Q}}$ and $\bm{\mathcal{P}}$, respectively even and odd under time-reversal (hereafter denoted $\mathcal{T}$). 
$\bm{\mathcal{C}}$ can be finite- or infinite-dimensional, depending on whether the system of interest is a single active {particle \cite{kumar2008active}}, or a spatially extended system described by an active field theory such as \cite{toner1995long,toner2005hydrodynamics}. $\bm{\mathcal{Q}}$ and $\bm{\mathcal{P}}$ need not be canonically conjugate variables nor even have the same number of components. The stochastic equations of motion describing the \emph{thermal equilibrium} dynamics of $\bm{\mathcal{C}}$ are\cite{ma1975sk,chaikin1995principles,zwanzig2001nonequilibrium,lau2007state}
\begin{equation}
\label{eq:genlang}
\partial_t\bm{\mathcal{C}}=-(\bm{\Gamma}+\bm{\mathcal{W}})\cdot{\nabla_{\bm{\mathcal{C}}} H}+T\nabla_{\bm{\mathcal{C}}} \cdot \bm{\mathcal{W}}+\bm{\xi}
\end{equation}
%
where $H(\bm{\mathcal{C}})$ is the effective Hamiltonian, 
\begin{equation}
\label{eq:gennoise}
\langle\bm{\xi}(t)\bm{\xi}(t')\rangle=2T\bm{\Gamma}\delta(t-t'),
\end{equation}
$\bm{\Gamma}$ is a symmetric matrix of dissipative couplings between the variables and $\bm{\mathcal{W}}$ is an antisymmetric matrix of reactive couplings. Terms involving $\bm{\mathcal{W}}$ must have, component by component, the same signature under $\mathcal{T}$ as $\partial_t\bm{\mathcal{C}}$, and those involving $\bm{\Gamma}$ must have the opposite $\mathcal{T}$-signature. Thus the $\bm{\mathcal{Q}} \bm{\mathcal{Q}}$ and $\bm{\mathcal{P}} \bm{\mathcal{P}}$ components of $\bm{\Gamma}$ must themselves be even under $\mathcal{T}$, while the $\bm{\mathcal{Q}} \bm{\mathcal{P}}$ and $\bm{\mathcal{P}} \bm{\mathcal{Q}}$  components must be odd. The case of specific interest to us is where $\bm{\mathcal{Q}}$ consists of a spatial part ${\bf X}$ and the chemical coordinate $n$, and $\bm{\mathcal{P}}$ has only a spatial part ${\bf P}$. Then the components $\bm{\Gamma}_{{\bf X}{\bf X}}$, $\bm{\Gamma}_{{\bf X}n}$, $\bm{\Gamma}_{\bf{P}\bf{P}}$, and $\Gamma_{nn}$ must be even under $\mathcal{T}$, and $\bm{\Gamma}_{{\bf X}{\bf P}}$ and $\bm{\Gamma}_{{\bf P}n}$ must be odd. Since we are not considering the possibility of an external field that breaks {$\mathcal{T}$}, such as a magnetic field \cite{casimir1945onsager,de2013non}, this implies that $\bm{\Gamma}_{{\bf X}{\bf P}}$, and $\bm{\Gamma}_{{\bf P}n}$ should themselves be odd in ${\bf P}$. 
When ${\bf \Gamma}$ depends on ${\bm{\mathcal{C}}}$ the noise \eqref{eq:gennoise} is multiplicative. For the steady-state distribution to be $e^{-H/T}$, we must then include in \eqref{eq:genlang} the additional drift $T (\nabla_{\bm{\mathcal{C}}} \cdot {\bf \Gamma} - \alpha \bsf{g}\cdot \nabla_{\bm{\mathcal{C}}} \bsf{g})$
where $\bsf{g} \cdot \bsf{g} = 2 {\bf \Gamma}$ \cite{lau2007state}, and $\alpha\in[0,1]$ parameterizes the noise interpretation. Similarly, $\mathcal{W}_{{\bf P}{\bf P}}$ should be odd under {$\mathcal{T}$} and therefore suitably ${\bf P}$-dependent.  The term $T{\nabla}_{\bm{\mathcal{C}}}\cdot \bm{\mathcal{W}}$ in \eqref{eq:genlang}, which emerges in standard derivations of generalized Langevin equations \cite{mori1965transport,ma1975sk,chaikin1995principles,mazenko2008nonequilibrium}, is required for the steady state to be $e^{-H/T}$. In familiar Langevin equations \cite{hohenberg1977theory} for dynamics at equilibrium this derivative vanishes, but in appendices \ref{prefspHall1} and \ref{subsubapp:velalign} we will present natural instances where it is nonzero. 

{In \cite{ramaswamy2017active}, the active terms were obtained by starting from a description in which the physical momentum is entrained by an off-diagonal \emph{dissipative} coupling to a momentum formally conjugate to $n$, taking the limit of vanishing inertia for this ``chemical momentum'', eliminating it in favour of the chemical force, and finally holding the latter fixed at a nonzero value. We demonstrate in appendix \ref{2derivation} that every active model we consider can be obtained this way as well. Time-reversal symmetry of course dictates that active systems with a \emph{reactive} coupling between ${\bf P}$ and $n$ can be obtained by the above dissipative entrainment of the physical velocity by the chemical one while those where this coupling is \emph{dissipative} require a reversible, anti-symmetric coupling between the physical and chemical velocities.}

Our aim is to arrive at dynamical equations for the mechanical degrees of freedom in the presence of a chemical driving $\Delta \mu$ in the Hamiltonian which we now write as 
\begin{equation}
\label{eq:Deltamudef}
H = H_0 + n \Delta\mu.
\end{equation}
To do this, we will now ignore all dissipative terms in the $X_i$ equations. We will also assume that neither $\bm{\Gamma}$ nor $\bm{\mathcal{W}}$ nor $H_0$ depend on $n$. In the absence of driving, the discrete character of $n$ which counts fuel molecules {consumed}, and the kinetic barrier for a single chemical reaction, should be incorporated by including in $H_0$ a term periodic in $n$, but we shall ignore this detail in our coarse-grained description. This leads to the simplified active equations of motion
\begin{equation}
\label{genXeqn}
\partial_t X_i=-\mathcal{W}_{X_iP_j}\frac{\partial H}{\partial P_j}
\end{equation}
\begin{equation}
\label{genPeqn}
\partial_t P_i=-\Gamma_{P_iP_j}\frac{\partial H}{\partial P_j}-\Gamma_{P_in}\Delta \mu-\mathcal{W}_{P_iX_j}\frac{\partial H}{\partial X_j}-\mathcal{W}_{P_in}\Delta \mu-\mathcal{W}_{P_iP_j}\frac{\partial H}{\partial P_j}+\xi_{P_i}.
\end{equation}
where $\Gamma_{P_in}$ is odd in $P_i$ by symmetry and $\mathcal{W}_{P_in}$ is even under {$\mathcal{T}$} and therefore, apart from constants, can only involve $X_i$ and even functions of ${\bf P}$. The noise correlator now has the usual form
\begin{equation}
\langle\xi_{P_i}(t)\xi_{P_j}(t')\rangle=2T\Gamma_{P_iP_j}\delta(t-t').
\end{equation}
Equations \eqref{genXeqn} and \eqref{genPeqn}  constitute the stochastic generalisation of the nonequilibrium thermodynamical derivation of the active matter equations proposed in \cite{julicher1997modeling}. In \cite{julicher1997modeling}, however, the effects of interest entered primarily through a reactive coupling of $n$ to ${\bf P}$; we will see that a dissipative coupling between the two quantities also introduces qualitatively distinct physics.  


In the next section we use this approach to motivate the dynamical equations both for active particles and active field theories. 

\section{Applications} \label{sec:applic}
\subsection{Reactive coupling between physical momentum and chemical variable} \label{sub:reversiblechemforce}
{\subsubsection{Self-propelled dimers}} \label{subsub:spdimers}
As our first example, we consider a dimer \cite{kumar2008active, baule2008exact} with centre-of mass position and momentum $X$ and $P$, and relative coordinate and momentum $x$ and $p$. We do not allow any dissipative coupling between $P$ and $p$ and  reactive coupling between $p$ and $n$, but assume, in the notation of section \ref*{sec:gendev}, a reversible off-diagonal kinetic coefficient $\mathcal{W}_{P n}=-\zeta x$. Then
 from \eqref{genXeqn} and \eqref{genPeqn} we immediately obtain the mechanical equations of motion at a constant chemical driving $\partial_n H = \Delta\mu$:
\begin{equation}
\label{mod1pos}
\dot{X}={\partial_P {H}}; \, \dot{x}={\partial_p {H}}
\end{equation}
\begin{equation}
\label{mod1mom}
\dot{P}+\Gamma \partial_P {H} =-{\partial_X H} +\zeta\Delta\mu{ x}+\xi_P
\end{equation}
\begin{equation}
\label{mod1aux}
\dot{p}+\gamma \partial_p {H} =- {\partial_x{H}}+{\xi}_p
\end{equation}
where we have taken $\Gamma_{PP}=\Gamma$ and $\Gamma_{pp}=\gamma$, and the thermal Gaussian white noises $\xi_P$ and $\xi_p$ have strengths $2 k_B T \Gamma$ and $2 k_BT \gamma$ respectively.
The chemical coordinate $n$ of course does not appear in the coupled $X-x$ dynamics, but makes its presence felt only through the active driving proportional to $\Delta\mu$, which cannot be absorbed within any redefinition of the Hamiltonian or the friction coefficient \cite{ramaswamy2017active}. 
If, additionally, we assume $x$ enters $H$ as $kx^2/2$, and drop inertia in both the equations by formally setting masses to zero, we obtain the equation for an active Ornstein-Uhlenbeck particle \cite{uhlenbeck1930theory,fodor2016far} with $(\zeta \Delta \mu / \Gamma) x$ as the coloured noise, with correlation time $\gamma/k$, in the $X$ dynamics, but with an additive white noise $\xi_P$ as well. {A centrosymmetric bistable potential for $x$ with deep wells or, in higher dimensions, a wine-bottle form with a deep trough, is equivalent to the ABP \cite{ebeling1999active}.} Within our single-bath stochastic description 
it is inconsistent to set $\xi_P$ to $0$ without also setting {$\xi_p$} to $0$ in \eqref{mod1aux}. 
The presence of translational diffusion $\xi_P$ in the dynamics of $X$ distinguishes our model, and the resulting stationary state, qualitatively from that of \cite{fodor2016far}. On the other hand, \textit{if we simply abolish  $\xi_P$ in \eqref{mod1mom}, then -- still working with masses set to zero -- for the case where the internal coordinate $x$ is harmonically bound, $\partial_x H = kx$}, this system of equations is equivalent to the second-order stochastic dynamics   
\begin{equation}
\label{almost_eq}
{\Gamma \gamma \over k} \ddot{X} = - \left(\Gamma + {\gamma \over k} \partial_X^2{H}\right) \dot{X} - \partial_X H + {\zeta\Delta\mu \over k}\xi_p
\end{equation}
with Gaussian white noise $\xi_p$. For the special case in which $H$ is a quadratic function of $X$, $\partial_X^2 H = K =$ constant, {and} (\ref{almost_eq}) reduces to the dynamics of a Brownian damped harmonic oscillator with mass ${\Gamma \gamma / k}$ and spring constant $K$, with constant damping coefficient $\Gamma_{eff} = \Gamma + {\gamma K/ k}$ at thermal equilibrium at temperature $T_{eff}=T(\zeta\Delta\mu/k)^2\gamma / \Gamma_{eff}$. In general, as already remarked by \cite{,fodor2016far}, \eqref{almost_eq} represents a nonequilibrium dynamics, whose departure from an effective thermal equilibrium is governed by the anharmonicity of $H$ and not by $\Delta \mu$. However, the above analysis applies only if the additive noise $\xi_P$ in (\ref{mod1mom}) is set to zero. This limit is inconsistent if the noise terms in both equations of motion originate in a common bath, but is in general permissible and important as a case where the magnitude of the effective temperature is governed by the strength of the activity, in a manner analogous to \cite{ramaswamy2000nonequilibrium,Prost1996}. Further, this reduces to a different equilibrium model \cite{,maggi2014generalized,koumakis2014directed} if the $\ddot{X}$ term on the left-hand side of \eqref{almost_eq} is ignored, which corresponds to the unified-coloured-noise approximation \cite{jung1987dynamical,hanggi1995p}{.} 

\subsubsection{Field theory of active polar rods on a substrate: Toner-Tu equation} \label{subsub:polarrodsTT}
We will now use \eqref{genXeqn} and \eqref{genPeqn} as applied to fields to obtain the coarse-grained dynamical equations for a collection of active polar rods on a substrate, thus situating the Toner-Tu theory of flocks\cite{toner1995long,toner1998flocks,Toner2012reanalysis} in the framework of spatial degrees of freedom forced by chemical driving. The rods are described by fields $\rho({\bf r}, t)$ for the density, ${\bf g}({\bf r}, t)$ for the momentum density (or velocity ${\bf v} = {\bf g}/\rho$) and ${\bf w}({\bf r}, t)$ for the polarisation, i.e., the collective vectorial orientation of the rods. The substrate on which the system resides serves as a momentum sink, resulting in a nonzero damping even for a spatially uniform velocity. It also offers a fixed frame of reference, breaking Galilean invariance and thus permitting a constant field-independent contribution to the components $\mathcal{W}_{w_i({\bf x}), g_j({\bf x}')}$ of the matrix of reversible kinetic coefficients. The result is a reactive coupling $\dot{\bf w}\propto {\bf v}$.  
To see this heuristically, note that in two-fluid models \cite{landau1959course}, every phase is advected by its own velocity and not by the velocity of the other phase. A system on a substrate can be regarded as a two-fluid model, viewed from the frame of reference of the second fluid which is treated as a rigid object. 

At a single-particle level, the polarization can be viewed, as in section \ref{subsub:spdimers}, as the relative coordinate of a dimer made of two dissimilar particles. If its centre of mass moves with velocity ${\bf v}$ \emph{relative to an ambient medium, which could simply be a fixed substrate}, the drag experienced by the two ends will in general be unequal. This differential drag implies that the \emph{relative} velocity will also experience a drag proportional to the centre-of-mass velocity, whose components along and normal to the dimer stretch/compress and rotate the dimer. The total force on the relative coordinate ${\bf w}$ is then $-(1/2)(\gamma_1-\gamma_2) {\bf v} - (1/4) (\gamma_1+\gamma_2)\dot{\bf w} +$ conservative forces, if the dimer consists of a pair of equal masses with drag coefficients $\gamma_1$ and $\gamma_2$. 
In a description where inertia is ignored, the rate of change of the polarisation then contains the ``weathercock effect'' \cite{brotto2013hydrodynamics,kumar2014flocking}, a contribution $\Lambda {\bf v}$ where $\Lambda = 2(\gamma_1-\gamma_2)/(\gamma_1+\gamma_2)$, although arising in this approach as the ratio of dissipative coefficients, is a \emph{reversible} kinetic coefficient. 
\footnote{In a \emph{bulk} polar liquid crystal \cite{kung2006hydrodynamics} momentum conservation rules out this coupling, allowing  only $\nabla^2 {\bf v}$ at leading order in gradients and fields. The reader may wonder how the Poisson bracket -- in which the reversible coupling originates -- of the same pair of variables change by introducing a substrate. The answer is that it's not the same pair of variables. The velocity with respect to the substrate is like the relative velocity of the two components of a binary fluid.}

Extending this notion to a many-particle system, the \emph{passive} equations of motion for polar rods on a substrate, with a free-energy $H_0=F[\rho,{\bf w}]+{\bf g}^2/2\rho$, are 
\begin{equation}
\label{eq:cont}
\partial_t\rho=-\nabla\cdot{\bf g}
\end{equation}
\begin{equation}
\partial_t{\bf w}+{\bf v}\cdot\nabla{\bf w}+\boldsymbol{\Omega}\cdot{\bf w }=-\Gamma_w\frac{\delta F}{\delta{\bf w}}+\Lambda{\bf v}+\lambda{\bf w}\cdot{\bsf A}+\boldsymbol{\xi}^w
\end{equation}
\begin{equation}
\label{momeq}
\partial_t{\bf g}+\nabla\cdot({\bf vg})=-\rho\nabla\frac{\delta F}{\delta \rho}-\Gamma{\bf v}+(\nabla{\bf w})\cdot\frac{\delta F}{\delta{\bf w}}+\nabla\cdot\left[{\bf w}\frac{\delta F}{\delta{\bf w}}\right]^A+\lambda \nabla\cdot\left[{\bf w}\frac{\delta F}{\delta{\bf w}}\right]^S-\Lambda\frac{\delta F}{\delta{\bf w}}+\boldsymbol{\xi}_g
\end{equation}
where the noise correlations are given by $\langle\boldsymbol{\xi}^w({\bf x}, t)\boldsymbol{\xi}^w({\bf x}', t')\rangle=2T\Gamma_w{\bsf I}\delta({\bf x}-{\bf x}')\delta(t-t')$ and 
\begin{equation}
\langle\boldsymbol{\xi}_g({\bf x}, t)\boldsymbol{\xi}_g({\bf x}', t')\rangle=2T{\bsf I}\Gamma\delta({\bf x}-{\bf x}')\delta(t-t'),
\end{equation}
and $\boldsymbol{\Omega}$ and ${\bsf A}$ are the anti-symmetric and symmetric parts of the velocity gradient tensor respectively. 

As advertised, we now write down the corresponding dynamical equation in the presence of a chemical driving $\Delta\mu$ in the Hamiltonian, now written as {$H=H_0+\int_{\bf r}n({\bf r})\Delta\mu({\bf r})$}, considering a reactive coupling between the momentum ${\bf g}({\bf r}, t)$ and the chemical variable $n({\bf r},t)$. This appears in the ${\bf g}$ equation through the term
\begin{equation}
\label{eq:mom_chem_rev_term}
\int_{{\bf r}'} \mathcal{W}_{{\bf g}({\bf r}), n({\bf r}')}\frac{\delta H}{\delta n({\bf r}')}, 
\end{equation}
where the vector $\mathcal{W}_{{\bf g}({\bf r}), n({\bf r}')}$ has to be even under time-reversal. Therefore, to the lowest order in gradients,
\begin{equation}
\label{eq:mom_chem_rev_Onsager}
\mathcal{W}_{{\bf g}({\bf r}), n({\bf r}')}=-\zeta_w{\bf w}{({\bf r})}\delta({\bf r}-{\bf r}')
\end{equation}
where {$\zeta_{w}$} is a phenomenological kinetic coefficient{\footnote{At first order in gradients and with one field, we only have $\nabla\rho$. At first order in gradients and two fields, we have three terms with two ${\bf w}$ and one gradient, and three terms with two ${\bf v}$ and one gradient. Terms like ${\bf w}\cdot\nabla{\bf v}$ are also allowed by the spatial symmetries of this system which lacks Galilean invariance, but by time-reversal symmetry \emph{can not} arise from a reactive coupling between $n$ and ${\bf g}$. Instead, they arise through the dissipative coupling between $n$ and ${\bf g}$, $-\Gamma_{gn}\delta H/\delta n$.  $\Gamma_{gn}$ can contain all vectors containing an odd power of ${\bf v}$ {and, in} particular, contains terms such as ${\bf w}\cdot\nabla{\bf v}$.}}

Inserting the coupling \eqref{eq:mom_chem_rev_term} and \eqref{eq:mom_chem_rev_Onsager} into \eqref{momeq}, creating an active state by requiring $\delta H / \delta n({\bf r}) = \Delta \mu $, i.e., imposing a spatially uniform, constant, nonzero reactant-product chemical-potential difference $\Delta \mu$, and discarding inertia, we get
\begin{equation}
\Gamma{\bf v}=\zeta_w\Delta\mu{\bf w}-\nabla\rho-\Lambda\frac{\delta F}{\delta{\bf w}}+\boldsymbol{\xi}_g, 
\end{equation}
so that the continuity equation \eqref{eq:cont} becomes
\begin{equation}
\partial_t\rho=-\nabla\cdot\left(\frac{\zeta_w\Delta\mu}{\Gamma}\rho{\bf w}\right)+\frac{\Lambda}{\Gamma}\nabla\cdot\left(\rho\frac{\delta F}{\delta{\bf w}}\right)-\nabla\cdot\left[\frac{\rho}{\Gamma}(-\nabla\rho+\boldsymbol{\xi}_g)\right]
\end{equation}
and the polarisation equation is 
\begin{equation}
\partial_t{\bf w}=\left(\frac{(\lambda-1)\zeta_w}{2\Gamma}\right)\Delta\mu{\bf w}\cdot\nabla{\bf w}-\frac{\lambda}{\Gamma}{\bf w}\cdot\nabla\nabla\rho+\left(\frac{(\lambda-1)\zeta_p}{4\Gamma}\right)\Delta\mu\nabla{\bf w}^2+\frac{1}{\Gamma}\nabla\rho\cdot\nabla{\bf w}-\frac{\Lambda}{\Gamma}\nabla\rho+\left(\frac{\Lambda\zeta_w}{\Gamma}\right)\Delta\mu{\bf w}-\left(\Gamma_w+\frac{\Lambda^2}{\Gamma}\right)\frac{\delta F}{\delta{\bf w}}+\boldsymbol{\xi}
\end{equation}
where the noise correlator, to the zeroth order in gradients, is
$
\langle\boldsymbol{\xi}({\bf x},t)\boldsymbol{\xi}({\bf x}',t')\rangle =2T\left[\Gamma_w+({\Lambda^2}/{\Gamma})\right]\delta({\bf x}-{\bf x}')\delta(t-t').
$
We see thus that our general framework generates coupled equations for the polar order parameter ${\bf w}$ and the density $\rho$ with the two characteristic features of the Toner-Tu equation \cite{toner1995long,toner1998flocks}: 1. a self-advective nonlinearity and 2. a {number} current proportional to the polarisation \emph{with an independent coefficient}. {We discuss in greater detail in \ref{subsub:velalignTT} why the simultaneous presence of a self-advective coefficient and a particle current proportional to polarisation with an independent coefficient inevitably breaks detailed balance}.
Note that the noise in the polarisation and concentration equation are not independent since a part of the noise in the ${\bf w}$ equation is derived from the noise in the momentum equation. However, this cross-correlation is $\mathcal{O}(\nabla)$ and turns out to be irrelevant compared to the wavevector-independent noise in the ${\bf w}$ -- the correct low frequency and small wavevector scaling of all correlation functions can be obtained by completely ignoring the noise in the density equation (except perpendicular to the wavevector, but in that case ${\bf w}$ and $\rho$ are linearly decoupled).

%

\subsection{Dissipative coupling between physical momentum and chemical variable}
\label{sub:irreversiblechemforce}
\subsubsection{Particle with preferred speed}
\label{Hallpart}
Distinct from the AOUP we discussed in section \ref{subsub:spdimers}, a particle can be active by possessing a preferred speed obtained through the interplay of a negative linear damping and a positive nonlinear one \cite{erdmann2000brownian,ebeling1999active,schweitzer1998complex,ebeling1999w}. We now show how a dissipative coupling between the physical momentum and the chemical coordinate can create a realisation of such a particle. Working for simplicity in one dimension, consider a particle with position and momentum $X,P$, with equation of motion 
\begin{equation}
\dot{X}=P; \,\,\,\,\,\,\,\,\, \dot{P}=-\partial_XH-\Gamma_{PP} P-\Gamma_{Pn}\partial_n H+\xi_P=-\partial_XH-\Gamma_{PP} P+\zeta\Delta\mu P+\xi_P
\end{equation}
in the notation of section \ref{sec:gendev}, where we have taken $\Gamma_{Pn}=-\zeta P$. If we now take $\Gamma_{PP}=\gamma_{1}+\gamma_2P^2$, we obtain 
\begin{equation}
\label{Hallparticle}
\dot{P}=-\partial_XH-(\gamma_1-\zeta\Delta\mu+\gamma_2P^2)P+\xi_P
\end{equation}
Note that the $P$-independent part of the friction is now liberated from the noise strength because of the $\Delta\mu$-dependent term, which can even make it negative. The noise is multiplicative with the correlator {$\langle\xi_P(0)\xi_P(t)\rangle=2T(\gamma_{1}+\gamma_2p^2)\delta(t)$}. This implies that \eqref{Hallparticle} must be interpreted using the anti-{\ito} interpretation. In a general interpretation, one must add a term $2T\gamma_2(1-\alpha)P$ to the R.H.S. of \eqref{Hallparticle}, where $\alpha$ parameterizes the interpretation, with $\alpha=0$ for {\ito}, $1/2$ for Stratonovich, and $1$ for anti-{\ito} \cite{lau2007state}.
Despite superficial appearances, however, this noise-induced drift can not turn the $P$-independent part of the friction negative. In all interpretations, the probability distribution of $P$ is Maxwellian if $\Delta \mu =0$, and the particle is not self-propelled. If $\zeta >\gamma_{1}/ \Delta \mu $, the zero-velocity state is on average unstable, and the particle moves with preferred velocity $\pm \sqrt{(\zeta \Delta \mu - \gamma_{1})/\gamma_2}$ in the limit of low noise. We have thus succeeded in producing a model self-propelled particle with inertia within our consistent stochastic framework.

{The reader may wonder whether the requirement of equilibrium dynamics when a nonzero average $\partial_n H$ is not imposed places any restriction on the magnitude of $\zeta$. It does not: the coupled dissipative dynamics of $P$ and $n$ is given by
\begin{equation}
\nonumber
\partial_t\begin{pmatrix}P\\n\end{pmatrix}=\begin{pmatrix}\gamma_1+\gamma_2 P^2 & -\zeta P\\-\zeta P &\Gamma_{nn}\end{pmatrix}\begin{pmatrix}\partial_PH\\\partial_nH\end{pmatrix}+\begin{pmatrix}\xi_P\\\xi_n \end{pmatrix}+ \text{NID}
\end{equation}
where NID denotes noise-induced drift. The friction matrix must be positive definite. This yields the condition $\Gamma_{nn}(\gamma_1+\gamma_2P^2)/P^2>\zeta^2$. This condition is obviously fulfilled for small $P$, while at large $P$ it requires $\Gamma_{nn}\gamma_2>\zeta^2$. However, $\Gamma_{nn}$ doesn't enter into the equation for $P$ and therefore, this condition does not pose any restriction for $\zeta$ -- it simply provides a lower limit to the friction coefficient in the $n$ equation which becomes immaterial once $\partial_n H$ is externally held constant. The noise-induced drift that is required for this system to be in equilibrium can also be calculated from this (see earlier footnote). After a little algebra, we see that the noise-induced drift in the $P$ equation is, as expected, $2T\gamma_2(1-\alpha)P$.
}
\subsubsection{Apolar active particle}
\label{apoact}
Macroscopic self-driven systems with alignment but no polarity -- active nematics -- have received a great deal of attention in the literature \cite{simha2002hydrodynamic}, but studies of the single-particle behaviour have focused on polar particles. A single apolar active particle executes uncorrelated random back-and-forth movement preferentially along an axis, as seen in certain elongated cells \cite{gruler1999nematic} or in a vertically vibrated rod lying on a surface \cite{kumar2014flocking}. This amounts to the kinetic energy being unequally partitioned between the components of the velocity along and transverse to the rod. We show that this apparently two-temperature dynamics emerges naturally from our single-bath framework. 

Consider a particle with position ${\bf X}$, momentum ${\bf P}$ and intrinsic anisotropy characterised by the apolar tensor ${\bsf Q}$, and chemical coordinate $n$, governed by a Hamiltonian $H$. We take a simple relaxational model for ${\bsf Q}$: 
\begin{equation}
\label{eq:Q_rot_diff}
\dot{{\bsf Q}}=-\Gamma_{QQ}{\delta H \over \delta{\bsf Q}}+\xi_Q
\end{equation}
with $\langle\xi_Q(0)\xi_Q(t)\rangle=2T\Gamma_{QQ}\delta(t)$, reducing to simple rotational diffusion in the limit where $\delta H / \delta{\bsf Q} \propto {\bsf Q}$. ${\bsf Q}$ can enter the dissipative cross-kinetic coefficient between ${\bf P}$ and $n$ in combination with ${\bf P}$:
\begin{equation}
\label{eq:apolar_particle}
\dot{{\bf X}}={\bf P}; \,\,\,\,\,\,\,\,\, \dot{{\bf P}}=-\partial_{\bf X}H-\Gamma_{PP} \partial_{\bf P}H-\Gamma_{Pn} \partial_n H +\xi_P=-\partial_{\bf X}H-  -\Gamma_{PP} \partial_{\bf P}H   -\zeta\Delta\mu {\bsf Q}\cdot{\bf P}+\xi_P
\end{equation}
where we have taken $\Gamma_{Pn}=\zeta{\bsf Q}\cdot{\bf P}$. This implies that the momentum is damped anisotropically while the noise correlator $\langle\xi_P(t)\xi_P(t')\rangle=2T \Gamma_{PP} \delta(t-t')$ is isotropic. In other words, this is like having an anisotropic temperature at the scale of individual particles \cite{mishra2010dynamics}, yet achieved within our isothermal framework where the driving is imposed only through $\Delta \mu$. The inertia-free limit of the dynamics, more familiar from active nematic theories {\cite{SRAditiToner2003nematic}}, can be obtained by adiabatic elimination of ${\bf P}$ from \eqref{eq:Q_rot_diff} and \eqref{eq:apolar_particle}. We will explore the phase behaviour and other statistical properties of {collections} of such particles elsewhere. 
\footnote{Note that though the kinetic coefficient explicitly depends on ${\bf P}$, which implies that the noise cross-correlation in the coupled $n-{\bf P}$ equation is multiplicative, there is no noise induced drift in either equation even when there is no external driving force on $n$. This is due to the symmetric traceless character of ${\bsf Q}$.}

\subsubsection{Active field theory with velocity alignment} \label{subsub:velalignTT}
In section \ref{subsub:polarrodsTT}, we had derived Toner-Tu equations for a system of polar rods, whose orientation vector dictated their preferred velocity via the active motility which arose as a reactive kinetic coefficient between the momentum density and the chemical variable. However, there are driven systems without an independent polarity where nevertheless a spontaneous transition occurs to a state with non-zero mean velocity. A particularly striking example of this is the zero-resistance state in microwave-driven 2DEG systems in a magnetic field \cite{mani2002,zudov2003evidence}. The Toner-Tu-like theory of such a system \cite{alicea2005transition} is best viewed as an active extension of the Navier-Stokes equations (modified by contact with a substrate), not an active version of polar liquid crystal dynamics, as there is no underlying orientational degree of freedom. 

As in \eqref{Hallparticle}, the crucial ingredient in such a theory is a negative linear friction which has to enter as a off-diagonal dissipative coupling between the momentum density field ${\bf g}({\bf r}, t)$ and $n({\bf r}, t)$. That is, the standard Navier-Stokes equation (together with substrate damping) is modified as
\begin{equation}
\label{deneq2}
\partial_t\rho=-\nabla\cdot{\bf g}
\end{equation}
\begin{equation}
\label{momeq2}
\partial_t{\bf g}+\nabla\cdot({\bf vg})=-\rho\nabla\frac{\delta F}{\delta \rho}-\Gamma_{gg}{\bf v}+[\eta\nabla^2  + (\zeta + \eta/3) \nabla \nabla \cdot ]{\bf v} -\Gamma_{gn}\frac{\delta H}{\delta n}+\boldsymbol{\xi}_g
\end{equation}
with $H=F[\rho]+\int_{\bf x}{\bf g}^2/2\rho+n\Delta\mu$, where $F$ is the purely $\rho$-dependent part of the free-energy, and
$
\langle\boldsymbol{\xi}_g({\bf r}, t)\boldsymbol{\xi}_g({\bf r}', t')\rangle=2T{\bsf I}[\Gamma_{gg}- \eta\nabla^2  - (\zeta + \eta/3)\nabla \nabla] \delta({\bf r}-{\bf r}')\delta(t-t').
$
Again, following \eqref{Hallparticle}, we take $\Gamma_{gg}=\gamma_1+\gamma_2 {\bf v}\cdot{\bf v}$ and {take\footnote{At next order in gradients, $\Gamma_{gn}$ admits a contribution proportional to $\nabla^2{\bf v}$ which, if it enters with a negative prefactor, implies a negative effective viscosity coefficient, spontaneously generating velocity gradients \cite{dunkel2013fluid}, controlled at larger wavenumbers by ``hyperviscosity'', i.e., a stabilising damping of order $\nabla^4$. The result is a vector variant of the Swift-Hohenberg equation \cite{swift1977hydrodynamic} which, when the linear friction coefficient is also {negative, presumably} displays regular modulated states such as stripes, as well as spatiotemporal chaos.} $\Gamma_{gn}=-\zeta{\bf v}$.}

The nonlinear damping $\propto \gamma_2$ again implies that the noise is multiplicative and \eqref{momeq2}{, as it stands,} has to be interpreted using an anti-{\ito} interpretation. In any other interpretation, a noise-induced drift $2(1-\alpha)T{\bf v}/\rho$ has to added. Note however, that unlike the $\zeta$ term, which when the chemical field is driven with a constant force, i.e. $\partial_nH =\Delta\mu$, is liberated from noise strength, this term is intimately connected to the noise and is required precisely to ensure that the steady state distribution of velocities {remains Maxwellian in the equilibrium limit} irrespective of the noise interpretation. Thus, while {$\zeta>\gamma_1/\Delta\mu$ linearly destabilises the zero velocity state, the noise-induced drift} has no such effect. The fluid acquires a preferred mean {speed with respect to the substrate} in this case which in the low temperature limit is {$[(\zeta\Delta\mu-\gamma_1)/\gamma_2]^{1/2}$}. This therefore is an active variant of the dynamics of a fluid in contact with a momentum sink or substrate \cite{ramaswamy1982linear}. It may be argued that this is still distinct from the Toner-Tu equation since the coefficient of the advective terms in the equations for number and momentum density still has a coefficient $1$. The modification of this requires a \emph{further active} contribution, which due to time-reversal symmetry of the coupled unforced ${\bf g}-n$ dynamics, has to enter not through the dissipative coupling at all but via $\mathcal{W}_{{\bf g}({\bf r})n({\bf r}')}$ [see \eqref{eq:mom_chem_rev_term}], which by symmetry, can contain all even vectors and, in particular, can contain terms such as $\lambda[\nabla\cdot({\bf gv})]({\bf r})\delta({\bf r}-{\bf r}')$ which leads to the term $\lambda\Delta\mu\nabla\cdot({\bf vg})$ in the ${\bf g}$ equation. 

{One may wonder whether a non-unit advective coefficient has to neccessarily break detailed balance. After all. in equilibrium, the advection term in the momentum equation arises from the reactive $\mathcal{W}_{g_i({\bf r})g_j({\bf r}')}$ coupling. The coefficient of this coupling is forced to be $1$ because of Galilean invariance (and momentum conservation). However, for fluids on substrates, which neither conserve momentum nor have Galilean invaraince, it may be argued that the antisymmetric coupling $\mathcal{W}_{g_i({\bf r})g_j({\bf r}')}$, where ${\bf g}$ is now the momentum \emph{relative to the substrate}, can have a non-unit coefficient $\lambda$. This would still result in a stochastic differential equation with a current-free steady-state with a distribution $e^{-H/T}$ i.e., this would not break detailed balance. However, looking at the form of $\mathcal{W}_{g_i({\bf r})g_j({\bf r}')}$ in detail \cite{kim1991equations}, we find that beyond the advective term, it also generates another term $\propto \rho\nabla({\bf v}^2/2)$. In standard hydrodynamics, this second term (whose presence in the equations of motion would have implied a lack of momentum conservation and Galiliean invariance) is \emph{exactly} cancelled by a term arising from the $\mathcal{W}_{{\bf g}({\bf r})\rho({\bf r}')}$ coupling. Therefore, multiplying the reactive coupling $\mathcal{W}_{g_i({\bf r})g_j({\bf r}')}$ by $\lambda$, without modifying the $\mathcal{W}_{{\bf g}({\bf r})\rho({\bf r}')}$ coupling leads to an extra term $(\lambda-1)\rho\nabla({\bf v}^2/2)$ in addition to an advective nonlinearity $\lambda\nabla\cdot({\bf vg})$. The \emph{absence} of this extra nonliearity with a related coefficient (with the relation being preserved under renormalisation) leads to the breaking of detailed balance. If we seek to remove this second nonlinearity with a symmetry-related coefficient by also multiplying the $\mathcal{W}_{{\bf g}({\bf r})\rho({\bf r}')}$ coupling by $\lambda$, we obtain again obtain a model which obeys detailed balance but in which the $\nabla\cdot({\bf vg})$ and the $\rho\nabla\delta F/\delta\rho$ terms in the momentum equation and the $\nabla\cdot{\bf g}$ term in the density are all multiplied by the \emph{same} coefficient $\lambda$. 
\footnote{{If in addition, the friction coefficient $\Gamma_{gg}$ is set to $0$, this model posseses Galilean invariance and conserves momentum. That is, multiplying both the $\mathcal{W}_{g_i({\bf r})g_j({\bf r}')}$ and the $\mathcal{W}_{{\bf g}({\bf r})\rho({\bf r}')}$ couplings by $\lambda$ does not spoil either momentum conservation or Galilean invariance.}}
Thus, velocity and density advection at different speeds, in the absence of other symmetry-related nonlinearities, requires breaking of detailed balance.
}


{It must be emphasised however that a model with a negative linear friction and \emph{unit} advective coefficient 
\footnote{{A model with the mass current proportional to ${\bf g}$, the advective nolinearity and the pressure term are all multiplied by $\lambda$ or one in which the advective nonlinearity $\lambda\nabla\cdot({\bf vg})$ is accompanied by $(\lambda-1)\rho\nabla({\bf v}^2/2)$ also belong to the same universality class.}}
still belongs to the same universality class as Toner-Tu \cite{toner1998flocks,toner2005hydrodynamics,Toner2012reanalysis} what is crucial is not that the advective coefficient can have an arbitrary value, but that it is present at all in a model that can spontaneously break symmetry. For such a system, any non-zero value of the advective nonlinearity, including $1$, leads to long range order in two dimensions. Moreover, in at least the  ``Malthusian'' \cite{toner2012birth} variant of Toner-Tu {in two dimensions}, this vertex is actually protected from renormalisation by an exact symmetry (pseudo-Galilean invariance) which is analogous to the Galilean symmetry that protects this vertex in the Navier-Stokes equation.}

\section{Quantification of nonequilibrium behaviour}
\label{sec:entpo}
In this section we will discuss how to quantify the degree of nonequilibriumness in some of the models discussed in the last section. To accomplish this, we define the entropy production rate as
\begin{equation}
\label{entproddefn}
\sigma=\lim_{t\to\infty}\frac{1}{t}\mathcal{S}
\end{equation}
{where the entropy production} \cite{lebowitz1999gallavotti}
\begin{equation}
\mathcal{S}=\left\langle\ln\frac{\mathcal{P}}{\mathcal{P}^R}\right\rangle
\end{equation}
is given by the Kullback-Leibler (KL) divergence \cite{kullback1951information} that measures the distinguishability of the probability weight associated with a noise realisation to generate a forward trajectory $\mathcal{P}$ and the weight {$\mathcal{P}^R$ of the noise-realisation required to obtain the time-reversed counterpart of that trajectory}. The angle brackets denote an average over noise realisations. However, under suitable ergodicity assumptions, which we implicitly use throughout the paper, the average over noise realisations can be replaced by the average over a single infinitely long noise realisation and, therefore, the angle brackets can be dropped \cite{fodor2016far,nardini2017entropy}.

The entropy production rate that we have defined distinguishes a forward trajectory from its time-reversal counterpart. In equilibrium, time-reversal symmetry implies that the two should be indistinguishable. Therefore, $\sigma$ provides a measure of how far this system is out of equilibrium. This measure crucially depends on the definition of the time-reversal operation, however.
In most cases the requirement that the dynamics in the limit $\Delta\mu=0$ should have detailed balance fixes the definition of this operation  but, {when distinct time reversal schemes {give} zero entropy production in {the} $\Delta\mu=0$ limit, {more than one equilibrium limit exists} (appendix \ref{appE:2dof}}). 
Therefore, in such cases, the answer to the question ``how far from equilibrium is an active system" is not unique

\subsection{Nonequilibrium dynamics of active particles}
We will now quantify the degree of nonequilibriumness of active polar particles discussed in \ref{subsub:spdimers}.
The first step in calculating the entropy production rate and all the subsequent analysis consists in writing the path probability of the particle. Since the noises in both \eqref{mod1mom} and \eqref{mod1aux} are Gaussian, the 
{path probability weight \cite{onsager1953fluctuations} takes the form $\mathcal{P}=e^{-\mathcal{A}}$, where, for the} {inertia-less limit of} {the equations of motion \eqref{mod1mom} and \eqref{mod1aux}, the stochastic action in Onsager-Machlup \cite{onsager1953fluctuations,cugliandolo2017rules} form is} 
\begin{equation}
\label{act1}
\mathcal{A}=\int_{t_i}^{t_f} dt\left[\frac{1}{4T\Gamma}\left(\Gamma\dot{X}+\frac{\partial{H}}{\partial X}-\upsilon x\right)^2+\frac{1}{4T\gamma}\left(\gamma\dot{x}+kx\right)^2-\alpha\left(\frac{\partial^2{H}}{\partial X^2}+1\right)\right], 
\end{equation}
where we have defined $\upsilon\equiv\zeta\Delta\mu$ and taken $\partial H/\partial x=kx$. The term with $\alpha\in[0,1]$ comes from the Jacobian for the variable transformation from the noise $(\xi_P,\xi_p)$ to $(X, x)$, {with $\alpha=0, 1/2, 1$ corresponding respectively to the {\ito}, Stratonovich and anti-{\ito} discretisations }. 
{Let us define $\mathcal{T}_1$ to be the time-reversal operation in the normal sense of that term, i.e., without a flip of the polarity or relative coordinate, i.e., $\mathcal{T}_1x=x; \mathcal{T}_1X=X$}

The weight associated with the time reverse of this, when the time-reversal operation $\mathcal{T}_1$ \emph{does not} involve a polarity-flip, i.e., $\mathcal{T}_1x=x; \mathcal{T}_1X=X$, is given by the action
\begin{equation}
\mathcal{T}_1\mathcal{A}=\mathcal{A}_R=\int_{t_i}^{t_f} dt\left[\frac{1}{4T\Gamma}\left(-\Gamma\dot{X}+\frac{\partial{H}}{\partial X}-\upsilon x\right)^2+\frac{1}{4T\gamma}\left(-\gamma\dot{x}+kx\right)^2-(1-\alpha)\left(\frac{\partial^2{H}}{\partial X^2}+1\right)\right]
\end{equation}
Therefore, the entropy production is 
\begin{equation}
\label{ent_prod}
\mathcal{S}_1=\mathcal{A}_R-\mathcal{A}=-\frac{1}{T}[H(t_i)-H(t_f)+\frac{\upsilon}{T}\int_{t_i}^{t_f} dt\dot{X}x
\end{equation}
where to obtain the last equality we have used the generalised {\ito} formula for the time derivative, which for any function $G[q_i(t)]$, where $q_i(t)$ {obey first-order dynamics with an additive white noise}, is
$
\dot{G}=\dot{q_i}\partial_{q_i}G+\frac{(1-2\alpha)}{2}D_{ij}(q_k)\partial_{q_i}\partial_{q_j} G,
$
where $D_{ij}$ is the noise correlation matrix. Since the interpretation does not change under $\mathcal{T}_1$ for $\alpha=1/2$, and the {\ito} formula reduces to the standard chain rule, we will only use the Stratonovich interpretation from now on. However, our results are, of course, invariant under the change of interpretations.
We now use \eqref{entproddefn} to calculate the entropy production rate. Since $H$ is bounded, the contribution from it vanishes in $\sigma$. Replacing the time-average in the second term in \eqref{ent_prod} with an average over the stationary measure we obtain
\begin{equation} 
\label{entroprod_noflip}
\sigma_1=\frac{\upsilon}{T}\langle\dot{X}x\rangle.
\end{equation}
For a potential {quadratic} in $X$, $\partial H/\partial X=KX$, this reduces to
\begin{equation} \label{sigma_quad_noflip}
\sigma_1=\frac{\upsilon^2}{\Gamma k+\gamma K}
\end{equation}

If one defines a time-reversal operation $\mathcal{T}_2$ that does involve a polarity flip, i.e., $\mathcal{T}_2X=X; \mathcal{T}_2x=-x$, the time-reversed action is 
\begin{equation}
\mathcal{T}_2\mathcal{A}=\mathcal{A}_R=\int_{t_i}^{t_f} dt\left[\frac{1}{4T\Gamma}\left(-\Gamma\dot{X}+\frac{\partial{H}}{\partial X}+\upsilon x\right)^2+\frac{1}{4T\gamma}\left(-\gamma\dot{x}+kx\right)^2-(1-\alpha)\left(\frac{\partial^2{H}}{\partial X^2}+1\right)\right]
\end{equation}
from which we can calculate the entropy production rate
\begin{equation}
\label{entroprod_flip}
\sigma_{2}=\frac{\upsilon}{\Gamma T}\left\langle\frac{\partial H}{\partial X}x\right\rangle.
\end{equation}
For $H$ {quadratic} in $X$, this yields
\begin{equation} \label{sigma_quad_flip}
\sigma_{2}=\frac{\upsilon^2 K\gamma}{k\Gamma(\Gamma k+\gamma K)}
\end{equation}
As pointed out {in \cite{pietzonka2017entropy}, the sum of the two entropy productions is a potential independent constant $\upsilon^2/(k\Gamma)$. Note that \eqref{sigma_quad_noflip} and \eqref{sigma_quad_flip} coincide when the relaxation time of $X$ equals the correlation time of $x$, i.e., when $\Gamma/K = \gamma/k$. Intriguingly, this is precisely the condition of critical damping for the equivalent damped harmonic oscillator limit of \eqref{almost_eq}.} Note that both definitions of the time-reversal operation, $\mathcal{T}_1$ and $\mathcal{T}_2$ lead to a non-vanishing entropy production for a quadratic $H$, in contrast to \cite{fodor2016far}. {The reason, as we have already pointed out, is that the centre of mass $X$ is in general subjected to an additive white noise, so that the particle samples a statistical steady state rather different from that in the special case considered in \cite{fodor2016far}.} However, it is important to note that this non-zero entropy production only emerges when one considers the {\textit{joint} path} probability distribution of $X$ and $x$. {We now show that the entropy-production rate vanishes for the dynamics in a quadratic potential when only the dynamics of $X$ is recorded. 
To see this let us rewrite the dynamics for $X$ as
$
\Gamma\dot{X}=-\partial_XH+\theta
$
where $\theta$ is a zero-mean Gaussian noise with the correlation
$
\langle\theta(t)\theta(s)\rangle=2T\Upsilon(t-s)
$.
The path measure is determined by the kernel {$\mathcal{G}$} defined by
$
\int ds{\mathcal{G}}(t-s)\Upsilon(s-r)=\delta(t-r)
$
We now define a stochastic action over an infinite time-interval $-\infty<t<\infty$:
\begin{equation}
\label{cnaction}
\mathcal{A}=\frac{1}{4T}\int_{-\infty}^{\infty} dt\int_{-\infty}^{\infty}  ds \,\theta(s)\mathcal{G}(t-s)\theta(t) =\frac{1}{4T}\int_{-\infty}^{\infty} \frac{d\omega}{2\pi}\theta_{-\omega}\mathcal{G}_\omega\theta_\omega.
\end{equation}
For a process driven by a combination of two uncorrelated noise sources, one white and the other coloured, we take $\theta=\eta+\upsilon x$ where $\eta$ and $x$ are uncorrelated to each other and obey 
$
\langle \eta_{\omega}\eta_{\omega'} \rangle=2T\Gamma 2 \pi \delta(\omega + \omega')
$
and
\begin{equation}
\langle x_{\omega}x_{\omega'} \rangle =\frac{2T\gamma}{k^2+\gamma^2\omega^2}  2 \pi \delta(\omega + \omega').
\end{equation}
However, now we only record the dynamics of $X$ and not the auxiliary variable.
The noise correlation is 
\begin{equation}
2T\Upsilon_\omega=2T\left(\Gamma+\frac{\upsilon^2\gamma}{k^2+\gamma^2\omega^2}\right)
\end{equation} 
and
\begin{equation}
\label{cnwncorr}
\mathcal{G}_\omega=\left(\frac{k^2+\gamma^2\omega^2}{\Gamma k^2+\Gamma\gamma^2\omega^2+\upsilon^2\gamma}\right)
\end{equation}
We now subtract $1/\Gamma$ {($= \mathcal{G}_{\omega}|_{\upsilon = 0}$) from $\mathcal{G}_\omega$ to obtain
\begin{equation}
\tilde{\mathcal{G}}_\omega \equiv \mathcal{G}_\omega - \frac{1}{\Gamma} = -{{\upsilon^2\gamma}/{\Gamma^2} \over \gamma^2\omega^2+k^2+\upsilon^2{\gamma}/{\Gamma}}
\end{equation}}
We will henceforth only consider $\tilde{\mathcal{G}}_\omega$ since {it must contain }the entire entropy production. We now expand it in a series in {$\omega^2$}:
\begin{equation}
\label{expan}
\tilde{\mathcal{G}}_\omega=\frac{\upsilon^2\gamma}{\Gamma^2}\left[-\left(\frac{\Gamma}{\Gamma k^2+\upsilon^2\gamma}\right)+\gamma^2\omega^2\left(\frac{\Gamma}{\Gamma k^2+\upsilon^2\gamma}\right)^2-\gamma^4\omega^4\left(\frac{\Gamma}{\Gamma k^2+\upsilon^2\gamma}\right)^3+...\right]
\end{equation}
This expansion can not, in general, be naturally truncated at any order, and is therefore only of formal value. If we assume that derivatives of the fields at all orders vanish as $\to\pm\infty$, we can transform to the field coordinates (we will ignore the Jacobian contribution since that remains even in time) and to {real} time to obtain the action 
\begin{multline}
\mathcal{A}=\frac{1}{4T}\frac{\upsilon^2\gamma}{\Gamma^2}\int dt\bigg[-\left(\frac{\Gamma}{\Gamma k^2+\upsilon^2\gamma}\right)\left(\Gamma\dot{X}+\partial_XH\right)^2+\gamma^2 \left(\frac{\Gamma}{\Gamma k^2+\upsilon^2\gamma}\right)^2\left\{\partial_t\left(\Gamma\dot{X}+\partial_XH\right)\right\}^2 \\-\gamma^4\left(\frac{\Gamma}{\Gamma k^2+\upsilon^2\gamma}\right)^3\left\{\partial_t^2\left(\Gamma\dot{X}+\partial_XH\right)\right\}^2+...\bigg].
\end{multline}
Noting that under the time-reversal operation $\mathcal{T}$, $\mathcal{T}X\to X$, and all total derivative terms vanish since the fields and all their derivatives vanish at the boundaries, we find (after repeated integration by parts) the KL divergence 
\begin{equation}
\mathcal{T}\mathcal{A}-\mathcal{A}=\frac{\upsilon^2\gamma}{T\Gamma}\int dt\bigg[\gamma^2 \left(\frac{\Gamma}{\Gamma k^2+\upsilon^2\gamma}\right)^2(\partial_t^3X\partial_XH)+\gamma^4\left(\frac{\Gamma}{\Gamma k^2+\upsilon^2\gamma}\right)^3(\partial_t^5X\partial_XH)+...\bigg]
\end{equation}
Now, if $H$ is harmonic in $X$, $\partial_XH {\propto} X$. We see that in this case the KL divergence vanishes term by term. Each term is of the form $(\partial_t^{2n+1}X)X$, where $n=1,2,3...$. After integrating by parts $n$ times, this becomes $\partial_t^{n+1}X\partial_t^nX=(1/2)\partial_t(\partial_t^n X)^2$. This vanishes since $\partial_t^n X$ vanishes at $t\to\pm\infty$. Therefore, the entropy production due to a particle in a harmonic trap vanishes when the dynamics of the auxiliary variable is not kept track of. This is the case even if we explicitly consider translational diffusion. More generally, this result holds for any arbitrary (Gaussian) noise. $\mathcal{G}_\omega$ can always be expanded in the form of \eqref{expan} as
$
\mathcal{G}_\omega=\nu_1+\nu2_2\gamma^2\omega^2+\nu_3\gamma^4\omega^4+...
$
and the later steps follow trivially, with only the coefficient of each term now being $\nu_i$.}

{This validates our assertion that if one wants to know whether a particle in a harmonic trap is in equilibrium or not, one must record the dynamics of the auxiliary variable. Recording just the position of the particle is insufficient to distinguish between equilibrium and nonequilibrium dynamics. 
\footnote{Another equivalent way to obtain this result is to marginalise the distribution $\mathcal{P}(X, x)=e^{-\mathcal{A}}=\mathcal{P}_X\mathcal{P}_x\mathcal{P}_{Xx}=e^{-(\mathcal{A}_X+\mathcal{A}_x+ \mathcal{A}_{Xx})}$ with respect to $x$ i.e., obtain $\mathcal{P}(X)|_{x}=\mathcal{P}_X\langle\mathcal{P}_{Xx}\rangle_{\mathcal{P}_x}$ and use this to calculate the entropy production.}}
This also remains true even if one considers underdamped particles or particles driven by arbitrarily coloured Gaussian noise (appendix  \ref{appinertia}).

While we have calculated the expressions for the entropy production rate of a single polar particle, it is conceptually interesting to relate it to the violation of a fluctuation-dissipation relation. Further, since correlation and response functions are {quantities that are accessible and relevant to experiment}, such a relation would provide a way to directly measure the entropy production rate. 
Thus, we now calculate a Harada-Sasa {relation (HSR)} for the entropy production \cite{harada2005equality, fodor2016far, nardini2017entropy,teramoto2005microscopic,harada2006energy,yamada2015unified}. This additionally allows us to access a \emph{frequency-resolved} entropy production which {directly shows us the timescales associated with the processes contributing to entropy production.}

\subsubsection{HSR when the time-reversal operation doesn't involve a polarity flip}
We first derive the HSR for the case in which the time-reversal operation does not involve a polarity flip.
{We use the MSRJD \cite{martin1973pc,janssen1976lagrangean,janssen1979field,de1976c,de1978c,aron2010symmetries} formalism, which is particularly well adapted to calculating response functions, }to write the  action in \eqref{act1} in the Stratonovich interpretation 
\footnote{Note however, that the MSRJD and OM actions are completely equivalent and one can go from the OM action to the MSRJD one via a  Hubbard-Stratonovich transformation while one can obtain the OM action from the MSRJD one by simply integrating out the response fields.}
\begin{equation}
\label{action_OM_forward}
\mathcal{A}=\int dt\left[T(\Gamma\tilde{X}^2+\gamma\tilde{x}^2)-\frac{1}{2}\left(\frac{\partial^2{H}}{\partial X^2}+1\right)-i\tilde{X}\left(\Gamma\dot{X}+\frac{\partial{H}}{\partial X}-\upsilon x\right)-i\tilde{x}\left(\gamma\dot{x}+kx\right)\right]
\end{equation}
where we have introduced the response fields $\tilde{X}$ and $\tilde{x}$.

{Letting $H \to H-h(t)X(t)$, and proceeding as in ref. \cite{lau2007state}, we see that the response function
\begin{equation}
\mathcal{R}_{XX}(t-t') \equiv \left.\frac{\delta\langle X(t)\rangle}{\delta h(t')}\right\rvert_{h=0}=
\frac{1}{2T\Gamma}\left\langle X(t)\left[\Gamma\partial_{t'} X(t')+\frac{\partial{H_0}}{\partial X}(t')-\upsilon x(t')\right]\right\rangle_0
\end{equation}
where the subscript $0$ on the the angle brackets implies that the average is computed at $h=0$. 
{Let us now perform the exchange $t\leftrightarrow t'$, and subtract the resulting expression from the original. Note: the difference between $\langle X(t)(\partial{H}/\partial X)(t') \rangle_0$ and its $t\leftrightarrow t'$ counterpart vanishes, regardless of the form of $H(X)$, for $t \to t'$, the limit of interest below. If $H$ is quadratic in $X$ it vanishes for all $t$ and $t'$. Anticipating the $t \to t'$ limit that we will take below, we therefore neglect this term and write} 
\begin{equation}
\label{FDR}
\frac{1}{2\Gamma T}\left[\Gamma\langle X(t)\partial_{t'}X(t')-X(t')\partial_tX(t)\rangle_0+\upsilon\langle X(t')x(t)-X(t)x(t')\rangle_0\right]=\mathcal{R}(t,t')-\mathcal{R}(t',t).
\end{equation}
The $\langle X(t)X(t')\rangle$ correlation function is invariant under time translations. Using this, 
differentiating with respect to $t$ and passing to the limit $t \to 0$, we get 
\begin{equation}
-\frac{\Gamma}{T}\lim_{t\to 0}\partial_t\left[T(\mathcal{R}(t)-\mathcal{R}(-t))+\partial_tC_{XX}(t)\rangle\right]=\frac{\upsilon\langle\dot{X}x\rangle}{T}=\sigma_1
\end{equation}
where $C_{XX}(t)$ is the $XX$ correlator, and the last equality follows from \eqref{entroprod_noflip}. {Note that if we had retained the $\langle X(t)(\partial{H}/\partial X)(t') \rangle_0$ term, it would have vanished at this stage for any $H$}.
Writing the Fourier transform of $(\mathcal{R}(t)-\mathcal{R}(-t))$ as $2i\chi''_{XX}(\omega)$, we get, in the frequency domain,}
\begin{equation}
\int^\infty_{-\infty}\frac{d\omega}{2\pi}\frac{\omega\Gamma}{T}[\omega C_{XX}(\omega)-2T\chi''_{XX}(\omega)]=\sigma_1=\int^\infty_{-\infty}\sigma_1^\omega d\omega
\end{equation}
{Explicitly, for the case where $H$ is quadratic in $X$,}
\begin{equation}
C_{XX}(\omega)= \frac{2T\Gamma}{\omega^2\Gamma^2+K^2}+\frac{2T\gamma\upsilon^2}{(\omega^2\gamma^2+k^2)(\omega^2\Gamma^2+k^2)}
\end{equation}
{and}
\begin{equation}
\chi''_{XX}(\omega)=\frac{\Gamma\omega}{\omega^2\Gamma^2+K^2}
\end{equation}
{so that}  
\begin{equation}
\omega C_{XX}(\omega)-2T\chi''_{XX}(\omega)=2T\omega \frac{\gamma\upsilon^2}{(\omega^2\gamma^2+k^2)(\omega^2\Gamma^2+K^2)}
\end{equation}
{Of course 
\begin{equation}
\int^\infty_{-\infty}\frac{d\omega}{2\pi}\frac{\omega\gamma_{11}}{T}[\omega C_{XX}(\omega)-2T\chi''_{XX}(\omega)]=\frac{\upsilon^2}{\Gamma k+\gamma K}
\end{equation}
which matches the entropy production we calculated earlier, but more important is the spectral decomposition of the entropy production:}  
\begin{equation}
\label{sigomega_noflip}
\sigma_1^\omega=\frac{\omega^2\Gamma\gamma\upsilon^2}{\pi(\omega^2\gamma^2+k^2)(\omega^2\Gamma^2+K^2)}
\end{equation}
{Note that $\sigma_1^\omega$ is appreciable only at intermediate values of $\omega$, vanishing as $\omega^2$ for $\omega\to 0$ and as $1/\omega^2$ as $\omega\to\infty$. The physical meaning of this distribution across frequencies merits some attention. Recall that $\gamma/k$ is the correlation time of the coloured noise. For example, let $\gamma/k \gg \Gamma/K$, so that the correlations of the coloured noise are significant during the process of relaxation in the potential. By \textit{high} frequency we mean $\omega \gg K/\Gamma$, which can be reinterpreted as probing timescales $1/\omega$ on which you explore a mean-square distance $T/\Gamma (1/\omega)$ smaller than the steady state positional variance $T/K$. That is, translational diffusion hasn't had a chance to act. It is this limit that our model is connected to that of \cite{fodor2016far}. By \textit{low} frequency we mean $\omega \gamma/k \ll 1$, so one doesn't see the ``colour'' of the ``noise'' $x$, which then simply adds to the existing white noise in 
the inertia-less version of \eqref{mod1mom}, just shifting the temperature of the overdamped oscillator coordinate $X$. Note that $\sigma_1^{\omega}$ will display a broad plateau between the well-separated frequencies $k/\gamma$ and $K/\Gamma$. Crucially from the point of view of experimental or numerical test of these ideas, the asymptotic vanishing of \eqref{sigomega_noflip} at low and high frequencies implies that data that (a) sample $X$ and $x$ at time intervals much longer than $\gamma/k$, or (b) are limited to a duration much smaller than $\Gamma/K$, will give the impression of time-reversibility.} 

{Our treatment so far takes all white noise strengths to be governed by a single bath temperature $T$. To turn our model into the traditional AOUP \cite{fodor2016far}, we must allow the temperature associated with the additive white noise in the equation for $X$ to go to zero. The calculation presented above would then trivially give an infinite entropy production because backstepping of $X$ is totally ruled out in the absence of the noise $\eta$ in \eqref{mod1mom}. One physically appealing way to reconcile this with the effective time-reversal invariance of the AOUP in a harmonic potential \cite{fodor2016far} is given in appendix \ref{nfdt2}.}

\subsubsection{HSR when the time-reversal operation involves a polarity flip}

In this case, we rewrite the equation for $X$ as
\be\bigg(\dot{X}-\frac{\upsilon}{\Gamma}x\bigg)=-\frac{1}{\Gamma}\frac{\partial H}{\partial X}+\tilde{\eta}\ee
where
$\langle \tilde{\eta}(t)\tilde{\eta}(t')\rangle=({2T}/\Gamma)\delta(t-t')$. Define $(\dot{X}-({\upsilon}/\Gamma)x)={V}$, the relative velocity. Let us define the response function
${\mathcal{R}_{VX}={\delta \langle V(t)\rangle}/{\delta h(t')}}$
where we have modified the Hamiltonian to {$H-hX$}. The change in action to first order in $h$ is 
\be\delta \mathcal{A}=-\frac{h}{2T}\int dt\bigg(\dot{X}-\frac{\upsilon}{\Gamma}x+\frac{1}{\Gamma}\frac{\partial H}{\partial X}\bigg).\ee
Recalling the definition of {$V$} and using the definition of the response function, we immediately obtain
\be2T\mathcal{R}_{VX}(t, t')=\left\langle V(t)\left(V(t')+\frac{1}{\Gamma}\frac{\partial H}{\partial X}(t')\right)\right\rangle.\ee
Note that $V$ and $X$ have opposite signs under time-reversal. Therefore, we now calculate the symmetric part of the response function
\begin{multline}2T[\mathcal{R}_{VX}(t, t')+\mathcal{R}_{VX}(t', t)]=\langle V(t) V(t')\rangle+\langle V(t') V(t)\rangle+\frac{1}{\Gamma}\bigg[\left\langle\dot{X}(t)\frac{\partial H}{\partial X}(t')\right\rangle+\left\langle\dot{X}(t')\frac{\partial H}{\partial X}(t)\right\rangle\bigg]\\-\frac{\upsilon}{\Gamma^2}\bigg[\left\langle x(t)\frac{\partial H}{\partial X}(t')\right\rangle+\left\langle x(t')\frac{\partial H}{\partial X}(t)\right\rangle\bigg].\end{multline}

If the system were in equilibrium, time-reversal symmetry would have resulted in the vanishing of the terms in {square brackets on the right-hand side}. This would have yielded the FDT
$ T[\mathcal{R}_{VX}(t, t')+\mathcal{R}_{VX}(t', t)]=\langle V(t) V(t')\rangle.$ {Out of equilibrium, the measure is not time-reversal symmetric. Nonetheless, for a quadratic potential, $\langle \dot{X}(t')X(t)\rangle+\langle \dot{X}(t)X(t')\rangle$ would of course be $0$. Even for a non-quadratic $H$, taking the limit $t-t'\to 0$ and using time-translation invariance so that the correlation functions are only functions of $t-t'$,} 
\be\lim_{t \to t'}\left\langle\dot{X}(t)\frac{\partial H}{\partial X}(t')\right\rangle=\langle\dot{H}\rangle=0.\ee Defining $ \mathcal{C}(t,t')=\langle V(t) V(t')\rangle,$
we finally obtain the Harada-Sasa relation
\be-\frac{\Gamma}{T}\lim_{(t-t')\to 0}\left[T\{\mathcal{R}(t, t')+\mathcal{R}(t', t)\}-\mathcal{C}(t,t')\right]=\frac{\upsilon}{\Gamma T}\left\langle x(t)\frac{\partial H}{\partial X}(t)\right\rangle=\sigma_2.\ee

{Evaluating $\mathcal{R}$ and $\mathcal{C}$ in the frequency domain for the case where $H$ is quadratic in $X$ so that $\partial_X H = K X$, we obtain 
$
\sigma_2 = \int^\infty_{-\infty}\sigma_2^\omega d\omega
$
with 
\be
\label{sigomegaflip}
\sigma_2^\omega={\upsilon^2K^2 \gamma\over \pi \Gamma} {1 \over k^2 + \gamma^2 \omega^2} {1 \over K^2+ {\Gamma^2 } \omega^2}
\ee
whose zero-frequency limit is nonzero and independent of the stiffness of the harmonic potential. Note that the frequency-resolved entropy production rates in the two cases \eqref{sigomega_noflip} and \eqref{sigomegaflip} have the same pole structures, and that their ratio is simply $ (\omega\Gamma/K)^2$.}  

{We now display the two different frequency-dependent entropy production rates for an AOUP in a harmonic potential. For this, we non-dimensionalise the frequency by $K/\Gamma$, i.e. $\tilde{\omega}=\omega\Gamma/K$, and the entropy production rates as 
$
\tilde{\sigma}^\omega_{1,2}=\sigma_{1,2}^\omega{\pi\gamma K^2}/{\upsilon^2\Gamma}
$, with $\tilde{\sigma}^\omega_1=\tilde{\omega}^2\tilde{\sigma}^\omega_2=\tilde{\omega}^2/\{(\tilde{\omega}^2+1)(\tilde{\omega}^2+\aleph^2)\}$, where the dimensionless constant $\aleph=k\Gamma/K\gamma$.}

We have established that there are two natural equilibrium limits available for the AOUP depending upon whether the polarity is flipped or not as a part of the time-reversal operation. They can be directly measured from two distinct violations of fluctuation-dissipation relation. However, this choice is available only because $x$ enters the equation for $X$ only through the active term and since in the $0$ activity limit the two equations decouple, one obtains two distinct equilibrium limits depending on whether $x$ is odd or even under time-reversal. More technically, this choice is the result of the active term breaking both time-reversal and $x\to -x$ symmetry in the model.
If $X$ and $x$ were coupled through equilibrium terms in addition to the active one, such that the $0$ activity limit did not possess this extra $x\to-x$ symmetry, this would not be the case -- the requirement that entropy production has to vanish in the absence of activity would fix the time-reversal characteristic of both $x$ and $X$. We explicitly demonstrate this in appendix \ref{appE:2dof}.

\subsubsection{entropy production of other active particle models}
We have also considered two other active particles: polar particles with a preferred speed and apolar particles. The entropy production for these two particles are
\begin{equation}
\sigma_P=\frac{\zeta\Delta\mu}{T}\left\langle\frac{P}{\Gamma(P^2)}\partial_XH\right\rangle,
\end{equation}
where $\Gamma(P^2)=\gamma_1+\gamma_3P^2$ and
\begin{equation}
\sigma_Q=-\frac{\zeta\Delta\mu}{\gamma T}\left\langle{\bf P}\cdot{\bsf Q}\cdot\partial_{\bf X}H\right\rangle,
\end{equation}
respectively. We have calculated the entropy production for the apolar particle in the overdamped limit.

\subsection{Nonequilibrium dynamics of active field theories}
\subsubsection{Entropy production for polar rods}
\label{HarSalabel}
We first compute entropy production for polar rods on a substrate which we discussed \ref{subsub:polarrodsTT}. However, we will now consider a simplified and rescaled version of those equations which retains its essential active features:
\begin{equation}
\dot{\rho}=-v\Delta\mu\nabla\cdot(\rho{\bf p})+\nabla^2\frac{\delta H}{\delta \rho}+\xi
\end{equation}
with $\langle \xi({\bf x}, t)\xi({\bf x}', t')\rangle=-2T\nabla^2\delta({\bf x}-{\bf x}')\delta(t-t')$ and

\begin{equation}
\dot{{\bf p}}+\lambda\Delta\mu{\bf p}\cdot\nabla{\bf p}=-\frac{\delta H}{\delta {\bf p}}+\boldsymbol{\xi}^p
\end{equation}
$\langle \boldsymbol{\xi}_p({\bf x}, t)\boldsymbol{\xi}_p({\bf x}', t')\rangle=2T{\bsf I}\delta({\bf x}-{\bf x}')\delta(t-t')$
Here, we have ignored the noise cross-correlation between $\rho$ and ${\bf p}$ and retained the active advection of the density and polarisation at different speeds. 
The Onsager-Machlup action, ignoring the contribution from the noise discretisation, which does not contribute to the time-antisymmetric part of the action, is
\begin{equation}
\mathcal{A}^{OM}=\frac{1}{4T}\int d{\bf x}dt\left[-\left\{\dot{\rho}+v\Delta\mu\nabla\cdot(\rho{\bf p})-\nabla^2\frac{\delta H}{\delta \rho}\right\}\nabla^{-2}\left\{\dot{\rho}+v\Delta\mu\nabla\cdot(\rho{\bf p})-\nabla^2\frac{\delta H}{\delta \rho}\right\}+\left(\dot{{\bf p}}+\lambda\Delta\mu{\bf p}\cdot\nabla{\bf p}+\frac{\delta H}{\delta {\bf p}}\right)^2\right]
\end{equation}
Since both $\mathcal{T}\rho=\rho$ and $\mathcal{T}{\bf p}={\bf p}$ the time antisymmetric part of this action is 
\begin{equation}
\mathcal{S}=-\frac{\Delta\mu}{T}\int dtd{\bf x}\left[{-v}\{\dot{\rho}\nabla^{-2}\nabla\cdot(\rho{\bf p})\}+{\lambda}\{\dot{{\bf p}}\cdot({\bf p}\cdot\nabla){\bf p}\}\right]
\end{equation}
We show in appendix \ref{hsaftapp} that the entropy production rate $\sigma=\lim_{t\to \infty}\mathcal{S}/t$ is related to the standard definitions of response and correlation function by a Harada-Sasa-like relation \cite{harada2005equality,fodor2016far} 
\begin{equation}
\sigma=\frac{1}{T}\lim_{t\to 0}\int d{\bf x}\partial_t\left[\nabla^{-2}\left(T[\mathcal{R}_{\rho\rho}(t)-\mathcal{R}_{\rho\rho}(-t)]+\partial_tC_{\rho\rho}\right)-\left(T[\mathcal{R}_{pp}(t)-\mathcal{R}_{pp}(-t)]+\partial_tC_{pp}\right)\right]
\end{equation}
where $\mathcal{R}_{ij}$ is the response of the field $j$ to a perturbation in the field $i$ and $C_{ij}$ is the correlation of the fields $i$ and $j$. 
%
%
%
We also show that in contrast to active model B \cite{nardini2017entropy}, the entropy production does not vanish in the $T\to 0$ limit even in a homogeneously polarised phase in appendix \ref{entotapp}.

\subsubsection{entropy production for the model with velocity-alignment}
We will now calculate the entropy-production of the active variant of the Navier-Stokes equation we discussed in section \ref{subsub:velalignTT}. For simplicity, let us set the viscosity $\eta=0$.
The action is
\begin{equation}
\mathcal{A}=\frac{1}{4T}\int dtd{\bf x}\frac{1}{\Gamma_{gg}(v)}\left[\partial_t{\bf g}+\nabla\cdot({\bf gv})+\rho\nabla\frac{\delta H}{\delta \rho}+\boldsymbol{\Gamma}_{gg}{\bf v}-\Delta\mu\zeta{\bf v}\right]^2
\end{equation}
since $\mathcal{T}{\bf g}=-{\bf g}$, the antisymmetric part of this action, which can not be expressed as a total derivative, is 
\begin{equation}
\label{anotherentropy}
\mathcal{S}=\frac{\Delta\mu}{T}\int dt d{\bf x}\frac{\zeta}{\Gamma_{gg}(v)}{\bf v}\cdot\left[\partial_t{\bf g}+\nabla\cdot({\bf gv})+\rho\nabla\frac{\delta H}{\delta \rho}\right].
\end{equation}
Note that there are two crucial requirements for non-zero entropy production: i. the presence of $\Delta\mu$ and ii. the multiplicativity of the noise. If the noise were not multiplicative, \eqref{anotherentropy} would have been a total derivative which would not lead to any entropy production. Of course, in that case the cubic velocity nonlinearity required for saturating the speed when the linear damping is negative, would also have to be an active term and that would have led to a non-zero entropy. 

\section{Conclusion} \label{conclusion}
{In this paper we first presented in some detail the stochastic
extension of the nonequilibrium thermodynamic construction
\cite{julicher1997modeling,kruse2005generic,joanny2007hydrodynamic,juelicher2007active,julicher2018hydrodynamic}
of the dynamical equations for active systems, i.e., systems in which detailed
balance is broken homogeneously at the level of the constituent degrees of freedom.
Through this treatment we saw that all active systems discussed in the literature can
be obtained from a corresponding equilibrium system in which the force on an internal
coordinate, identified as a chemical degree of freedom, is maintained at a nonzero
constant value, while all degrees of freedom are in contact with a \emph{common}
thermal bath.  This includes apparently ``two-temperature'' models, such as apolar
rods which we discuss as well as those studied in \cite{grosberg2015nonequilibrium},
and systems with non-mutual pair interactions \cite{saha2014clusters,
	liebchen2017phoretic, liebchen2016pattern}.}
{For active polar particles we then use the ratio of the experimentally measurable
large deviation functions for trajectories in position-polarisation space to
demonstrate that there are two \emph{a priori} equally valid answers to the question
how far a system is from equilibrium.}  Our approach requires some comment: active
systems are in general complicated and quantitative measurements of the total entropy
production rate, which is related to the total heat dissipation rate, are generally
impractical to carry out. What is desired instead is that, given a measurement of
certain quantities, to understand how different it is from a measurement of those
quantities if the system were in equilibrium. This limited question is answered by
the ``entropy production" we calculate though we show that even this may not have
unique answers in certain cases. {Our work is similar in spirit to that of \cite{gladrow2016broken, battle2016broken}, who try to find, in a configuration space spanned by experimentally measured quantities, a circulating dissipative current whose presence signifies the breaking of detailed balance. As long as one chooses two variables with the same sign under time-reversal, measuring the circulating current provides a direct quantitative measure of a system's distance
from equilibrium, and can be easily related to the measure we discuss
\cite{seifert2012stochastic}. We too take an objective, empirical approach to the extent of time-reversal breaking based on measurements on a subset of degrees of freedom, with no implied link to heat. Rather than a vectorial quantity such as a configuration-space current, we work with a scalar measure.} 

Finally, turning to field theories of polar active systems, we have clarified the two
equilibrium limits of the Toner-Tu equation -- polar liquid crystal and the
Navier-Stokes equation -- and explained the origin and the consequence of
nonequilibrium driving in both of them. Our work here clarifies a long-standing
confusion about the self-advective nonlinearity of the Toner-Tu equation, which we
show requires breaking of both detailed balance and Galilean invariance. 

While our work in this paper has been mainly clarificatory,  we have made testable
predictions regarding entropy production rate of polar particles. In particular, we
calculated the frequency-dependence of the entropy production in harmonic traps and
demonstrated that when only the position of the particle is measured, there is no way
to distinguish the dynamics from an equilibrium one. We look forward to tests of our
predictions on model active systems in the colloidal or granular domain.

{\section*{Acknowledgements} We thank Mike Cates, Suraj Shankar and M. Cristina Marchetti for illuminating discussions. LPD thanks the Department of Physics, Indian Institute of Science, for hospitality and support. SR was supported by a J C Bose Fellowship of the SERB (India) and the Tata Education and Development Trust.}

\setcounter{section}{0}
\setcounter{equation}{0}
\renewcommand{\theequation}{\thesection.\arabic{equation}}
\section*{Appendices} 
\label{Appendicesmark}
\renewcommand\thesection{\Alph{section}}
\renewcommand\thesubsection{\arabic{subsection}}
\renewcommand\thesubsubsection{\alph{subsubsection}}

\section{Active dynamics through the coupling of physical and chemical velocities}
\label{app:physchem}
\setcounter{equation}{0}
In this appendix, we follow \cite{ramaswamy2017active} and retain an extra chemical velocity variable $\Phi$, which is conjugate to the chemical coordinate $n$. We obtain active equations of motion for each system described in the main text without invoking any direct coupling between $n$ and other variables, but through an entrainment of the physical velocity by the chemical velocity or an antisymmetric reactive coupling between the chemical and the physical velocities. We take the dynamics of $n$ to be $\partial_t n=\Phi$. Our discussion highlights the fact that multiple thermodynamically consistent {dynamical models} lead to the same effective active equations of motion. {While active equations of motion for the models discussed \ref{sub:reversiblechemforce} have already been derived using this approach in \cite{ramaswamy2017active}, in this appendix we focus on the models discussed in \ref{sub:irreversiblechemforce}. Unlike the models discussed in \cite{ramaswamy2017active}, in which the coupling between the physical and chemical velocities were dissipative in nature, these models naturally require a \emph{reversible} coupling between physical and chemical velocities.}
\label{2derivation}

{\subsection{Reversible Coupling of physical and chemical velocities}} \label{sechall}
We now derive the equations of motion for {(i) an} active polar model with inertia and a negative linear damping which can be viewed as an active extension of the Drude model, {(ii) an} apolar variant of the AOUP and {(iii) a} field theory of a polar model with inertia and negative linear damping. 
\subsubsection{Active particle with preferred speed}
\label{prefspHall1}
{Start with the system of equations
\begin{equation}
\dot{X}=\partial_P H; \,  \dot{n}=\partial_\Phi H
\end{equation}
\begin{equation}
\label{peqn}
\dot{P}=-\Gamma_{11}\partial_P H + \Gamma_{12} \partial_{\Phi} H -{\partial_X H} +\xi^P=-\Gamma_{11}\partial_P H +\zeta P \partial_{\Phi} H-{\partial_X H}+\xi^P, 
\end{equation}
\begin{equation}
\label{phieq}
\dot{\Phi}=-\Gamma_{22}\partial_{\Phi} H - \Gamma_{12}\partial_P H + T\partial_P\Gamma_{12}-{\partial_n H} + \xi^\Phi=-\Gamma_{22}\partial_{\Phi} H - \zeta P \partial_P H + T\zeta - {\partial_n H} + \xi^\Phi
\end{equation}
with $\langle \xi^p(t)\xi^p(0)\rangle=2\Gamma_{11}T\delta(t)$ and $\langle \xi^\Phi(t)\xi^\Phi(0)\rangle=2\Gamma_{22}T\delta(t)$.  Here, we have coupled $P$ and $\Phi$ reversibly through a Poisson bracket $[P, \Phi] = \Gamma_{12}=\zeta P$.  
The term $T\partial_P\Gamma_{12} = T \zeta$ arises routinely in the projection-operator construction of generalized Langevin equations \cite{,zwanzig2001nonequilibrium,mori1965transport}, and its presence is formally required to ensure that the Gibbs-Boltzmann distribution is the stationary solution to the corresponding Fokker-Planck equation. In most other systems this term conveniently vanishes; here it does not.} 

{Ignoring inertia in \eqref{phieq}, we obtain
\begin{equation}
\partial_{\Phi} H=-\frac{\zeta P}{\Gamma_{22}}  \partial_P H+\frac{T\zeta}{\Gamma_{22}}-\frac{1}{\Gamma_{22}}\frac{\partial H}{\partial n}+\frac{1}{\Gamma_{22}}\xi^\Phi
\end{equation}
which allows us to eliminate $\partial_{\Phi} H$ from \eqref{peqn}, yielding, for the case where $H$ is quadratic in $P$ with $\partial_P H = P$,  
\be
\label{pfin}
\dot{P} 
= -\left(\Gamma_{11}+\frac{\zeta}{\Gamma_{22}}\frac{\partial H}{\partial n}-\frac{T\zeta^2}{\Gamma_{22}}\right)P-\frac{\zeta^2 P^2}{\Gamma_{22}}P-\frac{\partial H}{\partial X}+\eta
\ee
where 
\be
\label{multnoise}
\eta \equiv ({\zeta P}/{\Gamma_{22}})\xi^\Phi+\xi^P, \,  
\langle \eta(t)\eta(0)\rangle = 2T\left(\frac{\zeta^2 P^2}{\Gamma_{22}}+\Gamma_{11}\right) \delta(t), 
\ee
is a multiplicative noise in the $P$ equation, with the Stratonovich interpretation, as can be seen by carrying out the adiabatic elimination in detail \`a la \cite{baule2008exact,ryter1981brownian}. Alternatively one can check this ex post facto by confirming that the corresponding Fokker-Planck equation has the correct equilibrium solution. The term $-(T \zeta^2/\Gamma_{22})P$ in \eqref{pfin}, whose origin is the derivative of the $[P,\Phi]$ Poisson bracket, must now be understood as a noise-induced drift.} Now, we introduce activity by holding $\partial_n H=\Delta\mu$ constant. This leads to the equation of motion for $P$
\begin{equation}
\label{pnoneq}
\dot{P}=-\left(\Gamma_{11}+\frac{\zeta}{\Gamma_{22}}\Delta\mu-\frac{T\zeta^2}{\Gamma_{22}}+\frac{\zeta^2 P^2}{\Gamma_{22}}\right)P-\frac{\partial H}{\partial X}+\eta
\end{equation}
{Note that the $P$-independent part of the friction is now liberated from the noise strength because of the $\Delta\mu$-dependent term, which can even make it negative. Despite superficial appearances, the noise-induced drift $-(T \zeta^2/\Gamma_{22})P$ has no such effect, and arises simply as a consequence of the multiplicative noise \cite{lau2007state}. In an anti-{\ito} \cite{lau2007state} description, the noise-induced drift vanishes. In all interpretations, of course, the probability distribution of $P$ is Maxwellian if $\Delta \mu =0$, and the particle is not self-propelled. If $\zeta < - \Gamma_{11} \Gamma_{22} / \Delta \mu $, the zero-velocity state is on average unstable, and the particle moves with preferred velocity $\pm \sqrt{(|\zeta| \Delta \mu - \Gamma_{11} \Gamma_{22})/\zeta^2}$ in the limit of low noise. We have thus succeeded in producing a model self-propelled particle with inertia within this consistent stochastic framework.}

\subsubsection{Apolar AOUP} \label{subsubapp:apolar}

Take the coordinate like variables to be $({\bf X}, n, {\bsf Q})$ where ${\bsf Q}$ is {a traceless symmetric second-rank} tensor. We assume that ${\bsf Q}$ is autonomous and {obeys a purely relaxational stochastic dynamics}:
\begin{equation}
\Gamma_{33}\dot{{\bsf Q}}=-{\bsf Q}+\xi^Q
\end{equation}
with $\langle \xi^Q(t)\xi^Q(t')\rangle=2\Gamma_{33}T\delta(t-t')$. As before, the momenta conjugate to $({\bf X}, n)$ are $({\bf P},\Phi)$. Take the system of equations
\begin{equation}
\dot{{\bf X}}={\bf P}
\end{equation}
\begin{equation}
\label{peqn2}
\dot{{\bf P}}=-\Gamma_{11} {\bf P}+\boldsymbol{\Gamma}_{12} \Phi-\frac{\partial H}{\partial {\bf X}}+\boldsymbol{\xi}^P=-\Gamma_{11} {\bf P}+\zeta {\bsf Q}\cdot{\bf P} \Phi-\frac{\partial H}{\partial {\bf X}}+\boldsymbol{\xi}^P
\end{equation}
with $\langle \boldsymbol{\xi}^P(t)\boldsymbol{\xi}^P(t')\rangle=2\Gamma_{11}T{\bsf I}\delta(t-t')$, where ${\bsf I}$ is the identity tensor.
\begin{equation}
\dot{n}=\Phi
\end{equation}
\begin{equation}
\label{phieq2}
\dot{\Phi}=-\Gamma_{22}\Phi-\boldsymbol{\Gamma_{12}}\cdot{\bf P}-\frac{\partial H}{\partial n}+\xi^\Phi=-\Gamma_{22}\Phi-\zeta {\bf P}\cdot{\bsf Q}\cdot{\bf P}-\frac{\partial H}{\partial n}+\xi^\Phi
\end{equation}
Note that since ${\bsf Q}$ is traceless, $\nabla_{\bf P}\cdot({\bsf Q}\cdot{\bf P})$ is $0$ and therefore, the derivative of the Poisson bracket term is $0$.
Again overdamping \eqref{phieq2} and replacing $\Phi$ in \eqref{peqn2}, we obtain
\begin{multline}
\dot{\bf P}=-\left[\Gamma_{11}{\bsf I}+\frac{\zeta}{\Gamma_{22}}\frac{\partial H}{\partial n}{\bsf Q}+\frac{\zeta^2}{\Gamma_{22}}({\bsf Q}\cdot{\bf P})({\bsf Q}\cdot{\bf P})\right]\cdot{\bf P}-\frac{\partial H}{\partial {\bf x}}+\frac{\zeta}{\Gamma_{22}}{\bsf Q}\cdot{\bf P}\xi^\Phi+\boldsymbol{\xi}^p \\=-\left[\Gamma_{11}{\bsf I}+\frac{\zeta}{\Gamma_{22}}\frac{\partial H}{\partial n}{\bsf Q}+\frac{\zeta^2}{\Gamma_{22}}({\bsf Q}\cdot{\bf P})({\bsf Q}\cdot{\bf P})\right]\cdot{\bf P}-\frac{\partial H}{\partial {\bf X}}+\boldsymbol{\eta}
\end{multline}
with the noise correlator
$
\langle \boldsymbol{\eta}(t)\boldsymbol{\eta}(t')\rangle=2T\left[\Gamma_{11}{\bsf I}+({\zeta^2}/{\Gamma_{22}})({\bsf Q}\cdot{\bf P})({\bsf Q}\cdot{\bf P})\right]
$
This equation should again be interpreted using the Stratonovich interpretation. However, due to tracelessness of ${\bsf Q}$, no interpretation-dependent noise is required to transfer between interpretations. In other words, this is a multiplicative noise equation in which there is no possibility of noise-induced drift. Now holding $\partial_n H=\Delta\mu$ constant, we obtain a nonequilibrium ${\bsf Q}$-dependent damping:
\begin{equation}
\label{actpeqn2}
\dot{\bf P}=-\left[\Gamma_{11}{\bsf I}+\frac{\zeta\Delta\mu}{\Gamma_{22}}{\bsf Q}+\frac{\zeta^2}{\Gamma_{22}}({\bsf Q}\cdot{\bf P})({\bsf Q}\cdot{\bf P})\right]\cdot{\bf P}-\frac{\partial H}{\partial {\bf X}}+\boldsymbol{\eta}
\end{equation}
If we now take the damping of ${\bf P}$ in \eqref{peqn2} to depend on ${\bsf Q}$ instead of being isotropic, we can cancel the passive ${\bsf Q}$ dependent part of \eqref{actpeqn2}. Write the damping in \eqref{peqn2} as $-\boldsymbol{\Gamma}_P\cdot{\bf P}=-(\Gamma_{11}{\bsf I}-\zeta^2({\bsf Q}\cdot{\bf P})({\bsf Q}\cdot{\bf P})/\Gamma_{22})\cdot{\bf P}$ which yields the simplest apolar version of AOUP:
\begin{equation}
\label{actpeqn3}
\dot{\bf P}=-\left[\Gamma_{11}{\bsf I}+\frac{\zeta\Delta\mu}{\Gamma_{22}}{\bsf Q}\right]\cdot{\bf P}-\frac{\partial H}{\partial {\bf X}}+\boldsymbol{\eta}_1
\end{equation} 
with $\langle \boldsymbol{\eta}_1(t)\boldsymbol{\eta}_1(t')\rangle=2T\Gamma_{11}{\bsf I}$. Taking the overdamped limit is now trivial.

\subsubsection{Active field theory with velocity alignment} 
\label{subsubapp:velalign}
We now start with the Navier-Stokes equation with friction.
\begin{equation}
\label{deneq2}
\partial_t\rho=-\nabla\cdot{\bf g}
\end{equation}
\begin{equation}
\label{momeq221}
\partial_t{\bf g}+\nabla\cdot({\bf vg})=-\rho\nabla\frac{\delta F}{\delta \rho}-\Gamma{\bf v}+\eta\nabla^2{\bf v}+\boldsymbol{\xi}^g
\end{equation}
with 
$
\langle\boldsymbol{\xi}^g({\bf x}, t)\boldsymbol{\xi}^g({\bf x}', t')\rangle=2T{\bsf I}(\Gamma-\eta\nabla^2)\delta({\bf x}-{\bf x}')\delta(t-t'),
$
and $H_0=\int_x{\bf g}^2/2\rho+F[\rho]$. We then introduce the interaction with the chemical field by extending the hamiltonian $H=H_0+n\Delta\mu$ and  We consider the reactive coupling
\begin{equation}
\nonumber
 \{\Phi({\bf x}'), {\bf g}({\bf x})\}\frac{\delta {H}}{\delta \Phi({\bf x}')},
\end{equation}
between $\Phi$ and ${\bf g}$ in the momentum equation. 
To the lowest order,
$
\{\Phi({\bf x}'), {\bf g}({\bf x})\}=\alpha_1(\rho){\bf v}({\bf x})\delta({\bf x}-{\bf x}')
$
Note that the derivative of the Poisson bracket with ${\bf g}$ is not zero: $\delta\{\Phi({\bf x}'), {\bf g}({\bf x})\}/\delta{\bf g}({\bf x}')\neq 0$.

We now write the dynamical equation for $\Phi$:
\begin{equation}
\partial_t\Phi=-\gamma\Phi-\frac{\delta F}{\delta n}+\int d{\bf x}'\left[\{{\bf g}({\bf x}'),\Phi({\bf x})\}\frac{\delta {H}}{\delta{\bf g}({\bf x}')}+T\frac{\delta\{\Phi({\bf x}'), {\bf g}({\bf x})\}}{\delta{\bf g}({\bf x}')}\right]+\xi^\Phi
\end{equation}
where the noise correlation is $\langle\xi^\Phi({\bf x}, t),\xi^\Phi({\bf x}', t')\rangle=2T\gamma\delta({\bf x}-{\bf x}')\delta(t-t')$. The finite-temperature contribution to the reactive piece is evaluated as
\begin{equation}
T\frac{\delta\{\Phi({\bf x}'), {\bf g}({\bf x})\}}{\delta{\bf g}({\bf x}')}=T\alpha_1(\rho)\frac{1}{\rho({\bf x})}\delta({\bf x}-{\bf x}')
\end{equation}
while
\begin{equation}
\{{\bf g}({\bf x}'),\Phi({\bf x})\}\frac{\delta {H}}{\delta{\bf g}({\bf x}')}=-\alpha_1(\rho){\bf v}({\bf x})\cdot{\bf v}({\bf x})\delta({\bf x}-{\bf x}')
\end{equation}
Finally, taking the overdamped limit in the $\Phi$ equation, we can solve for $\Phi$ as
\begin{equation}
\Phi=-\frac{1}{\gamma}\frac{\delta H}{\delta n}-\frac{\alpha_1(\rho)}{\gamma}{\bf v}^2+T\frac{\alpha_1(\rho)}{\rho\gamma}+\frac{1}{\gamma}\xi^\Phi
\end{equation}
\begin{equation}
\int d{\bf x}'\{\Phi({\bf x}'), {\bf g}({\bf x})\}\frac{\delta H}{\delta \Phi({\bf x}')}=\alpha_1(\rho){\bf v}\Phi=-\left(\frac{\alpha_1(\rho)}{\gamma}\frac{\delta H}{\delta n}-T\frac{\alpha_1^2(\rho)}{\rho\gamma}\right){\bf v}-\frac{\alpha_1^2(\rho)}{\gamma}{\bf v}^2{\bf v}+\frac{\alpha_1(\rho)}{\gamma}{\bf v}\xi^\Phi
\end{equation}
Finally, noting that $\delta H/\delta n=\Delta\mu$ is the constant chemical drive, and using \eqref{deneq2} we have the Toner-Tu equation:
\begin{multline}
\label{TTvel1}
\rho(\partial_t{\bf v}+{\bf v}\cdot\nabla\cdot{\bf v})=-\rho\nabla\frac{\delta F}{\delta \rho}-\left(\Gamma+\frac{\alpha_1(\rho)}{\gamma}\Delta\mu-T\frac{\alpha_1^2(\rho)}{\rho\gamma}\right){\bf v}-\frac{\alpha_1^2(\rho)}{\gamma}{\bf v}^2{\bf v}+\eta\nabla^2{\bf v}+\left(\boldsymbol{\xi}^g+\frac{\alpha_1(\rho)}{\gamma}{\bf v}\xi^\Phi\right)
\end{multline}
The noise is again multiplicative with the correlator 
\begin{multline}
\left\langle\left(\boldsymbol{\xi}^g+\frac{\alpha_1(\rho)}{\gamma}{\bf v}\xi^\Phi\right)({\bf x}, t)\left(\boldsymbol{\xi}^g+\frac{\alpha_1(\rho)}{\gamma}{\bf v}\xi^\Phi\right)({\bf x}', t')\right\rangle=\langle\boldsymbol{\xi}^v({\bf x}, t)\boldsymbol{\xi}^v({\bf x}', t')\rangle\\=2T[{\bsf I}(\Gamma-\eta\nabla^2)+\alpha_1(\rho)^2{\bf v}{\bf v}/\gamma]\delta({\bf x}-{\bf x}')\delta(t-t')
\end{multline}
This equation should also be interpreted in the Stratonovich interpretation, since it should have a steady state {$\exp{(-H_0/T)}$ when $\Delta\mu=0$}.

{\section{Activity through fixed reaction velocity}} 
\label{app:reactvel} 
\setcounter{equation}{0}
In the main text, we introduced activity through a fixed chemical potential difference between reactants and products, i.e. $\partial{H}/\partial n=\text{const.}$. This is in line with the traditional rationalisation for active matter equations where the chemical potential difference between a fuel (for instance, ATP of actomyosin networks in cells and sugar in bacterial systems) and its reaction by-products (for instance, ADP in actomyosin systems) is assumed to be constant. However, this choice is not essential -- indeed, one can imagine bacteria or birds or even artificial particles which have fixed reaction velocity and adjusts $\partial{H}/\partial n$ to hold that constant (i.e., uses as much or as little fuel as required to perform a fixed number of duty cycles in a second). We now demonstrate that an equation of the form \eqref{mod1mom} can be derived even by holding $\Phi$ constant -- in other words, similar effective equations result irrespective of whether microscopically a rate ($\Phi$) or a field ($\partial H/\partial n$) is held constant. As in section \ref{sec:gendev}, take the model (with $\mathscr{Q}=(X,n,x)$ and $\mathscr{P}=(P,\Phi,p)$)
\begin{equation}
\label{co-od}
\dot{\boldsymbol{\mathscr{Q}}}=\frac{\partial {H}}{\partial \boldsymbol{\mathscr{P}}} \;\;\;\;\;\;\;\;\;\;\;\;
\dot{\mathscr{P}}=-\frac{\partial {H}}{\partial \boldsymbol{\mathscr{Q}} }-\boldsymbol{\Gamma}\cdot\frac{\partial {H}}{\partial \boldsymbol{\mathscr{P}}}+\boldsymbol{\xi}
\end{equation}
where the noise $\boldsymbol{\xi}=({\xi}^P, \xi^\Phi, {\xi}^p)$ has a correlation
$
\langle\boldsymbol{\xi}(t)\boldsymbol{\xi}(t')\rangle=2T\boldsymbol{\Gamma}^S\delta(t-t').
$
and the friction matrix is 
\begin{equation}
\boldsymbol{\Gamma}=\begin{pmatrix} \Gamma & -\gamma_{12}{x} &0\\ -\gamma_{12}{ x}&\gamma_{22} &0\\ 0 & 0 &\gamma\end{pmatrix}.
\end{equation}
After taking the overdamped limit, we now take $\Phi$ to be held constant instead of $\partial{H}/\partial n$. This leads to the equations-of motion
\begin{equation}
\label{mom3}
\Gamma\dot{X}=-\frac{\partial {H}}{\partial { X}}+\Phi\gamma_{12}{x}+{\xi}^P
\end{equation}
\begin{equation}
\label{aux3}
\gamma\dot{x}=- {kx}+{\xi}^p
\end{equation}
where we have assumed that ${H}$ is quadratic in $x$ (i.e $\partial_xH=kx$). This has the same form as the overdamped version of \eqref{mod1mom} and \eqref{mod1aux} except for the coefficient of the active coupling.

\section{Inertial active particles}
\label{appinertia} 
\setcounter{equation}{0}
In the main text, we have only considered the dynamics of overdamped active particles. In this appendix, we briefly consider the underdamped case.
The explicit underdamped versions of the equations of motion \eqref{mod1mom} and \eqref{mod1aux} are
\begin{equation}
\label{mod2momapp1}
\ddot{X}+\Gamma\dot{X}+\frac{\partial {H}}{\partial {X}}-\upsilon{ x}={\eta}
\end{equation}
\begin{equation}
\label{mod2auxapp1}
\gamma\dot{x}+kx={\xi}^p
\end{equation}
with the noise correlations 
$
\langle\eta(t)\eta(t')\rangle=2T\Gamma\delta(t-t')
$ and
$
\langle\xi^p(t)\xi^p(t')\rangle=2T\gamma\delta(t-t').
$
The action is 
\begin{equation}
\mathcal{A}=\frac{1}{4T}\int dt \frac{1}{\Gamma}(\ddot{X}+\Gamma\dot{X}+\frac{\partial {H}}{\partial {X}}-\upsilon{ x})^2+\frac{1}{\gamma}(\gamma\dot{x}+x)^2
\end{equation}
Reversing $t\to-t$ while holding $x$ fixed, we calculate the KL divergence which is 
\begin{equation}
\mathcal{S}_1=\frac{\upsilon}{T}\langle\dot{X}x\rangle
\end{equation}
where we have as usual assumed ergodicity and replaced a time-average with an ensemble average. If we measure the KL divergence conditioned on $x\to -x$ under time-reversal, we obtain
\begin{equation}
\mathcal{S}_2=\frac{\upsilon}{\Gamma T}\langle x(\ddot{X}+\partial_XH)\rangle
\end{equation}
The sum $\mathcal{S}_1+\mathcal{S}_2=\upsilon^2\langle x^2\rangle/T\Gamma=\upsilon^2/\Gamma k$ as in the case without inertia. 

We now show that our conclusion that a nonvanishing entropy production for a particle in a harmonic trap is only possible when the dynamics of the auxiliary variable $x$ is explicitly tracked is valid even in the underdamped limit. In this case, \eqref{mod2momapp1} is modified to
\begin{equation}
\label{maineq2}
\ddot{X}+\Gamma\dot{X}+\partial_XH=\eta
\end{equation}
The action case for a general noise is
\begin{multline}
\mathcal{A}=\frac{1}{4T}\int dt\bigg[
\nu_1\left(\ddot{X}+\Gamma\dot{X}+\partial_XH\right)^2+\nu_2\gamma^2 \left\{\partial_t\left(\ddot{X}+\Gamma\dot{X}+\partial_XH\right)\right\}^2 +\nu_3\gamma^4\left\{\partial_t^2\left(\ddot{X}+\Gamma\dot{X}+\partial_XH\right)\right\}^2+...\bigg]
\end{multline}
using the earlier definitions. The time-asymmetric part of this can be trivially calculated (integrating by parts repeatedly)
\begin{equation}
\mathcal{S}=\frac{1}{T}\int dt\left[-\nu_1(\dot{X})(\ddot{X}+\partial_XH)+\nu_2(\partial_t^3X)(\ddot{X}+\partial_XH)- \nu_3(\partial_t^5X)(\ddot{X}+\partial_XH) +... \right]
\end{equation}
Thus, inertia only contributes terms $(\partial_t^{2n+3}X)X$. This is just $(1/2)\partial_t(\partial_t^{n+1}X)^2$ (integrating by parts $n+1$ times). All these terms vanish irrespective of the potential. We have already shown that the terms with $\partial_XH$ vanish if $H$ is harmonic in $X$. Therefore, the presence of inertia has no effect on the entropy production for any noise when one only records the particle position.

\section{Entropy production due to particle driven by pure coloured noise}
\label{app:purecolour} 
\setcounter{equation}{0}
In this appendix, we calculate the entropy production rate of a particle in a harmonic potential and driven by a monochromatic coloured noise using \eqref{cnaction} and demonstrate that the {result is equivalent to \cite{fodor2016far}}. THe equation of motion of such a particle is $\Gamma\dot{X}+\partial_x H=\upsilon\xi$ with the  noise correlation
\begin{equation}
\langle\xi_\omega\xi_{-\omega}\rangle=\frac{2T\gamma}{1+\gamma^2\omega^2},
\end{equation}
and
therefore,
\begin{equation}
\mathcal{G}_\omega=\frac{1+\gamma^2\omega^2}{\upsilon^2\gamma}
\end{equation}
This leads to the action
\begin{equation}
\mathcal{A}=\frac{1}{4T\upsilon^2\gamma}\int d\omega\eta_{-\omega}(1+\gamma^2\omega^2)\eta_\omega=\frac{1}{4T\upsilon^2\gamma}\int dt(\eta(t)^2-\gamma^2\ddot{\eta}(t)\eta(t))=\frac{1}{4T\upsilon^2\gamma}\int dt[\eta(t)^2+\gamma^2(\dot{\eta}(t))^2]
\end{equation}
which upon transformation to the $X$ variables is
\begin{equation}
\mathcal{A}=\frac{1}{4T\upsilon^2\gamma}\int dt[(\Gamma\dot{X}+\partial_XH)^2+\gamma^2(\Gamma\ddot{X}+\dot{X}\partial_X^2H)^2].
\end{equation}
Adding only total derivative terms, this can be written as
\begin{equation}
\mathcal{A}=\frac{1}{4T\upsilon^2\gamma}\int dt[\gamma\Gamma\ddot{X}+(
\Gamma+
\gamma\partial_X^2H)\dot{X}+\partial_XH]^2
\end{equation}

The time-reversed action is
\begin{equation}
\mathcal{A}^R=\frac{1}{4T\upsilon^2\gamma}\int dt[\gamma\Gamma\ddot{X}-(
\Gamma+
\gamma\partial_X^2H)\dot{X}+\partial_XH]^2
\end{equation}
and the path-dependent antisymmetric part of this action is
\begin{equation}
\mathcal{S}=\frac{\Gamma\gamma}{2T\upsilon^2}\int dt \dot{X}^3\partial_X^3H
\end{equation}
This the result for the KL divergence {obtained in \cite{fodor2016far}}.
\section{Entropy production and time-reversal in general two degrees of freedom systems}
\label{appE:2dof} 
\setcounter{equation}{0}
We show that two distinct equilibrium limits are possible for AOUP/ ABP only because the active term {-- in addition to breaking time-reversal symmetry --} is also the only coupling between between $X$ and $x$. In other words, activity in this case breaks two different symmetries -- (a) time-reversal and (b) independent $X\to-X$ and $x\to -x$ symmetries. When the active term is present, the only symmetry is under joint $X\to -X$ and $x\to -x$. (if the active term is proportional to $x^2$ instead of $x$, thus not breaking this symmetry, there is no difference between entropy production irrespective of whether $x\to x$ or $-x$ under $t\to -t$). If the passive system does not have independent $X\to -X$ and $x\to -x$ symmetry, only one of the two measues vanish with $\Delta\mu$. (Strictly, the important symmetry is not under $X\to -X$ but under $x\to -x$. That is the symmetry whose presence in the passive limit of ABP/ AOUP and absence in the active case leads to two distinct equilibrium limits, when polarity is flipped and when it isn't). For example, if an ABP Hamiltonian contained a coupling ${\bf p}\cdot{\bf r}$ where ${\bf p}$ is the polarisation and ${\bf r}$ is the position, then the "entropy production rate" calculated by flipping ${\bf p}$ will not vanish in the $\Delta\mu\to 0$ limit but the one calculated without flipping ${\bf p}$ will.

Take the phenomenological equations

\be\dot{X}=-\Gamma\frac{\partial H}{\partial X}-\gamma_{12}\frac{\partial H}{\partial x}+\zeta\Delta\mu x+\xi_X\ee
\be\dot{x}=-\gamma_{22}\frac{\partial H}{\partial x}-\gamma_{12}\frac{\partial H}{\partial X}+\xi_x\ee
with the noise correlators
$\langle\xi_X(t)\xi_X(t')\rangle=2T\Gamma\delta(t-t')$, $\langle\xi_x(t)\xi_x(t')\rangle=2T\gamma_{22}\delta(t-t')$ and $\langle\xi_X(t)\xi_x(t')\rangle=2T\gamma_{12}\delta(t-t')$.
This has the action
\begin{multline}\mathcal{A}=\frac{1}{4T(\Gamma\gamma_{22}-\gamma_{12}^2)}\int dt \gamma_{22}\left[\dot{X}+\Gamma\frac{\partial H}{\partial X}+\gamma_{12}\frac{\partial H}{\partial x}-\zeta\Delta\mu x\right]^2+\Gamma\left[\dot{x}+\gamma_{22}\frac{\partial H}{\partial x}+\gamma_{12}\frac{\partial H}{\partial X}\right]^2\\-2\gamma_{12}\left[\dot{X}+\Gamma\frac{\partial H}{\partial X}+\gamma_{12}\frac{\partial H}{\partial x}-\zeta\Delta\mu x\right]\left[\dot{x}+\gamma_{22}\frac{\partial H}{\partial x}+\gamma_{12}\frac{\partial H}{\partial X}\right]\end{multline}

Implement two different time reversal protocols (i) $t\to -t$, $X\to X$, $x\to x$ (ii) $t\to -t$, $X\to X$, $x\to -x$.
Note that in case (ii) in the absence of $\Delta\mu$ the off-diagonal kinetic coefficient in the x equation is reversible and should have the opposite sign in the equilibrium limit.  
For (i), the time-reversed action is
\begin{multline}A^R_{(i)}=\frac{1}{4T(\Gamma\gamma_{22}-\gamma_{12}^2)}\int dt \gamma_{22}\left[-\dot{X}+\Gamma\frac{\partial H}{\partial X}+\gamma_{12}\frac{\partial H}{\partial x}-\zeta\Delta\mu x\right]^2+\Gamma\left[-\dot{x}+\gamma_{22}\frac{\partial H}{\partial x}+\gamma_{12}\frac{\partial H}{\partial X}\right]^2\\-2\gamma_{12}\left[-\dot{X}+\Gamma\frac{\partial H}{\partial X}+\gamma_{12}\frac{\partial H}{\partial x}-\zeta\Delta\mu x\right]\left[-\dot{x}+\gamma_{22}\frac{\partial H}{\partial x}+\gamma_{12}\frac{\partial H}{\partial X}\right]\end{multline}

Defining $\sigma_{(i)}=\lim_{t\to\infty}({\mathcal{A}^R_{(i)}-\mathcal{A}})/{t}$, we get
\begin{multline}\sigma_{(i)}=\lim_{t\to\infty}\frac{1}{t}\frac{1}{T}\int dt\bigg[-\left(\dot{X}\frac{\partial H}{\partial X}+\dot{x}\frac{\partial H}{\partial x}\right)+\frac{\zeta\Delta\mu}{(\Gamma\gamma_{22}-\gamma_{12}^2)}\left(\gamma_{22}\dot{X}x-\gamma_{12}\dot{x}x\right)\bigg]\\=\lim_{t\to\infty}\frac{1}{t}\frac{1}{T}\int dt\bigg[-\dot{H}+\frac{\zeta\Delta\mu}{(\Gamma\gamma_{22}-\gamma_{12}^2)}\left(\gamma_{22}\dot{X}x-\gamma_{12}\dot{x^2}\right)\bigg]=\frac{\gamma_{22}\zeta\Delta\mu}{T(\Gamma\gamma_{22}-\gamma_{12}^2)}\langle\dot{X}x\rangle\end{multline}
where the last line replaces a time-average by an ensemble average.

For (ii), the time-reversed action is
\begin{multline}\mathcal{A}^R_{(ii)}=\frac{1}{4T(\Gamma\gamma_{22}-\gamma_{12}^2)}\int dt \gamma_{22}\left[-\dot{X}+\Gamma\frac{\partial H}{\partial X}-\gamma_{12}\frac{\partial H}{\partial x}+\zeta\Delta\mu x\right]^2+\Gamma\left[\dot{x}-\gamma_{22}\frac{\partial H}{\partial x}+\gamma_{12}\frac{\partial H}{\partial X}\right]^2\\-2\gamma_{12}\left[-\dot{X}+\Gamma\frac{\partial H}{\partial X}-\gamma_{12}\frac{\partial H}{\partial x}+\zeta\Delta\mu x\right]\left[\dot{x}-\gamma_{22}\frac{\partial H}{\partial x}+\gamma_{12}\frac{\partial H}{\partial X}\right]\end{multline}

The ``entropy production rate'' can be formally calculated using the earlier definition. Note, this should be simply interpreted as the ratio of probabilities of the forward process and a backward process conditioned on $x$ being reversed.
\be \sigma_{(ii)}=\lim_{t\to\infty}\frac{1}{t}\frac{1}{T}\int dt\bigg[-\left(\dot{X}\frac{\partial H}{\partial X}+\dot{x}\frac{\partial H}{\partial x}\right)+\zeta\Delta\mu\frac{\partial H}{\partial X}x-\gamma_{12}\frac{\partial H}{\partial X}\frac{\partial H}{\partial x}+\frac{\gamma_{12}}{(\Gamma\gamma_{22}-\gamma_{12}^2)}(\dot{X}\dot{x}-\zeta\Delta\mu\dot{x}x)\bigg]\ee

Taking the limit and converting time-averages to ensemble averages we obtain
\be\sigma_{(ii)}=\frac{\zeta\Delta\mu}{T}\left\langle\frac{\partial H}{\partial X}x\right\rangle+\gamma_{12}\left(\frac{1}{(\Gamma\gamma_{22}-\gamma_{12}^2)}\langle\dot{X}\dot{x}\rangle-\left\langle\frac{\partial H}{\partial X}\frac{\partial H}{\partial x}\right\rangle\right)\ee

Note, as expected $\sigma_{(ii)}$, unlike $\sigma_{(i)}$, does not vanish in the $\Delta\mu\to 0$ limit. It is nonetheless a measure of departure from thermal equilibrium, but the \emph{extra} parameter governing that departure, when $\Delta\mu=0$, is the difference between the value $\gamma_{12}$ assigned to the off-diagonal kinetic coefficient $\gamma_{21}$, and the value it should have, namely, $-\gamma_{12}$. In the special case when $\gamma_{12}=0$, we obtain the more familiar forms
$\sigma_{(i)}={\zeta\Delta\mu}\langle\dot{X}x\rangle/{T\Gamma}$
and
$\sigma_{(ii)}={\zeta\Delta\mu}\left\langle({\partial H}/{\partial X})x\right\rangle/{T}$, 
both of which vanish in the $\Delta\mu\to 0$ limit. The sum of these two is, as expected,
${(\zeta\Delta\mu)^2\langle x^2\rangle/}{T\Gamma}$. Thus, $\Delta\mu=0$ leads to two different equally physically meaningful equilibrium limits only when $\Delta\mu$ performs a double duty -- breaking time-reversal symmetry and providing the only coupling between $X$ and $x$. 

When the passive couplings between $X$ and $x$ are anti-symmetric instead, we have the set of equations
\be\dot{X}=-\Gamma\frac{\partial H}{\partial X}-\gamma_{12}\frac{\partial H}{\partial x}+\zeta\Delta\mu x+\xi_X\ee
\be\dot{x}=-\gamma_{22}\frac{\partial H}{\partial x}+\gamma_{12}\frac{\partial H}{\partial X}+\xi_x\ee
with the noise correlators
$\langle\xi_X(t)\xi_X(t')\rangle=2T\Gamma\delta(t-t'),$ $\langle\xi_x(t)\xi_x(t')\rangle=2T\gamma_{22}\delta(t-t')$, $\langle\xi_X(t)\xi_x(t')\rangle=0$.
The stochastic Onsager-Machlup action is
\be\mathcal{A}=\frac{1}{4T}\int dt \frac{1}{\Gamma}\left[\dot{X}+\Gamma\frac{\partial H}{\partial X}+\gamma_{12}\frac{\partial H}{\partial x}-\zeta\Delta\mu x\right]^2+\frac{1}{\gamma_{22}}\left[\dot{x}+\gamma_{22}\frac{\partial H}{\partial x}-\gamma_{12}\frac{\partial H}{\partial X}\right]^2\ee

The first time reversal protocol is (i) $t\to -t$, $X\to X$, $x\to x$ which results in the time-reversed action

\be\mathcal{A}^R_{(i)}=\frac{1}{4T}\int dt \frac{1}{\Gamma}\left[-\dot{X}+\Gamma\frac{\partial H}{\partial X}+\gamma_{12}\frac{\partial H}{\partial x}-\zeta\Delta\mu x\right]^2+\frac{1}{\gamma_{22}}\left[-\dot{x}+\gamma_{22}\frac{\partial H}{\partial x}-\gamma_{12}\frac{\partial H}{\partial X}\right]^2\ee

leading to the entropy production
\be\sigma_{(i)}=\lim_{t\to \infty}\frac{1}{t}\frac{1}{T}\int\bigg[-\left(\dot{X}\frac{\partial H}{\partial X}+\dot{x}\frac{\partial H}{\partial x}\right)-\gamma_{12}\left(\frac{1}{\Gamma}\dot{X}\frac{\partial H}{\partial x}-\frac{1}{\gamma_{22}}\dot{x}\frac{\partial H}{\partial X}\right)+\frac{\zeta\Delta\mu}{\Gamma}\dot{X}x\bigg]\ee
Again, assuming ergodicity,
\be\sigma_{(i)}=-\frac{\gamma_{12}}{T}\left\langle \frac{1}{\Gamma}\dot{X}\frac{\partial H}{\partial x}-\frac{1}{\gamma_{22}}\dot{x}\frac{\partial H}{\partial X}\right\rangle+\frac{\zeta\Delta\mu}{T\Gamma}\langle\dot{X}x\rangle\ee
is shown to be non zero in the limit $\Delta\mu\to 0$ as expected and goes to $0$ if $\gamma_{12}=0$. Similarly, for the time-reversal operation (ii) $t\to -t$, $X\to X$, $x\to -x$,
\be\mathcal{A}^R_{(ii)}=\frac{1}{4T}\int dt \frac{1}{\Gamma}\left[-\dot{X}+\Gamma\frac{\partial H}{\partial X}-\gamma_{12}\frac{\partial H}{\partial x}+\zeta\Delta\mu x\right]^2+\frac{1}{\gamma_{22}}\left[\dot{x}-\gamma_{22}\frac{\partial H}{\partial x}-\gamma_{12}\frac{\partial H}{\partial X}\right]^2\ee
leading to the entropy production
\be\sigma_{(ii)}=\lim_{t\to \infty}\frac{1}{t}\frac{1}{T}\int\bigg[-\left(\dot{X}\frac{\partial H}{\partial X}+\dot{x}\frac{\partial H}{\partial x}\right)+\zeta\Delta\mu\frac{\partial H}{\partial X}x\bigg]=\frac{\zeta\Delta\mu}{T}\left\langle\frac{\partial H}{\partial X}x\right\rangle.\ee
As expected, this vanishes when $\Delta\mu\to 0$.
Thus, when the passive dynamics is not $x\to -x$ invariant, only one ``time-reversal prescription'' vanishes in the $\Delta\mu\to 0$ limit and it follows that it is {the only} admissible prescription.

{\section{Nonequilibrium fluctuation-dissipation relation for an active particle in a harmonic trap}}
\label{nfdt2}
\setcounter{equation}{0}
A pure AOUP (i.e. one without translational diffusion) in a harmonic potential is an equilibrium problem with an effective temperature
\begin{equation}
T_{eff}=T\frac{\upsilon^2\gamma}{k(\Gamma k+K\gamma)}
\end{equation}
In the case of a harmonic potential, we can calculate the covariance matrix from \eqref{mod1mom} and \eqref{mod1aux} as
\begin{equation}
{\bsf C}=T_{eff}\begin{pmatrix}\frac{(1+\beta)k(\Gamma k+K\gamma)+\upsilon^2\gamma}{\gamma K\upsilon^2} && \frac{1}{\upsilon}\\ \frac{1}{\upsilon} &&\frac{\Gamma(\Gamma k+K\gamma)}{\gamma\upsilon^2}\end{pmatrix}
\end{equation}
The steady-state distribution function is $\propto e^{(-\phi/T_{eff})}$ where the potential
\begin{equation}
\phi=\frac{1}{2}{\bf q}\cdot{\bsf U}\cdot{\bf q}
\end{equation}
with ${\bsf U}=T_{eff}{\bsf C}^{-1}$ and ${\bf q}=(X, x)$. We now rewrite the dynamical equations in the form
\begin{equation}
\label{eom}
\dot{\bf q}=-{\bsf L}\cdot\frac{\partial\phi}{\partial{\bf q}}+\boldsymbol{\eta}
\end{equation}
where $\langle\boldsymbol{\eta}(t)\boldsymbol{\eta}(t')\rangle=2{\bsf D}T_{eff}\delta(t-t')$ with 
\begin{equation}
\nonumber
{\bsf D}=\begin{pmatrix}\frac{k(1+\beta)(\Gamma k+K\gamma)}{\upsilon^2\Gamma\gamma}&&0\\0&&\frac{k(\Gamma k+K\gamma)\Gamma}{\gamma^2\upsilon^2}\end{pmatrix} \,\,\,\,\,\, \text{and} \,\,
{\bsf L}=-{\bsf F}\cdot{\bsf C}/T_{eff} \,\,\,\,\,\, \text{where} \,\,
{\bsf F}=\begin{pmatrix}-\frac{K}{\Gamma} && \frac{\upsilon}{\Gamma}\\0 &&-\frac{k}{\gamma}\end{pmatrix}.
\end{equation}
Note that ${\bsf L}+{\bsf L}^T=2{\bsf D}$.

To calculate the response function we now modify $\phi$: $\phi^h=\phi^0-{\bf h}\cdot{\bf x}$. We define the response function as
\begin{equation}
\mathcal{R}_{ij}=\left.\frac{\delta x_i}{\delta h_j}\right\rvert_{h_i=0}=(-i\omega{\bsf I}-{\bsf F})^{-1}\cdot{\bsf L}
\end{equation}
where ${\bsf I}$ is the identity. Then,
\begin{equation}
\label{FDT1}
T_{eff}i[\mathcal{R}_{ji}(-\omega)-\mathcal{R}_{ij}(\omega)]=2T_{eff}\chi''_{ij}(\omega)=\omega\mathcal{C}_{ij}(\omega)
\end{equation}
where $\mathcal{C}_{ij}$ is the correlation matrix (note that $\int (d\omega/2\pi)\bm{\mathcal{C}}(\omega)={\bsf C}$). This relation holds for arbitrary $\beta$ and, in particular, also for $\beta=-1$ which corresponds to the zero translational diffusion case. For $\beta=-1$, the equation of motion becomes
\begin{equation}
\label{inertia}
\frac{\Gamma\gamma}{k}\ddot{X}=-\left(\frac{K\gamma}{k}+\Gamma\right)\dot{X}-KX+\frac{\upsilon}{k}\xi^p
\end{equation}
Adding a perturbing force to \eqref{inertia} results in the response function with an imaginary part
\begin{equation}
\label{resp}
\tilde{\chi}''_{XX0}(\omega)=\frac{k(k\Gamma+K\gamma)\omega}{(K^2+\Gamma^2\omega^2)(k^2+\gamma^2\omega^2)}
\end{equation}
and the fluctuation-dissipation theorem $ 2T_{eff}\tilde{\chi}''_{XX0}(\omega)=\omega{C}_{XX0}(\omega)$.

Note that despite the formal similarity to an equilibrium FDT, there is an important difference: in equilibrium, we would have added the perturbation directly to the Hamiltonian i.e., the perturbing force would have appeared directly in original equations of motion. We do not do that here. The force that appears in the equation of motion is ${\bsf L}\cdot{\bf h}$. However, the response function is defined not with respect to this force but purely with respect to ${\bf h}$. This FDT was first explicitly calculated by Eyink, Lebowitz and Spohn \cite{eyink1996hydrodynamics}, inspired by a more general result by Graham \cite{graham1977covariant}. Basically, this finds a force the response to which equals the correlation function in the steady state. This is distinct from the fluctuation-dissipation relation discussed in \cite{prost2009generalized} which amounts to adding perturbing forces to \emph{original} equations of motion and finding the correlation functions that the response to these forces are equal to. For linear dynamics, these variables whose correlators correspond to the response are given for our system as $({\bsf F}^{-1})^T\cdot{\bsf U}\cdot{\bf x}$. 

Finally, one might argue that the strength of noise should be immaterial for the response to an external physical force. Despite the response functions we calculate not being responses to physical external forces, let me examine this. Basically, let us calculate the response to a perturbation when the translational diffusion is 0 i.e. use \eqref{resp} as the response function even in the presence of translational diffusion. The $XX$ correlation function in the presence of translational diffusion has the value
\begin{equation}
\mathcal{C}_{XX}(\omega)=2T\frac{\upsilon^2\gamma+(1+\beta)[\Gamma(k^2+\gamma^2\omega^2)]}{(K^2+\Gamma^2\omega^2)(k^2+\gamma^2\omega^2)}
\end{equation}
This implies
\begin{equation}
\mathcal{C}_{XX}(\omega)-\frac{2T_{eff}}{\omega}\tilde{\chi}''_{XX0}(\omega)=\frac{1+\beta}{\Gamma}\frac{2T}{\omega^2+K^2/\Gamma^2}
\end{equation}
This implies that at short times i.e. when $\omega\gg K/\Gamma$, the measured correlation function will seem to satisfy a fluctuation-dissipation relation with the response function measured at $0$ translational diffusion, even when translational diffusion is present. This denotes the time-scale below which AOUP seems to have inertia even in the presence of translational diffusion.

\section{Harada-Sasa relation active field theory for polar rods}
\label{hsaftapp}
\setcounter{equation}{0}
In this appendix, we derive the Harada-Sasa relation for the field theory of active polar rods. To calculate the response functions, we add $-h_1\rho$ to $F_\rho$ and $-{\bf h}_2\cdot{\bf p}$ to $F_{\bf p}$. The action becomes
\begin{equation}
\mathcal{A}^{OM}=\frac{1}{4}\int\left[-\frac{1}{T}\left\{\dot{\rho}+v\nabla\cdot(\rho{\bf p})-\nabla^2\frac{\delta F}{\delta \rho}+\nabla^2h_1\right\}\nabla^{-2}\left\{\dot{\rho}+v\nabla\cdot(\rho{\bf p})-\nabla^2\frac{\delta F}{\delta \rho}+\nabla^2h_1\right\}+\frac{1}{T}\left(\dot{{\bf p}}+\lambda{\bf p}\cdot\nabla{\bf p}+\frac{\delta F}{\delta {\bf p}}-{\bf h}_2\right)^2\right]
\end{equation}
with the part linear in $h_1$ and ${\bf h}_2$ being
\begin{equation}
\delta\mathcal{A}^{OM}=-\frac{1}{2}\int \left[\frac{1}{T}h_1\left\{\dot{\rho}+v\nabla\cdot(\rho{\bf p})-\nabla^2\frac{\delta F}{\delta \rho}\right\}+\frac{1}{T}{\bf h}_2\cdot\left(\dot{{\bf p}}+\lambda{\bf p}\cdot\nabla{\bf p}+\frac{\delta F}{\delta {\bf p}}\right)\right]
\end{equation}
Directly evaluating the response functions from $\delta\mathcal{A}^{OM}$ we find 
%
%
\begin{equation}
\mathcal{R}_{\rho\rho}(t)-\mathcal{R}_{\rho\rho}(-t)=-\frac{1}{T}\partial_t\left\langle \rho({\bf x},t)\rho({\bf x}, 0)\right\rangle+\frac{1}{2T}\left\langle\rho({\bf x},t)\left[v\nabla\cdot(\rho{\bf p})-\nabla^2\frac{\delta F}{\delta \rho}\right]({\bf x}, 0)-\left[v\nabla\cdot(\rho{\bf p})-\nabla^2\frac{\delta F}{\delta \rho}\right]({\bf x}, t)\rho({\bf x},0)\right\rangle
\end{equation}
which implies 
\begin{equation}
T[\mathcal{R}_{\rho\rho}(t)-\mathcal{R}_{\rho\rho}(-t)]+\partial_tC_{\rho\rho}=\frac{1}{2}\left\langle\rho({\bf x},t)\left[v\nabla\cdot(\rho{\bf p})-\nabla^2\frac{\delta F}{\delta \rho}\right]({\bf x}, 0)-\left[v\nabla\cdot(\rho{\bf p})-\nabla^2\frac{\delta F}{\delta \rho}\right]({\bf x}, t)\rho({\bf x},0)\right\rangle
\end{equation}
Similarly, 
\begin{equation}
T[\mathcal{R}_{pp}(t)-\mathcal{R}_{pp}(-t)]+\partial_tC_{pp}=\frac{1}{2}\left\langle{\bf p}({\bf x}, t)\cdot\left(\lambda{\bf p}\cdot\nabla{\bf p}+\frac{\delta F}{\delta {\bf p}}\right)({\bf x},0)-\left(\lambda{\bf p}\cdot\nabla{\bf p}+\frac{\delta F}{\delta {\bf p}}\right)({\bf x},t)\cdot{\bf p}({\bf x}, 0)\right\rangle
\end{equation}
The entropy production rate can now be expressed in terms of the response and correlation functions using the definition and the expression for the entropy production rate in \ref{HarSalabel}:
{
\begin{equation}
\sigma=\frac{1}{T}\lim_{t\to 0}\int d{\bf x}\partial_t\left[\nabla^{-2}\left(T[\mathcal{R}_{\rho\rho}(t)-\mathcal{R}_{\rho\rho}(-t)]+\partial_tC_{\rho\rho}\right)-\left(T[\mathcal{R}_{pp}(t)-\mathcal{R}_{pp}(-t)]+\partial_tC_{pp}\right)\right].
\end{equation}
This is the Harada-Sasa relation for the Toner-Tu model.
Defining $[\mathcal{R}_{\rho\rho}(t)-\mathcal{R}_{\rho\rho}(-t)]=2i\chi''_{\rho\rho}({\bf q},\omega)$ and $[\mathcal{R}_{pp}(t)-\mathcal{R}_{pp}(-t)]=2i\chi''_{pp}({\bf q},\omega)$, where ${\bf q}$ is the wavevector, we find an expression for the frequency and wavevector resolved entropy production $\sigma=\int d\omega d{\bf q}\sigma_{\omega{\bf q}}$,
\begin{equation}
\sigma_{\omega{\bf q}}=\frac{\omega}{(2\pi)^{d+1}T}\left[q^2\{\omega C_{\rho\rho}({\bf q},\omega)-2T\chi''_{\rho\rho}({\bf q},\omega)\}+\{\omega C_{pp}({\bf q},\omega)-2T\chi''_{pp}({\bf q},\omega)\}\right]
\end{equation}
This expression opens up the possibility pf quantitatively measuring the departure of a flock from equilibrium through difficult, but in principle feasible correlation and response measurements. }

\section{Entropy production in the homogeneous polarised phase in the $T\to 0$ limit}
\label{entotapp}
\setcounter{equation}{0}

In active model B, entropy production is small within homogeneous phases. We now show that this is not the case for the Toner-Tu model. We expand both $\rho$ and ${\bf p}$ in noise-strength in the weak-noise limit. 
$
\rho=\rho_0+\sqrt{T}\rho_1+T\rho_2+...
$; and 
$
{\bf p}={\bf p}_0+\sqrt{T}{\bf p}_1+T{\bf p}_2+...
$
Assuming (though the form of the free-energy does not matter)
\begin{equation}
F=\int d{\bf x}\left[\frac{A}{2}\rho^2+\frac{\alpha}{2}p^2+\frac{\beta}{4}p^4+\ell{\bf p}\cdot\nabla\rho+\frac{K}{2}(\nabla{\bf p})^2\right]
\end{equation}
we obtain the equations of motion

\begin{equation}
\dot{\rho}_0=-v\nabla\cdot(\rho_0{\bf p}_0)+\nabla^2\mu_0
\end{equation}
\begin{equation}
\dot{\rho}_1=-v\nabla\cdot(\rho_1{\bf p}_0+\rho_0{\bf p}_1)+\nabla^2\mu_1+\xi
\end{equation}
\begin{equation}
\dot{{\bf p}}_0+\lambda{\bf p}_0\cdot\nabla{\bf p}_0=-{\bf H}_0
\end{equation}
\begin{equation}
\dot{{\bf p}}_1+\lambda({\bf p}_1\cdot\nabla{\bf p}_0+{\bf p}_0\cdot\nabla{\bf p}_1)=-{\bf H}_1+\boldsymbol{\xi}^p
\end{equation}
where $
{\bf H}_0=\alpha{\bf p}_0+\beta p^2{\bf p}_0+\ell\nabla\rho_0-K\nabla^2{\bf p}_0,
$
$
{\bf H}_1=\alpha{\bf p}_1+3\beta p^2{\bf p}_1+\ell\nabla\rho_1-K\nabla^2{\bf p}_1,
$
$
\mu_0=A\rho_0-\ell\nabla\cdot{\bf p}_0,
$ and
$
\mu_1=A\rho_1-\ell\nabla\cdot{\bf p}_1
$
The action, to zeroth order in noise-strength, is
\begin{equation}
\mathcal{A}^{OM}=\int dtd{\bf x}[-\{\dot{\rho}_1+v\nabla\cdot(\rho_1{\bf p}_0+\rho_0{\bf p}_1)-\nabla^2\mu_1\}\nabla^{-2}\{\dot{\rho}_1+v\nabla\cdot(\rho_1{\bf p}_0+\rho_0{\bf p}_1-\nabla^2\mu_1\}+\{\dot{{\bf p}}_1+\lambda({\bf p}_1\cdot\nabla{\bf p}_0+{\bf p}_0\cdot\nabla{\bf p}_1)+{\bf H}_1\}^2]
\end{equation}
We now assume that both $\rho_0$ and ${\bf p}_0$ relaxes to constant profiles. Even then, there is a part of entropy production rate that is independent of noise strength:
\begin{equation}
\sigma_0=v\langle\dot{\rho}_1\nabla^{-2}(\rho_0\nabla\cdot{\bf p}_1+{\bf p}_0\cdot\nabla\rho_1\rangle-\lambda\langle\dot{{\bf p}}_1\cdot({\bf p}_0\cdot\nabla){\bf p}_1\rangle+O(\sqrt{T})
\end{equation}
This implies that even homogeneous phases have an entropy-production rate at $O(T_1^0)$ unlike in active scalar models (model $A$ or $B$). This is to be expected on symmetry grounds.

\bibliographystyle{apsrev4-1}
\bibliography{bibfile}

\begin{thebibliography}{87}%
\makeatletter
\providecommand \@ifxundefined [1]{%
 \@ifx{#1\undefined}
}%
\providecommand \@ifnum [1]{%
 \ifnum #1\expandafter \@firstoftwo
 \else \expandafter \@secondoftwo
 \fi
}%
\providecommand \@ifx [1]{%
 \ifx #1\expandafter \@firstoftwo
 \else \expandafter \@secondoftwo
 \fi
}%
\providecommand \natexlab [1]{#1}%
\providecommand \enquote  [1]{``#1''}%
\providecommand \bibnamefont  [1]{#1}%
\providecommand \bibfnamefont [1]{#1}%
\providecommand \citenamefont [1]{#1}%
\providecommand \href@noop [0]{\@secondoftwo}%
\providecommand \href [0]{\begingroup \@sanitize@url \@href}%
\providecommand \@href[1]{\@@startlink{#1}\@@href}%
\providecommand \@@href[1]{\endgroup#1\@@endlink}%
\providecommand \@sanitize@url [0]{\catcode `\\12\catcode `\$12\catcode
  `\&12\catcode `\#12\catcode `\^12\catcode `\_12\catcode `\%12\relax}%
\providecommand \@@startlink[1]{}%
\providecommand \@@endlink[0]{}%
\providecommand \url  [0]{\begingroup\@sanitize@url \@url }%
\providecommand \@url [1]{\endgroup\@href {#1}{\urlprefix }}%
\providecommand \urlprefix  [0]{URL }%
\providecommand \Eprint [0]{\href }%
\providecommand \doibase [0]{http://dx.doi.org/}%
\providecommand \selectlanguage [0]{\@gobble}%
\providecommand \bibinfo  [0]{\@secondoftwo}%
\providecommand \bibfield  [0]{\@secondoftwo}%
\providecommand \translation [1]{[#1]}%
\providecommand \BibitemOpen [0]{}%
\providecommand \bibitemStop [0]{}%
\providecommand \bibitemNoStop [0]{.\EOS\space}%
\providecommand \EOS [0]{\spacefactor3000\relax}%
\providecommand \BibitemShut  [1]{\csname bibitem#1\endcsname}%
\let\auto@bib@innerbib\@empty
\bibitem [{\citenamefont {Ramaswamy}(2010)}]{ramaswamy2010mechanics}%
  \BibitemOpen
  \bibfield  {author} {\bibinfo {author} {\bibfnamefont {S.}~\bibnamefont
  {Ramaswamy}},\ }\href {\doibase 10.1146/annurev-conmatphys-070909-104101}
  {\bibfield  {journal} {\bibinfo  {journal} {Annual Review of Condensed Matter
  Physics}\ }\textbf {\bibinfo {volume} {1}},\ \bibinfo {pages} {323} (\bibinfo
  {year} {2010})},\ \Eprint
  {http://arxiv.org/abs/https://doi.org/10.1146/annurev-conmatphys-070909-104101}
  {https://doi.org/10.1146/annurev-conmatphys-070909-104101} \BibitemShut
  {NoStop}%
\bibitem [{\citenamefont {Marchetti}\ \emph {et~al.}(2013)\citenamefont
  {Marchetti}, \citenamefont {Joanny}, \citenamefont {Ramaswamy}, \citenamefont
  {Liverpool}, \citenamefont {Prost}, \citenamefont {Rao},\ and\ \citenamefont
  {Simha}}]{marchetti2013hydrodynamics}%
  \BibitemOpen
  \bibfield  {author} {\bibinfo {author} {\bibfnamefont {M.~C.}\ \bibnamefont
  {Marchetti}}, \bibinfo {author} {\bibfnamefont {J.-F.}\ \bibnamefont
  {Joanny}}, \bibinfo {author} {\bibfnamefont {S.}~\bibnamefont {Ramaswamy}},
  \bibinfo {author} {\bibfnamefont {T.~B.}\ \bibnamefont {Liverpool}}, \bibinfo
  {author} {\bibfnamefont {J.}~\bibnamefont {Prost}}, \bibinfo {author}
  {\bibfnamefont {M.}~\bibnamefont {Rao}}, \ and\ \bibinfo {author}
  {\bibfnamefont {R.~A.}\ \bibnamefont {Simha}},\ }\href@noop {} {\bibfield
  {journal} {\bibinfo  {journal} {Reviews of Modern Physics}\ }\textbf
  {\bibinfo {volume} {85}},\ \bibinfo {pages} {1143} (\bibinfo {year}
  {2013})}\BibitemShut {NoStop}%
\bibitem [{\citenamefont {Fodor}\ \emph {et~al.}(2016)\citenamefont {Fodor},
  \citenamefont {Nardini}, \citenamefont {Cates}, \citenamefont {Tailleur},
  \citenamefont {Visco},\ and\ \citenamefont {van Wijland}}]{fodor2016far}%
  \BibitemOpen
  \bibfield  {author} {\bibinfo {author} {\bibfnamefont {{\'E}.}~\bibnamefont
  {Fodor}}, \bibinfo {author} {\bibfnamefont {C.}~\bibnamefont {Nardini}},
  \bibinfo {author} {\bibfnamefont {M.~E.}\ \bibnamefont {Cates}}, \bibinfo
  {author} {\bibfnamefont {J.}~\bibnamefont {Tailleur}}, \bibinfo {author}
  {\bibfnamefont {P.}~\bibnamefont {Visco}}, \ and\ \bibinfo {author}
  {\bibfnamefont {F.}~\bibnamefont {van Wijland}},\ }\href@noop {} {\bibfield
  {journal} {\bibinfo  {journal} {Physical Review Letters}\ }\textbf {\bibinfo
  {volume} {117}},\ \bibinfo {pages} {038103} (\bibinfo {year}
  {2016})}\BibitemShut {NoStop}%
\bibitem [{\citenamefont {J{\"u}licher}\ \emph {et~al.}(1997)\citenamefont
  {J{\"u}licher}, \citenamefont {Ajdari},\ and\ \citenamefont
  {Prost}}]{julicher1997modeling}%
  \BibitemOpen
  \bibfield  {author} {\bibinfo {author} {\bibfnamefont {F.}~\bibnamefont
  {J{\"u}licher}}, \bibinfo {author} {\bibfnamefont {A.}~\bibnamefont
  {Ajdari}}, \ and\ \bibinfo {author} {\bibfnamefont {J.}~\bibnamefont
  {Prost}},\ }\href@noop {} {\bibfield  {journal} {\bibinfo  {journal} {Reviews
  of Modern Physics}\ }\textbf {\bibinfo {volume} {69}},\ \bibinfo {pages}
  {1269} (\bibinfo {year} {1997})}\BibitemShut {NoStop}%
\bibitem [{\citenamefont {Kruse}\ \emph {et~al.}(2005)\citenamefont {Kruse},
  \citenamefont {Joanny}, \citenamefont {J{\"u}licher}, \citenamefont {Prost},\
  and\ \citenamefont {Sekimoto}}]{kruse2005generic}%
  \BibitemOpen
  \bibfield  {author} {\bibinfo {author} {\bibfnamefont {K.}~\bibnamefont
  {Kruse}}, \bibinfo {author} {\bibfnamefont {J.-F.}\ \bibnamefont {Joanny}},
  \bibinfo {author} {\bibfnamefont {F.}~\bibnamefont {J{\"u}licher}}, \bibinfo
  {author} {\bibfnamefont {J.}~\bibnamefont {Prost}}, \ and\ \bibinfo {author}
  {\bibfnamefont {K.}~\bibnamefont {Sekimoto}},\ }\href@noop {} {\bibfield
  {journal} {\bibinfo  {journal} {The European Physical Journal E}\ }\textbf
  {\bibinfo {volume} {16}},\ \bibinfo {pages} {5} (\bibinfo {year}
  {2005})}\BibitemShut {NoStop}%
\bibitem [{\citenamefont {Joanny}\ \emph {et~al.}(2007)\citenamefont {Joanny},
  \citenamefont {J{\"u}licher}, \citenamefont {Kruse},\ and\ \citenamefont
  {Prost}}]{joanny2007hydrodynamic}%
  \BibitemOpen
  \bibfield  {author} {\bibinfo {author} {\bibfnamefont {J.-F.}\ \bibnamefont
  {Joanny}}, \bibinfo {author} {\bibfnamefont {F.}~\bibnamefont
  {J{\"u}licher}}, \bibinfo {author} {\bibfnamefont {K.}~\bibnamefont {Kruse}},
  \ and\ \bibinfo {author} {\bibfnamefont {J.}~\bibnamefont {Prost}},\
  }\href@noop {} {\bibfield  {journal} {\bibinfo  {journal} {New Journal of
  Physics}\ }\textbf {\bibinfo {volume} {9}},\ \bibinfo {pages} {422} (\bibinfo
  {year} {2007})}\BibitemShut {NoStop}%
\bibitem [{\citenamefont {J{\"u}licher}\ \emph {et~al.}(2007)\citenamefont
  {J{\"u}licher}, \citenamefont {Kruse}, \citenamefont {Prost},\ and\
  \citenamefont {Joanny}}]{juelicher2007active}%
  \BibitemOpen
  \bibfield  {author} {\bibinfo {author} {\bibfnamefont {F.}~\bibnamefont
  {J{\"u}licher}}, \bibinfo {author} {\bibfnamefont {K.}~\bibnamefont {Kruse}},
  \bibinfo {author} {\bibfnamefont {J.}~\bibnamefont {Prost}}, \ and\ \bibinfo
  {author} {\bibfnamefont {J.-F.}\ \bibnamefont {Joanny}},\ }\href@noop {}
  {\bibfield  {journal} {\bibinfo  {journal} {Physics Reports}\ }\textbf
  {\bibinfo {volume} {449}},\ \bibinfo {pages} {3} (\bibinfo {year}
  {2007})}\BibitemShut {NoStop}%
\bibitem [{\citenamefont {J{\"u}licher}\ \emph {et~al.}(2018)\citenamefont
  {J{\"u}licher}, \citenamefont {Grill},\ and\ \citenamefont
  {Salbreux}}]{julicher2018hydrodynamic}%
  \BibitemOpen
  \bibfield  {author} {\bibinfo {author} {\bibfnamefont {F.}~\bibnamefont
  {J{\"u}licher}}, \bibinfo {author} {\bibfnamefont {S.~W.}\ \bibnamefont
  {Grill}}, \ and\ \bibinfo {author} {\bibfnamefont {G.}~\bibnamefont
  {Salbreux}},\ }\href@noop {} {\bibfield  {journal} {\bibinfo  {journal}
  {Reports on Progress in Physics}\ }\textbf {\bibinfo {volume} {81}} (\bibinfo
  {year} {2018})}\BibitemShut {NoStop}%
\bibitem [{\citenamefont {Pietzonka}\ and\ \citenamefont
  {Seifert}(2017)}]{pietzonka2017entropy}%
  \BibitemOpen
  \bibfield  {author} {\bibinfo {author} {\bibfnamefont {P.}~\bibnamefont
  {Pietzonka}}\ and\ \bibinfo {author} {\bibfnamefont {U.}~\bibnamefont
  {Seifert}},\ }\href@noop {} {\bibfield  {journal} {\bibinfo  {journal}
  {Journal of Physics A: Mathematical and Theoretical}\ }\textbf {\bibinfo
  {volume} {51}},\ \bibinfo {pages} {01LT01} (\bibinfo {year}
  {2017})}\BibitemShut {NoStop}%
\bibitem [{\citenamefont {Neri}\ \emph {et~al.}(2017)\citenamefont {Neri},
  \citenamefont {Rold{\'a}n},\ and\ \citenamefont
  {J{\"u}licher}}]{neri2017statistics}%
  \BibitemOpen
  \bibfield  {author} {\bibinfo {author} {\bibfnamefont {I.}~\bibnamefont
  {Neri}}, \bibinfo {author} {\bibfnamefont {{\'E}.}~\bibnamefont
  {Rold{\'a}n}}, \ and\ \bibinfo {author} {\bibfnamefont {F.}~\bibnamefont
  {J{\"u}licher}},\ }\href@noop {} {\bibfield  {journal} {\bibinfo  {journal}
  {Physical Review X}\ }\textbf {\bibinfo {volume} {7}},\ \bibinfo {pages}
  {011019} (\bibinfo {year} {2017})}\BibitemShut {NoStop}%
\bibitem [{\citenamefont {Shankar}\ and\ \citenamefont
  {Marchetti}(2018)}]{shankar2018hidden}%
  \BibitemOpen
  \bibfield  {author} {\bibinfo {author} {\bibfnamefont {S.}~\bibnamefont
  {Shankar}}\ and\ \bibinfo {author} {\bibfnamefont {M.~C.}\ \bibnamefont
  {Marchetti}},\ }\href@noop {} {\bibfield  {journal} {\bibinfo  {journal}
  {arXiv preprint arXiv:1804.03099}\ } (\bibinfo {year} {2018})}\BibitemShut
  {NoStop}%
\bibitem [{\citenamefont {Pigolotti}\ \emph {et~al.}(2017)\citenamefont
  {Pigolotti}, \citenamefont {Neri}, \citenamefont {Rold{\'a}n},\ and\
  \citenamefont {J{\"u}licher}}]{pigolotti2017generic}%
  \BibitemOpen
  \bibfield  {author} {\bibinfo {author} {\bibfnamefont {S.}~\bibnamefont
  {Pigolotti}}, \bibinfo {author} {\bibfnamefont {I.}~\bibnamefont {Neri}},
  \bibinfo {author} {\bibfnamefont {{\'E}.}~\bibnamefont {Rold{\'a}n}}, \ and\
  \bibinfo {author} {\bibfnamefont {F.}~\bibnamefont {J{\"u}licher}},\
  }\href@noop {} {\bibfield  {journal} {\bibinfo  {journal} {Physical Review
  Letters}\ }\textbf {\bibinfo {volume} {119}},\ \bibinfo {pages} {140604}
  (\bibinfo {year} {2017})}\BibitemShut {NoStop}%
\bibitem [{\citenamefont {Kumar}\ \emph {et~al.}(2015)\citenamefont {Kumar},
  \citenamefont {Soni}, \citenamefont {Ramaswamy},\ and\ \citenamefont
  {Sood}}]{kumar2015anisotropic}%
  \BibitemOpen
  \bibfield  {author} {\bibinfo {author} {\bibfnamefont {N.}~\bibnamefont
  {Kumar}}, \bibinfo {author} {\bibfnamefont {H.}~\bibnamefont {Soni}},
  \bibinfo {author} {\bibfnamefont {S.}~\bibnamefont {Ramaswamy}}, \ and\
  \bibinfo {author} {\bibfnamefont {A.~K.}\ \bibnamefont {Sood}},\ }\href@noop
  {} {\bibfield  {journal} {\bibinfo  {journal} {Physical Review E}\ }\textbf
  {\bibinfo {volume} {91}},\ \bibinfo {pages} {030102} (\bibinfo {year}
  {2015})}\BibitemShut {NoStop}%
\bibitem [{\citenamefont {Schimansky-Geier}\ \emph {et~al.}(1995)\citenamefont
  {Schimansky-Geier}, \citenamefont {Mieth}, \citenamefont {Ros{\'e}},\ and\
  \citenamefont {Malchow}}]{schimansky1995structure}%
  \BibitemOpen
  \bibfield  {author} {\bibinfo {author} {\bibfnamefont {L.}~\bibnamefont
  {Schimansky-Geier}}, \bibinfo {author} {\bibfnamefont {M.}~\bibnamefont
  {Mieth}}, \bibinfo {author} {\bibfnamefont {H.}~\bibnamefont {Ros{\'e}}}, \
  and\ \bibinfo {author} {\bibfnamefont {H.}~\bibnamefont {Malchow}},\
  }\href@noop {} {\bibfield  {journal} {\bibinfo  {journal} {Physics Letters
  A}\ }\textbf {\bibinfo {volume} {207}},\ \bibinfo {pages} {140} (\bibinfo
  {year} {1995})}\BibitemShut {NoStop}%
\bibitem [{\citenamefont {Ebeling}\ \emph {et~al.}(1999)\citenamefont
  {Ebeling}, \citenamefont {Schweitzer},\ and\ \citenamefont
  {Tilch}}]{ebeling1999active}%
  \BibitemOpen
  \bibfield  {author} {\bibinfo {author} {\bibfnamefont {W.}~\bibnamefont
  {Ebeling}}, \bibinfo {author} {\bibfnamefont {F.}~\bibnamefont {Schweitzer}},
  \ and\ \bibinfo {author} {\bibfnamefont {B.}~\bibnamefont {Tilch}},\
  }\href@noop {} {\bibfield  {journal} {\bibinfo  {journal} {BioSystems}\
  }\textbf {\bibinfo {volume} {49}},\ \bibinfo {pages} {17} (\bibinfo {year}
  {1999})}\BibitemShut {NoStop}%
\bibitem [{\citenamefont {Romanczuk}\ \emph {et~al.}(2012)\citenamefont
  {Romanczuk}, \citenamefont {B{\"a}r}, \citenamefont {Ebeling}, \citenamefont
  {Lindner},\ and\ \citenamefont {Schimansky-Geier}}]{romanczuk2012active}%
  \BibitemOpen
  \bibfield  {author} {\bibinfo {author} {\bibfnamefont {P.}~\bibnamefont
  {Romanczuk}}, \bibinfo {author} {\bibfnamefont {M.}~\bibnamefont {B{\"a}r}},
  \bibinfo {author} {\bibfnamefont {W.}~\bibnamefont {Ebeling}}, \bibinfo
  {author} {\bibfnamefont {B.}~\bibnamefont {Lindner}}, \ and\ \bibinfo
  {author} {\bibfnamefont {L.}~\bibnamefont {Schimansky-Geier}},\ }\href@noop
  {} {\bibfield  {journal} {\bibinfo  {journal} {The European Physical Journal
  Special Topics}\ }\textbf {\bibinfo {volume} {202}},\ \bibinfo {pages} {1}
  (\bibinfo {year} {2012})}\BibitemShut {NoStop}%
\bibitem [{\citenamefont {Maggi}\ \emph {et~al.}(2014)\citenamefont {Maggi},
  \citenamefont {Paoluzzi}, \citenamefont {Pellicciotta}, \citenamefont
  {Lepore}, \citenamefont {Angelani},\ and\ \citenamefont
  {Di~Leonardo}}]{maggi2014generalized}%
  \BibitemOpen
  \bibfield  {author} {\bibinfo {author} {\bibfnamefont {C.}~\bibnamefont
  {Maggi}}, \bibinfo {author} {\bibfnamefont {M.}~\bibnamefont {Paoluzzi}},
  \bibinfo {author} {\bibfnamefont {N.}~\bibnamefont {Pellicciotta}}, \bibinfo
  {author} {\bibfnamefont {A.}~\bibnamefont {Lepore}}, \bibinfo {author}
  {\bibfnamefont {L.}~\bibnamefont {Angelani}}, \ and\ \bibinfo {author}
  {\bibfnamefont {R.}~\bibnamefont {Di~Leonardo}},\ }\href@noop {} {\bibfield
  {journal} {\bibinfo  {journal} {Physical Review Letters}\ }\textbf {\bibinfo
  {volume} {113}},\ \bibinfo {pages} {238303} (\bibinfo {year}
  {2014})}\BibitemShut {NoStop}%
\bibitem [{\citenamefont {Koumakis}\ \emph {et~al.}(2014)\citenamefont
  {Koumakis}, \citenamefont {Maggi},\ and\ \citenamefont
  {Di~Leonardo}}]{koumakis2014directed}%
  \BibitemOpen
  \bibfield  {author} {\bibinfo {author} {\bibfnamefont {N.}~\bibnamefont
  {Koumakis}}, \bibinfo {author} {\bibfnamefont {C.}~\bibnamefont {Maggi}}, \
  and\ \bibinfo {author} {\bibfnamefont {R.}~\bibnamefont {Di~Leonardo}},\
  }\href@noop {} {\bibfield  {journal} {\bibinfo  {journal} {Soft matter}\
  }\textbf {\bibinfo {volume} {10}},\ \bibinfo {pages} {5695} (\bibinfo {year}
  {2014})}\BibitemShut {NoStop}%
\bibitem [{\citenamefont {Toner}\ and\ \citenamefont
  {Tu}(1995)}]{toner1995long}%
  \BibitemOpen
  \bibfield  {author} {\bibinfo {author} {\bibfnamefont {J.}~\bibnamefont
  {Toner}}\ and\ \bibinfo {author} {\bibfnamefont {Y.}~\bibnamefont {Tu}},\
  }\href@noop {} {\bibfield  {journal} {\bibinfo  {journal} {Physical Review
  Letters}\ }\textbf {\bibinfo {volume} {75}},\ \bibinfo {pages} {4326}
  (\bibinfo {year} {1995})}\BibitemShut {NoStop}%
\bibitem [{\citenamefont {Toner}\ and\ \citenamefont
  {Tu}(1998)}]{toner1998flocks}%
  \BibitemOpen
  \bibfield  {author} {\bibinfo {author} {\bibfnamefont {J.}~\bibnamefont
  {Toner}}\ and\ \bibinfo {author} {\bibfnamefont {Y.}~\bibnamefont {Tu}},\
  }\href@noop {} {\bibfield  {journal} {\bibinfo  {journal} {Physical Review
  E}\ }\textbf {\bibinfo {volume} {58}},\ \bibinfo {pages} {4828} (\bibinfo
  {year} {1998})}\BibitemShut {NoStop}%
\bibitem [{\citenamefont {Toner}\ \emph {et~al.}(2005)\citenamefont {Toner},
  \citenamefont {Tu},\ and\ \citenamefont
  {Ramaswamy}}]{toner2005hydrodynamics}%
  \BibitemOpen
  \bibfield  {author} {\bibinfo {author} {\bibfnamefont {J.}~\bibnamefont
  {Toner}}, \bibinfo {author} {\bibfnamefont {Y.}~\bibnamefont {Tu}}, \ and\
  \bibinfo {author} {\bibfnamefont {S.}~\bibnamefont {Ramaswamy}},\ }\href@noop
  {} {\bibfield  {journal} {\bibinfo  {journal} {Annals of Physics}\ }\textbf
  {\bibinfo {volume} {318}},\ \bibinfo {pages} {170} (\bibinfo {year}
  {2005})}\BibitemShut {NoStop}%
\bibitem [{\citenamefont {Sandford}\ \emph {et~al.}(2017)\citenamefont
  {Sandford}, \citenamefont {Grosberg},\ and\ \citenamefont
  {Joanny}}]{sandford2017pressure}%
  \BibitemOpen
  \bibfield  {author} {\bibinfo {author} {\bibfnamefont {C.}~\bibnamefont
  {Sandford}}, \bibinfo {author} {\bibfnamefont {A.~Y.}\ \bibnamefont
  {Grosberg}}, \ and\ \bibinfo {author} {\bibfnamefont {J.-F.}\ \bibnamefont
  {Joanny}},\ }\href@noop {} {\bibfield  {journal} {\bibinfo  {journal}
  {Physical Review E}\ }\textbf {\bibinfo {volume} {96}},\ \bibinfo {pages}
  {052605} (\bibinfo {year} {2017})}\BibitemShut {NoStop}%
\bibitem [{\citenamefont {Harada}\ and\ \citenamefont
  {Sasa}(2005)}]{harada2005equality}%
  \BibitemOpen
  \bibfield  {author} {\bibinfo {author} {\bibfnamefont {T.}~\bibnamefont
  {Harada}}\ and\ \bibinfo {author} {\bibfnamefont {S.-i.}\ \bibnamefont
  {Sasa}},\ }\href@noop {} {\bibfield  {journal} {\bibinfo  {journal} {Physical
  Review Letters}\ }\textbf {\bibinfo {volume} {95}},\ \bibinfo {pages}
  {130602} (\bibinfo {year} {2005})}\BibitemShut {NoStop}%
\bibitem [{\citenamefont {Wang}\ \emph {et~al.}(2016)\citenamefont {Wang},
  \citenamefont {Kawaguchi}, \citenamefont {Sasa},\ and\ \citenamefont
  {Tang}}]{wang2016entropy}%
  \BibitemOpen
  \bibfield  {author} {\bibinfo {author} {\bibfnamefont {S.-W.}\ \bibnamefont
  {Wang}}, \bibinfo {author} {\bibfnamefont {K.}~\bibnamefont {Kawaguchi}},
  \bibinfo {author} {\bibfnamefont {S.-i.}\ \bibnamefont {Sasa}}, \ and\
  \bibinfo {author} {\bibfnamefont {L.-H.}\ \bibnamefont {Tang}},\ }\href@noop
  {} {\bibfield  {journal} {\bibinfo  {journal} {Physical Review Letters}\
  }\textbf {\bibinfo {volume} {117}},\ \bibinfo {pages} {070601} (\bibinfo
  {year} {2016})}\BibitemShut {NoStop}%
\bibitem [{\citenamefont {Brand}\ \emph {et~al.}(2014)\citenamefont {Brand},
  \citenamefont {Pleiner},\ and\ \citenamefont {Sven{\v{s}}ek}}]{Brand2014}%
  \BibitemOpen
  \bibfield  {author} {\bibinfo {author} {\bibfnamefont {H.~R.}\ \bibnamefont
  {Brand}}, \bibinfo {author} {\bibfnamefont {H.}~\bibnamefont {Pleiner}}, \
  and\ \bibinfo {author} {\bibfnamefont {D.}~\bibnamefont {Sven{\v{s}}ek}},\
  }\href {\doibase 10.1140/epje/i2014-14083-4} {\bibfield  {journal} {\bibinfo
  {journal} {The European Physical Journal E}\ }\textbf {\bibinfo {volume}
  {37}},\ \bibinfo {pages} {83} (\bibinfo {year} {2014})}\BibitemShut {NoStop}%
\bibitem [{\citenamefont {Ramaswamy}(2017)}]{ramaswamy2017active}%
  \BibitemOpen
  \bibfield  {author} {\bibinfo {author} {\bibfnamefont {S.}~\bibnamefont
  {Ramaswamy}},\ }\href@noop {} {\bibfield  {journal} {\bibinfo  {journal}
  {Journal of Statistical Mechanics: Theory and Experiment}\ }\textbf {\bibinfo
  {volume} {2017}},\ \bibinfo {pages} {054002} (\bibinfo {year}
  {2017})}\BibitemShut {NoStop}%
\bibitem [{\citenamefont {Kumar}\ \emph {et~al.}(2008)\citenamefont {Kumar},
  \citenamefont {Ramaswamy},\ and\ \citenamefont {Rao}}]{kumar2008active}%
  \BibitemOpen
  \bibfield  {author} {\bibinfo {author} {\bibfnamefont {K.~V.}\ \bibnamefont
  {Kumar}}, \bibinfo {author} {\bibfnamefont {S.}~\bibnamefont {Ramaswamy}}, \
  and\ \bibinfo {author} {\bibfnamefont {M.}~\bibnamefont {Rao}},\ }\href@noop
  {} {\bibfield  {journal} {\bibinfo  {journal} {Physical Review E}\ }\textbf
  {\bibinfo {volume} {77}},\ \bibinfo {pages} {020102} (\bibinfo {year}
  {2008})}\BibitemShut {NoStop}%
\bibitem [{\citenamefont {Ma}\ and\ \citenamefont {Mazenko}(1975)}]{ma1975sk}%
  \BibitemOpen
  \bibfield  {author} {\bibinfo {author} {\bibfnamefont {S.-k.}\ \bibnamefont
  {Ma}}\ and\ \bibinfo {author} {\bibfnamefont {G.~F.}\ \bibnamefont
  {Mazenko}},\ }\href {\doibase 10.1103/PhysRevB.11.4077} {\bibfield  {journal}
  {\bibinfo  {journal} {Phys. Rev. B}\ }\textbf {\bibinfo {volume} {11}},\
  \bibinfo {pages} {4077} (\bibinfo {year} {1975})}\BibitemShut {NoStop}%
\bibitem [{\citenamefont {Chaikin}\ and\ \citenamefont
  {Lubensky}(1995)}]{chaikin1995principles}%
  \BibitemOpen
  \bibfield  {author} {\bibinfo {author} {\bibfnamefont {P.~M.}\ \bibnamefont
  {Chaikin}}\ and\ \bibinfo {author} {\bibfnamefont {T.~C.}\ \bibnamefont
  {Lubensky}},\ }\href@noop {} {\emph {\bibinfo {title} {Principles of
  condensed matter physics}}},\ Vol.~\bibinfo {volume} {1}\ (\bibinfo
  {publisher} {Cambridge university press Cambridge},\ \bibinfo {year}
  {1995})\BibitemShut {NoStop}%
\bibitem [{\citenamefont {Zwanzig}(2001)}]{zwanzig2001nonequilibrium}%
  \BibitemOpen
  \bibfield  {author} {\bibinfo {author} {\bibfnamefont {R.}~\bibnamefont
  {Zwanzig}},\ }\href@noop {} {\emph {\bibinfo {title} {Nonequilibrium
  Statistical Mechanics}}}\ (\bibinfo  {publisher} {Oxford University Press},\
  \bibinfo {year} {2001})\BibitemShut {NoStop}%
\bibitem [{\citenamefont {Lau}\ and\ \citenamefont
  {Lubensky}(2007)}]{lau2007state}%
  \BibitemOpen
  \bibfield  {author} {\bibinfo {author} {\bibfnamefont {A.~W.}\ \bibnamefont
  {Lau}}\ and\ \bibinfo {author} {\bibfnamefont {T.~C.}\ \bibnamefont
  {Lubensky}},\ }\href@noop {} {\bibfield  {journal} {\bibinfo  {journal}
  {Physical Review E}\ }\textbf {\bibinfo {volume} {76}},\ \bibinfo {pages}
  {011123} (\bibinfo {year} {2007})}\BibitemShut {NoStop}%
\bibitem [{\citenamefont {Casimir}(1945)}]{casimir1945onsager}%
  \BibitemOpen
  \bibfield  {author} {\bibinfo {author} {\bibfnamefont {H.~B.~G.}\
  \bibnamefont {Casimir}},\ }\href@noop {} {\bibfield  {journal} {\bibinfo
  {journal} {Reviews of Modern Physics}\ }\textbf {\bibinfo {volume} {17}},\
  \bibinfo {pages} {343} (\bibinfo {year} {1945})}\BibitemShut {NoStop}%
\bibitem [{\citenamefont {De~Groot}\ and\ \citenamefont
  {Mazur}(2013)}]{de2013non}%
  \BibitemOpen
  \bibfield  {author} {\bibinfo {author} {\bibfnamefont {S.~R.}\ \bibnamefont
  {De~Groot}}\ and\ \bibinfo {author} {\bibfnamefont {P.}~\bibnamefont
  {Mazur}},\ }\href@noop {} {\emph {\bibinfo {title} {Non-equilibrium
  thermodynamics}}}\ (\bibinfo  {publisher} {Courier Corporation},\ \bibinfo
  {year} {2013})\BibitemShut {NoStop}%
\bibitem [{\citenamefont {Mori}(1965)}]{mori1965transport}%
  \BibitemOpen
  \bibfield  {author} {\bibinfo {author} {\bibfnamefont {H.}~\bibnamefont
  {Mori}},\ }\href@noop {} {\bibfield  {journal} {\bibinfo  {journal} {Progress
  of theoretical physics}\ }\textbf {\bibinfo {volume} {33}},\ \bibinfo {pages}
  {423} (\bibinfo {year} {1965})}\BibitemShut {NoStop}%
\bibitem [{\citenamefont {Mazenko}(2008)}]{mazenko2008nonequilibrium}%
  \BibitemOpen
  \bibfield  {author} {\bibinfo {author} {\bibfnamefont {G.~F.}\ \bibnamefont
  {Mazenko}},\ }\href@noop {} {\emph {\bibinfo {title} {Nonequilibrium
  statistical mechanics}}}\ (\bibinfo  {publisher} {John Wiley \& Sons},\
  \bibinfo {year} {2008})\BibitemShut {NoStop}%
\bibitem [{\citenamefont {Hohenberg}\ and\ \citenamefont
  {Halperin}(1977)}]{hohenberg1977theory}%
  \BibitemOpen
  \bibfield  {author} {\bibinfo {author} {\bibfnamefont {P.~C.}\ \bibnamefont
  {Hohenberg}}\ and\ \bibinfo {author} {\bibfnamefont {B.~I.}\ \bibnamefont
  {Halperin}},\ }\href@noop {} {\bibfield  {journal} {\bibinfo  {journal}
  {Reviews of Modern Physics}\ }\textbf {\bibinfo {volume} {49}},\ \bibinfo
  {pages} {435} (\bibinfo {year} {1977})}\BibitemShut {NoStop}%
\bibitem [{\citenamefont {Baule}\ \emph {et~al.}(2008)\citenamefont {Baule},
  \citenamefont {Kumar},\ and\ \citenamefont {Ramaswamy}}]{baule2008exact}%
  \BibitemOpen
  \bibfield  {author} {\bibinfo {author} {\bibfnamefont {A.}~\bibnamefont
  {Baule}}, \bibinfo {author} {\bibfnamefont {K.~V.}\ \bibnamefont {Kumar}}, \
  and\ \bibinfo {author} {\bibfnamefont {S.}~\bibnamefont {Ramaswamy}},\
  }\href@noop {} {\bibfield  {journal} {\bibinfo  {journal} {Journal of
  Statistical Mechanics: Theory and Experiment}\ }\textbf {\bibinfo {volume}
  {2008}},\ \bibinfo {pages} {P11008} (\bibinfo {year} {2008})}\BibitemShut
  {NoStop}%
\bibitem [{\citenamefont {Uhlenbeck}\ and\ \citenamefont
  {Ornstein}(1930)}]{uhlenbeck1930theory}%
  \BibitemOpen
  \bibfield  {author} {\bibinfo {author} {\bibfnamefont {G.~E.}\ \bibnamefont
  {Uhlenbeck}}\ and\ \bibinfo {author} {\bibfnamefont {L.~S.}\ \bibnamefont
  {Ornstein}},\ }\href@noop {} {\bibfield  {journal} {\bibinfo  {journal}
  {Physical Review}\ }\textbf {\bibinfo {volume} {36}},\ \bibinfo {pages} {823}
  (\bibinfo {year} {1930})}\BibitemShut {NoStop}%
\bibitem [{\citenamefont {Ramaswamy}\ \emph {et~al.}(2000)\citenamefont
  {Ramaswamy}, \citenamefont {Toner},\ and\ \citenamefont
  {Prost}}]{ramaswamy2000nonequilibrium}%
  \BibitemOpen
  \bibfield  {author} {\bibinfo {author} {\bibfnamefont {S.}~\bibnamefont
  {Ramaswamy}}, \bibinfo {author} {\bibfnamefont {J.}~\bibnamefont {Toner}}, \
  and\ \bibinfo {author} {\bibfnamefont {J.}~\bibnamefont {Prost}},\
  }\href@noop {} {\bibfield  {journal} {\bibinfo  {journal} {Physical Review
  Letters}\ }\textbf {\bibinfo {volume} {84}},\ \bibinfo {pages} {3494}
  (\bibinfo {year} {2000})}\BibitemShut {NoStop}%
\bibitem [{\citenamefont {Prost}\ and\ \citenamefont
  {Bruinsma}(1996)}]{Prost1996}%
  \BibitemOpen
  \bibfield  {author} {\bibinfo {author} {\bibfnamefont {J.}~\bibnamefont
  {Prost}}\ and\ \bibinfo {author} {\bibfnamefont {R.}~\bibnamefont
  {Bruinsma}},\ }\href {http://stacks.iop.org/0295-5075/33/i=4/a=321}
  {\bibfield  {journal} {\bibinfo  {journal} {EPL (Europhysics Letters)}\
  }\textbf {\bibinfo {volume} {33}},\ \bibinfo {pages} {321} (\bibinfo {year}
  {1996})}\BibitemShut {NoStop}%
\bibitem [{\citenamefont {Jung}\ and\ \citenamefont
  {H{\"a}nggi}(1987)}]{jung1987dynamical}%
  \BibitemOpen
  \bibfield  {author} {\bibinfo {author} {\bibfnamefont {P.}~\bibnamefont
  {Jung}}\ and\ \bibinfo {author} {\bibfnamefont {P.}~\bibnamefont
  {H{\"a}nggi}},\ }\href@noop {} {\bibfield  {journal} {\bibinfo  {journal}
  {Physical Review A}\ }\textbf {\bibinfo {volume} {35}},\ \bibinfo {pages}
  {4464} (\bibinfo {year} {1987})}\BibitemShut {NoStop}%
\bibitem [{\citenamefont {{H{\"a}nggi}}\ and\ \citenamefont
  {{Jung}}(1995)}]{hanggi1995p}%
  \BibitemOpen
  \bibfield  {author} {\bibinfo {author} {\bibfnamefont {P.}~\bibnamefont
  {{H{\"a}nggi}}}\ and\ \bibinfo {author} {\bibfnamefont {P.}~\bibnamefont
  {{Jung}}},\ }\href@noop {} {\bibfield  {journal} {\bibinfo  {journal}
  {Advances in Chemical Physics}\ }\textbf {\bibinfo {volume} {89}},\ \bibinfo
  {pages} {239} (\bibinfo {year} {1995})}\BibitemShut {NoStop}%
\bibitem [{\citenamefont {Toner}(2012{\natexlab{a}})}]{Toner2012reanalysis}%
  \BibitemOpen
  \bibfield  {author} {\bibinfo {author} {\bibfnamefont {J.}~\bibnamefont
  {Toner}},\ }\href {\doibase 10.1103/PhysRevE.86.031918} {\bibfield  {journal}
  {\bibinfo  {journal} {Phys. Rev. E}\ }\textbf {\bibinfo {volume} {86}},\
  \bibinfo {pages} {031918} (\bibinfo {year} {2012}{\natexlab{a}})}\BibitemShut
  {NoStop}%
\bibitem [{\citenamefont {Landau}\ and\ \citenamefont
  {Lifshitz}(1959)}]{landau1959course}%
  \BibitemOpen
  \bibfield  {author} {\bibinfo {author} {\bibfnamefont {L.}~\bibnamefont
  {Landau}}\ and\ \bibinfo {author} {\bibfnamefont {E.}~\bibnamefont
  {Lifshitz}},\ }\href@noop {} {\emph {\bibinfo {title} {Course of theoretical
  physics. vol. 6: Fluid mechanics}}}\ (\bibinfo  {publisher} {London},\
  \bibinfo {year} {1959})\BibitemShut {NoStop}%
\bibitem [{\citenamefont {Brotto}\ \emph {et~al.}(2013)\citenamefont {Brotto},
  \citenamefont {Caussin}, \citenamefont {Lauga},\ and\ \citenamefont
  {Bartolo}}]{brotto2013hydrodynamics}%
  \BibitemOpen
  \bibfield  {author} {\bibinfo {author} {\bibfnamefont {T.}~\bibnamefont
  {Brotto}}, \bibinfo {author} {\bibfnamefont {J.-B.}\ \bibnamefont {Caussin}},
  \bibinfo {author} {\bibfnamefont {E.}~\bibnamefont {Lauga}}, \ and\ \bibinfo
  {author} {\bibfnamefont {D.}~\bibnamefont {Bartolo}},\ }\href@noop {}
  {\bibfield  {journal} {\bibinfo  {journal} {Physical Review Letters}\
  }\textbf {\bibinfo {volume} {110}},\ \bibinfo {pages} {038101} (\bibinfo
  {year} {2013})}\BibitemShut {NoStop}%
\bibitem [{\citenamefont {Kumar}\ \emph {et~al.}(2014)\citenamefont {Kumar},
  \citenamefont {Soni}, \citenamefont {Ramaswamy},\ and\ \citenamefont
  {Sood}}]{kumar2014flocking}%
  \BibitemOpen
  \bibfield  {author} {\bibinfo {author} {\bibfnamefont {N.}~\bibnamefont
  {Kumar}}, \bibinfo {author} {\bibfnamefont {H.}~\bibnamefont {Soni}},
  \bibinfo {author} {\bibfnamefont {S.}~\bibnamefont {Ramaswamy}}, \ and\
  \bibinfo {author} {\bibfnamefont {A.}~\bibnamefont {Sood}},\ }\href@noop {}
  {\bibfield  {journal} {\bibinfo  {journal} {Nature communications}\ }\textbf
  {\bibinfo {volume} {5}},\ \bibinfo {pages} {4688} (\bibinfo {year}
  {2014})}\BibitemShut {NoStop}%
\bibitem [{\citenamefont {Kung}\ \emph {et~al.}(2006)\citenamefont {Kung},
  \citenamefont {Marchetti},\ and\ \citenamefont
  {Saunders}}]{kung2006hydrodynamics}%
  \BibitemOpen
  \bibfield  {author} {\bibinfo {author} {\bibfnamefont {W.}~\bibnamefont
  {Kung}}, \bibinfo {author} {\bibfnamefont {M.~C.}\ \bibnamefont {Marchetti}},
  \ and\ \bibinfo {author} {\bibfnamefont {K.}~\bibnamefont {Saunders}},\
  }\href@noop {} {\bibfield  {journal} {\bibinfo  {journal} {Physical Review
  E}\ }\textbf {\bibinfo {volume} {73}},\ \bibinfo {pages} {031708} (\bibinfo
  {year} {2006})}\BibitemShut {NoStop}%
\bibitem [{\citenamefont {Erdmann}\ \emph {et~al.}(2000)\citenamefont
  {Erdmann}, \citenamefont {Ebeling}, \citenamefont {Schimansky-Geier},\ and\
  \citenamefont {Schweitzer}}]{erdmann2000brownian}%
  \BibitemOpen
  \bibfield  {author} {\bibinfo {author} {\bibfnamefont {U.}~\bibnamefont
  {Erdmann}}, \bibinfo {author} {\bibfnamefont {W.}~\bibnamefont {Ebeling}},
  \bibinfo {author} {\bibfnamefont {L.}~\bibnamefont {Schimansky-Geier}}, \
  and\ \bibinfo {author} {\bibfnamefont {F.}~\bibnamefont {Schweitzer}},\
  }\href@noop {} {\bibfield  {journal} {\bibinfo  {journal} {The European
  Physical Journal B-Condensed Matter and Complex Systems}\ }\textbf {\bibinfo
  {volume} {15}},\ \bibinfo {pages} {105} (\bibinfo {year} {2000})}\BibitemShut
  {NoStop}%
\bibitem [{\citenamefont {Schweitzer}\ \emph {et~al.}(1998)\citenamefont
  {Schweitzer}, \citenamefont {Ebeling},\ and\ \citenamefont
  {Tilch}}]{schweitzer1998complex}%
  \BibitemOpen
  \bibfield  {author} {\bibinfo {author} {\bibfnamefont {F.}~\bibnamefont
  {Schweitzer}}, \bibinfo {author} {\bibfnamefont {W.}~\bibnamefont {Ebeling}},
  \ and\ \bibinfo {author} {\bibfnamefont {B.}~\bibnamefont {Tilch}},\
  }\href@noop {} {\bibfield  {journal} {\bibinfo  {journal} {Physical Review
  Letters}\ }\textbf {\bibinfo {volume} {80}},\ \bibinfo {pages} {5044}
  (\bibinfo {year} {1998})}\BibitemShut {NoStop}%
\bibitem [{\citenamefont {Ebeling}(1999)}]{ebeling1999w}%
  \BibitemOpen
  \bibfield  {author} {\bibinfo {author} {\bibfnamefont {W.}~\bibnamefont
  {Ebeling}},\ }\href@noop {} {\bibfield  {journal} {\bibinfo  {journal}
  {BioSystems}\ }\textbf {\bibinfo {volume} {49}},\ \bibinfo {pages} {17}
  (\bibinfo {year} {1999})}\BibitemShut {NoStop}%
\bibitem [{\citenamefont {Simha}\ and\ \citenamefont
  {Ramaswamy}(2002)}]{simha2002hydrodynamic}%
  \BibitemOpen
  \bibfield  {author} {\bibinfo {author} {\bibfnamefont {R.~A.}\ \bibnamefont
  {Simha}}\ and\ \bibinfo {author} {\bibfnamefont {S.}~\bibnamefont
  {Ramaswamy}},\ }\href@noop {} {\bibfield  {journal} {\bibinfo  {journal}
  {Physical Review Letters}\ }\textbf {\bibinfo {volume} {89}},\ \bibinfo
  {pages} {058101} (\bibinfo {year} {2002})}\BibitemShut {NoStop}%
\bibitem [{\citenamefont {Gruler}\ \emph {et~al.}(1999)\citenamefont {Gruler},
  \citenamefont {Dewald},\ and\ \citenamefont {Eberhardt}}]{gruler1999nematic}%
  \BibitemOpen
  \bibfield  {author} {\bibinfo {author} {\bibfnamefont {H.}~\bibnamefont
  {Gruler}}, \bibinfo {author} {\bibfnamefont {U.}~\bibnamefont {Dewald}}, \
  and\ \bibinfo {author} {\bibfnamefont {M.}~\bibnamefont {Eberhardt}},\
  }\href@noop {} {\bibfield  {journal} {\bibinfo  {journal} {The European
  Physical Journal B-Condensed Matter and Complex Systems}\ }\textbf {\bibinfo
  {volume} {11}},\ \bibinfo {pages} {187} (\bibinfo {year} {1999})}\BibitemShut
  {NoStop}%
\bibitem [{\citenamefont {Mishra}(2010)}]{mishra2010dynamics}%
  \BibitemOpen
  \bibfield  {author} {\bibinfo {author} {\bibfnamefont {S.}~\bibnamefont
  {Mishra}},\ }\emph {\bibinfo {title} {Dynamics, order and fluctuations in
  active nematics: numerical and theoretical studies}},\ \href
  {http://etd.ncsi.iisc.ernet.in/handle/2005/832} {Ph.D. thesis},\ \bibinfo
  {school} {Indian Institute of Science} (\bibinfo {year} {2010}),\ \Eprint
  {http://arxiv.org/abs/http://etd.ncsi.iisc.ernet.in/handle/2005/832}
  {http://etd.ncsi.iisc.ernet.in/handle/2005/832} \BibitemShut {NoStop}%
\bibitem [{\citenamefont {Ramaswamy}\ \emph {et~al.}(2003)\citenamefont
  {Ramaswamy}, \citenamefont {Simha},\ and\ \citenamefont
  {Toner}}]{SRAditiToner2003nematic}%
  \BibitemOpen
  \bibfield  {author} {\bibinfo {author} {\bibfnamefont {S.}~\bibnamefont
  {Ramaswamy}}, \bibinfo {author} {\bibfnamefont {R.~A.}\ \bibnamefont
  {Simha}}, \ and\ \bibinfo {author} {\bibfnamefont {J.}~\bibnamefont
  {Toner}},\ }\href@noop {} {\bibfield  {journal} {\bibinfo  {journal}
  {Europhysics Letters}\ }\textbf {\bibinfo {volume} {62}},\ \bibinfo {pages}
  {196} (\bibinfo {year} {2003})}\BibitemShut {NoStop}%
\bibitem [{\citenamefont {Mani}\ \emph {et~al.}(2002)\citenamefont {Mani},
  \citenamefont {Smet}, \citenamefont {von Klitzing}, \citenamefont
  {Narayanamurti}, \citenamefont {Johnson},\ and\ \citenamefont
  {Umansky}}]{mani2002}%
  \BibitemOpen
  \bibfield  {author} {\bibinfo {author} {\bibfnamefont {R.~G.}\ \bibnamefont
  {Mani}}, \bibinfo {author} {\bibfnamefont {J.~H.}\ \bibnamefont {Smet}},
  \bibinfo {author} {\bibfnamefont {K.}~\bibnamefont {von Klitzing}}, \bibinfo
  {author} {\bibfnamefont {V.}~\bibnamefont {Narayanamurti}}, \bibinfo {author}
  {\bibfnamefont {W.~B.}\ \bibnamefont {Johnson}}, \ and\ \bibinfo {author}
  {\bibfnamefont {V.}~\bibnamefont {Umansky}},\ }\href
  {http://dx.doi.org/10.1038/nature01277} {\bibfield  {journal} {\bibinfo
  {journal} {Nature}\ }\textbf {\bibinfo {volume} {420}},\ \bibinfo {pages}
  {646 EP } (\bibinfo {year} {2002})}\BibitemShut {NoStop}%
\bibitem [{\citenamefont {Zudov}\ \emph {et~al.}(2003)\citenamefont {Zudov},
  \citenamefont {Du}, \citenamefont {Pfeiffer},\ and\ \citenamefont
  {West}}]{zudov2003evidence}%
  \BibitemOpen
  \bibfield  {author} {\bibinfo {author} {\bibfnamefont {M.}~\bibnamefont
  {Zudov}}, \bibinfo {author} {\bibfnamefont {R.}~\bibnamefont {Du}}, \bibinfo
  {author} {\bibfnamefont {L.}~\bibnamefont {Pfeiffer}}, \ and\ \bibinfo
  {author} {\bibfnamefont {K.}~\bibnamefont {West}},\ }\href@noop {} {\bibfield
   {journal} {\bibinfo  {journal} {Physical Review Letters}\ }\textbf {\bibinfo
  {volume} {90}},\ \bibinfo {pages} {046807} (\bibinfo {year}
  {2003})}\BibitemShut {NoStop}%
\bibitem [{\citenamefont {Alicea}\ \emph {et~al.}(2005)\citenamefont {Alicea},
  \citenamefont {Balents}, \citenamefont {Fisher}, \citenamefont
  {Paramekanti},\ and\ \citenamefont {Radzihovsky}}]{alicea2005transition}%
  \BibitemOpen
  \bibfield  {author} {\bibinfo {author} {\bibfnamefont {J.}~\bibnamefont
  {Alicea}}, \bibinfo {author} {\bibfnamefont {L.}~\bibnamefont {Balents}},
  \bibinfo {author} {\bibfnamefont {M.~P.}\ \bibnamefont {Fisher}}, \bibinfo
  {author} {\bibfnamefont {A.}~\bibnamefont {Paramekanti}}, \ and\ \bibinfo
  {author} {\bibfnamefont {L.}~\bibnamefont {Radzihovsky}},\ }\href@noop {}
  {\bibfield  {journal} {\bibinfo  {journal} {Physical Review B}\ }\textbf
  {\bibinfo {volume} {71}},\ \bibinfo {pages} {235322} (\bibinfo {year}
  {2005})}\BibitemShut {NoStop}%
\bibitem [{\citenamefont {Dunkel}\ \emph {et~al.}(2013)\citenamefont {Dunkel},
  \citenamefont {Heidenreich}, \citenamefont {Drescher}, \citenamefont
  {Wensink}, \citenamefont {B{\"a}r},\ and\ \citenamefont
  {Goldstein}}]{dunkel2013fluid}%
  \BibitemOpen
  \bibfield  {author} {\bibinfo {author} {\bibfnamefont {J.}~\bibnamefont
  {Dunkel}}, \bibinfo {author} {\bibfnamefont {S.}~\bibnamefont {Heidenreich}},
  \bibinfo {author} {\bibfnamefont {K.}~\bibnamefont {Drescher}}, \bibinfo
  {author} {\bibfnamefont {H.~H.}\ \bibnamefont {Wensink}}, \bibinfo {author}
  {\bibfnamefont {M.}~\bibnamefont {B{\"a}r}}, \ and\ \bibinfo {author}
  {\bibfnamefont {R.~E.}\ \bibnamefont {Goldstein}},\ }\href@noop {} {\bibfield
   {journal} {\bibinfo  {journal} {Physical Review Letters}\ }\textbf {\bibinfo
  {volume} {110}},\ \bibinfo {pages} {228102} (\bibinfo {year}
  {2013})}\BibitemShut {NoStop}%
\bibitem [{\citenamefont {Swift}\ and\ \citenamefont
  {Hohenberg}(1977)}]{swift1977hydrodynamic}%
  \BibitemOpen
  \bibfield  {author} {\bibinfo {author} {\bibfnamefont {J.}~\bibnamefont
  {Swift}}\ and\ \bibinfo {author} {\bibfnamefont {P.~C.}\ \bibnamefont
  {Hohenberg}},\ }\href@noop {} {\bibfield  {journal} {\bibinfo  {journal}
  {Physical Review A}\ }\textbf {\bibinfo {volume} {15}},\ \bibinfo {pages}
  {319} (\bibinfo {year} {1977})}\BibitemShut {NoStop}%
\bibitem [{\citenamefont {Ramaswamy}\ and\ \citenamefont
  {Mazenko}(1982)}]{ramaswamy1982linear}%
  \BibitemOpen
  \bibfield  {author} {\bibinfo {author} {\bibfnamefont {S.}~\bibnamefont
  {Ramaswamy}}\ and\ \bibinfo {author} {\bibfnamefont {G.~F.}\ \bibnamefont
  {Mazenko}},\ }\href@noop {} {\bibfield  {journal} {\bibinfo  {journal}
  {Physical Review A}\ }\textbf {\bibinfo {volume} {26}},\ \bibinfo {pages}
  {1735} (\bibinfo {year} {1982})}\BibitemShut {NoStop}%
\bibitem [{\citenamefont {Kim}\ and\ \citenamefont
  {Mazenko}(1991)}]{kim1991equations}%
  \BibitemOpen
  \bibfield  {author} {\bibinfo {author} {\bibfnamefont {B.}~\bibnamefont
  {Kim}}\ and\ \bibinfo {author} {\bibfnamefont {G.~F.}\ \bibnamefont
  {Mazenko}},\ }\href@noop {} {\bibfield  {journal} {\bibinfo  {journal}
  {Journal of statistical physics}\ }\textbf {\bibinfo {volume} {64}},\
  \bibinfo {pages} {631} (\bibinfo {year} {1991})}\BibitemShut {NoStop}%
\bibitem [{\citenamefont {Toner}(2012{\natexlab{b}})}]{toner2012birth}%
  \BibitemOpen
  \bibfield  {author} {\bibinfo {author} {\bibfnamefont {J.}~\bibnamefont
  {Toner}},\ }\href@noop {} {\bibfield  {journal} {\bibinfo  {journal}
  {Physical Review Letters}\ }\textbf {\bibinfo {volume} {108}},\ \bibinfo
  {pages} {088102} (\bibinfo {year} {2012}{\natexlab{b}})}\BibitemShut
  {NoStop}%
\bibitem [{\citenamefont {Lebowitz}\ and\ \citenamefont
  {Spohn}(1999)}]{lebowitz1999gallavotti}%
  \BibitemOpen
  \bibfield  {author} {\bibinfo {author} {\bibfnamefont {J.~L.}\ \bibnamefont
  {Lebowitz}}\ and\ \bibinfo {author} {\bibfnamefont {H.}~\bibnamefont
  {Spohn}},\ }\href@noop {} {\bibfield  {journal} {\bibinfo  {journal} {Journal
  of Statistical Physics}\ }\textbf {\bibinfo {volume} {95}},\ \bibinfo {pages}
  {333} (\bibinfo {year} {1999})}\BibitemShut {NoStop}%
\bibitem [{\citenamefont {Kullback}\ and\ \citenamefont
  {Leibler}(1951)}]{kullback1951information}%
  \BibitemOpen
  \bibfield  {author} {\bibinfo {author} {\bibfnamefont {S.}~\bibnamefont
  {Kullback}}\ and\ \bibinfo {author} {\bibfnamefont {R.~A.}\ \bibnamefont
  {Leibler}},\ }\href@noop {} {\bibfield  {journal} {\bibinfo  {journal} {The
  Annals of Mathematical Statistics}\ }\textbf {\bibinfo {volume} {22}},\
  \bibinfo {pages} {79} (\bibinfo {year} {1951})}\BibitemShut {NoStop}%
\bibitem [{\citenamefont {Nardini}\ \emph {et~al.}(2017)\citenamefont
  {Nardini}, \citenamefont {Fodor}, \citenamefont {Tjhung}, \citenamefont
  {Van~Wijland}, \citenamefont {Tailleur},\ and\ \citenamefont
  {Cates}}]{nardini2017entropy}%
  \BibitemOpen
  \bibfield  {author} {\bibinfo {author} {\bibfnamefont {C.}~\bibnamefont
  {Nardini}}, \bibinfo {author} {\bibfnamefont {{\'E}.}~\bibnamefont {Fodor}},
  \bibinfo {author} {\bibfnamefont {E.}~\bibnamefont {Tjhung}}, \bibinfo
  {author} {\bibfnamefont {F.}~\bibnamefont {Van~Wijland}}, \bibinfo {author}
  {\bibfnamefont {J.}~\bibnamefont {Tailleur}}, \ and\ \bibinfo {author}
  {\bibfnamefont {M.~E.}\ \bibnamefont {Cates}},\ }\href@noop {} {\bibfield
  {journal} {\bibinfo  {journal} {Physical Review X}\ }\textbf {\bibinfo
  {volume} {7}},\ \bibinfo {pages} {021007} (\bibinfo {year}
  {2017})}\BibitemShut {NoStop}%
\bibitem [{\citenamefont {Onsager}\ and\ \citenamefont
  {Machlup}(1953)}]{onsager1953fluctuations}%
  \BibitemOpen
  \bibfield  {author} {\bibinfo {author} {\bibfnamefont {L.}~\bibnamefont
  {Onsager}}\ and\ \bibinfo {author} {\bibfnamefont {S.}~\bibnamefont
  {Machlup}},\ }\href@noop {} {\bibfield  {journal} {\bibinfo  {journal}
  {Physical Review}\ }\textbf {\bibinfo {volume} {91}},\ \bibinfo {pages}
  {1505} (\bibinfo {year} {1953})}\BibitemShut {NoStop}%
\bibitem [{\citenamefont {Cugliandolo}\ and\ \citenamefont
  {Lecomte}(2017)}]{cugliandolo2017rules}%
  \BibitemOpen
  \bibfield  {author} {\bibinfo {author} {\bibfnamefont {L.~F.}\ \bibnamefont
  {Cugliandolo}}\ and\ \bibinfo {author} {\bibfnamefont {V.}~\bibnamefont
  {Lecomte}},\ }\href@noop {} {\bibfield  {journal} {\bibinfo  {journal}
  {Journal of Physics A: Mathematical and Theoretical}\ }\textbf {\bibinfo
  {volume} {50}},\ \bibinfo {pages} {345001} (\bibinfo {year}
  {2017})}\BibitemShut {NoStop}%
\bibitem [{\citenamefont {Teramoto}\ and\ \citenamefont
  {Sasa}(2005)}]{teramoto2005microscopic}%
  \BibitemOpen
  \bibfield  {author} {\bibinfo {author} {\bibfnamefont {H.}~\bibnamefont
  {Teramoto}}\ and\ \bibinfo {author} {\bibfnamefont {S.-i.}\ \bibnamefont
  {Sasa}},\ }\href@noop {} {\bibfield  {journal} {\bibinfo  {journal} {Physical
  Review E}\ }\textbf {\bibinfo {volume} {72}},\ \bibinfo {pages} {060102}
  (\bibinfo {year} {2005})}\BibitemShut {NoStop}%
\bibitem [{\citenamefont {Harada}\ and\ \citenamefont
  {Sasa}(2006)}]{harada2006energy}%
  \BibitemOpen
  \bibfield  {author} {\bibinfo {author} {\bibfnamefont {T.}~\bibnamefont
  {Harada}}\ and\ \bibinfo {author} {\bibfnamefont {S.-i.}\ \bibnamefont
  {Sasa}},\ }\href@noop {} {\bibfield  {journal} {\bibinfo  {journal} {Physical
  Review E}\ }\textbf {\bibinfo {volume} {73}},\ \bibinfo {pages} {026131}
  (\bibinfo {year} {2006})}\BibitemShut {NoStop}%
\bibitem [{\citenamefont {Yamada}\ and\ \citenamefont
  {Yoshimori}(2015)}]{yamada2015unified}%
  \BibitemOpen
  \bibfield  {author} {\bibinfo {author} {\bibfnamefont {K.}~\bibnamefont
  {Yamada}}\ and\ \bibinfo {author} {\bibfnamefont {A.}~\bibnamefont
  {Yoshimori}},\ }\href@noop {} {\bibfield  {journal} {\bibinfo  {journal}
  {Journal of the Physical Society of Japan}\ }\textbf {\bibinfo {volume}
  {84}},\ \bibinfo {pages} {044008} (\bibinfo {year} {2015})}\BibitemShut
  {NoStop}%
\bibitem [{\citenamefont {Martin}\ \emph {et~al.}(1973)\citenamefont {Martin},
  \citenamefont {Siggia},\ and\ \citenamefont {Rose}}]{martin1973pc}%
  \BibitemOpen
  \bibfield  {author} {\bibinfo {author} {\bibfnamefont {P.~C.}\ \bibnamefont
  {Martin}}, \bibinfo {author} {\bibfnamefont {E.~D.}\ \bibnamefont {Siggia}},
  \ and\ \bibinfo {author} {\bibfnamefont {H.~A.}\ \bibnamefont {Rose}},\
  }\href {\doibase 10.1103/PhysRevA.8.423} {\bibfield  {journal} {\bibinfo
  {journal} {Physical Review A}\ }\textbf {\bibinfo {volume} {8}},\ \bibinfo
  {pages} {423} (\bibinfo {year} {1973})}\BibitemShut {NoStop}%
\bibitem [{\citenamefont {Janssen}(1976)}]{janssen1976lagrangean}%
  \BibitemOpen
  \bibfield  {author} {\bibinfo {author} {\bibfnamefont {H.-K.}\ \bibnamefont
  {Janssen}},\ }\href@noop {} {\bibfield  {journal} {\bibinfo  {journal}
  {Zeitschrift f{\"u}r Physik B Condensed Matter}\ }\textbf {\bibinfo {volume}
  {23}},\ \bibinfo {pages} {377} (\bibinfo {year} {1976})}\BibitemShut
  {NoStop}%
\bibitem [{\citenamefont {Janssen}(1979)}]{janssen1979field}%
  \BibitemOpen
  \bibfield  {author} {\bibinfo {author} {\bibfnamefont {H.}~\bibnamefont
  {Janssen}},\ }in\ \href@noop {} {\emph {\bibinfo {booktitle} {Dynamical
  critical phenomena and related topics}}}\ (\bibinfo  {publisher} {Springer},\
  \bibinfo {year} {1979})\ pp.\ \bibinfo {pages} {25--47}\BibitemShut {NoStop}%
\bibitem [{\citenamefont {De~Dominicis}(1976)}]{de1976c}%
  \BibitemOpen
  \bibfield  {author} {\bibinfo {author} {\bibfnamefont {C.}~\bibnamefont
  {De~Dominicis}},\ }in\ \href@noop {} {\emph {\bibinfo {booktitle} {Journal de
  Physique Colloques}}},\ Vol.~\bibinfo {volume} {37}\ (\bibinfo {year}
  {1976})\ pp.\ \bibinfo {pages} {C1--247}\BibitemShut {NoStop}%
\bibitem [{\citenamefont {De~Dominicis}(1978)}]{de1978c}%
  \BibitemOpen
  \bibfield  {author} {\bibinfo {author} {\bibfnamefont {C.}~\bibnamefont
  {De~Dominicis}},\ }\href {\doibase 10.1103/PhysRevB.18.4913} {\bibfield
  {journal} {\bibinfo  {journal} {Physical Review B}\ }\textbf {\bibinfo
  {volume} {18}},\ \bibinfo {pages} {4913} (\bibinfo {year}
  {1978})}\BibitemShut {NoStop}%
\bibitem [{\citenamefont {Aron}\ \emph {et~al.}(2010)\citenamefont {Aron},
  \citenamefont {Biroli},\ and\ \citenamefont
  {Cugliandolo}}]{aron2010symmetries}%
  \BibitemOpen
  \bibfield  {author} {\bibinfo {author} {\bibfnamefont {C.}~\bibnamefont
  {Aron}}, \bibinfo {author} {\bibfnamefont {G.}~\bibnamefont {Biroli}}, \ and\
  \bibinfo {author} {\bibfnamefont {L.~F.}\ \bibnamefont {Cugliandolo}},\
  }\href@noop {} {\bibfield  {journal} {\bibinfo  {journal} {Journal of
  Statistical Mechanics: Theory and Experiment}\ }\textbf {\bibinfo {volume}
  {2010}},\ \bibinfo {pages} {P11018} (\bibinfo {year} {2010})}\BibitemShut
  {NoStop}%
\bibitem [{\citenamefont {Grosberg}\ and\ \citenamefont
  {Joanny}(2015)}]{grosberg2015nonequilibrium}%
  \BibitemOpen
  \bibfield  {author} {\bibinfo {author} {\bibfnamefont {A.}~\bibnamefont
  {Grosberg}}\ and\ \bibinfo {author} {\bibfnamefont {J.-F.}\ \bibnamefont
  {Joanny}},\ }\href@noop {} {\bibfield  {journal} {\bibinfo  {journal}
  {Physical Review E}\ }\textbf {\bibinfo {volume} {92}},\ \bibinfo {pages}
  {032118} (\bibinfo {year} {2015})}\BibitemShut {NoStop}%
\bibitem [{\citenamefont {Saha}\ \emph {et~al.}(2014)\citenamefont {Saha},
  \citenamefont {Golestanian},\ and\ \citenamefont
  {Ramaswamy}}]{saha2014clusters}%
  \BibitemOpen
  \bibfield  {author} {\bibinfo {author} {\bibfnamefont {S.}~\bibnamefont
  {Saha}}, \bibinfo {author} {\bibfnamefont {R.}~\bibnamefont {Golestanian}}, \
  and\ \bibinfo {author} {\bibfnamefont {S.}~\bibnamefont {Ramaswamy}},\
  }\href@noop {} {\bibfield  {journal} {\bibinfo  {journal} {Physical Review
  E}\ }\textbf {\bibinfo {volume} {89}},\ \bibinfo {pages} {062316} (\bibinfo
  {year} {2014})}\BibitemShut {NoStop}%
\bibitem [{\citenamefont {Liebchen}\ \emph {et~al.}(2017)\citenamefont
  {Liebchen}, \citenamefont {Marenduzzo},\ and\ \citenamefont
  {Cates}}]{liebchen2017phoretic}%
  \BibitemOpen
  \bibfield  {author} {\bibinfo {author} {\bibfnamefont {B.}~\bibnamefont
  {Liebchen}}, \bibinfo {author} {\bibfnamefont {D.}~\bibnamefont
  {Marenduzzo}}, \ and\ \bibinfo {author} {\bibfnamefont {M.~E.}\ \bibnamefont
  {Cates}},\ }\href@noop {} {\bibfield  {journal} {\bibinfo  {journal}
  {Physical Review Letters}\ }\textbf {\bibinfo {volume} {118}},\ \bibinfo
  {pages} {268001} (\bibinfo {year} {2017})}\BibitemShut {NoStop}%
\bibitem [{\citenamefont {Liebchen}\ \emph {et~al.}(2016)\citenamefont
  {Liebchen}, \citenamefont {Cates},\ and\ \citenamefont
  {Marenduzzo}}]{liebchen2016pattern}%
  \BibitemOpen
  \bibfield  {author} {\bibinfo {author} {\bibfnamefont {B.}~\bibnamefont
  {Liebchen}}, \bibinfo {author} {\bibfnamefont {M.~E.}\ \bibnamefont {Cates}},
  \ and\ \bibinfo {author} {\bibfnamefont {D.}~\bibnamefont {Marenduzzo}},\
  }\href@noop {} {\bibfield  {journal} {\bibinfo  {journal} {Soft Matter}\
  }\textbf {\bibinfo {volume} {12}},\ \bibinfo {pages} {7259} (\bibinfo {year}
  {2016})}\BibitemShut {NoStop}%
\bibitem [{\citenamefont {Gladrow}\ \emph {et~al.}(2016)\citenamefont
  {Gladrow}, \citenamefont {Fakhri}, \citenamefont {MacKintosh}, \citenamefont
  {Schmidt},\ and\ \citenamefont {Broedersz}}]{gladrow2016broken}%
  \BibitemOpen
  \bibfield  {author} {\bibinfo {author} {\bibfnamefont {J.}~\bibnamefont
  {Gladrow}}, \bibinfo {author} {\bibfnamefont {N.}~\bibnamefont {Fakhri}},
  \bibinfo {author} {\bibfnamefont {F.}~\bibnamefont {MacKintosh}}, \bibinfo
  {author} {\bibfnamefont {C.}~\bibnamefont {Schmidt}}, \ and\ \bibinfo
  {author} {\bibfnamefont {C.}~\bibnamefont {Broedersz}},\ }\href@noop {}
  {\bibfield  {journal} {\bibinfo  {journal} {Physical Review Letters}\
  }\textbf {\bibinfo {volume} {116}},\ \bibinfo {pages} {248301} (\bibinfo
  {year} {2016})}\BibitemShut {NoStop}%
\bibitem [{\citenamefont {Battle}\ \emph {et~al.}(2016)\citenamefont {Battle},
  \citenamefont {Broedersz}, \citenamefont {Fakhri}, \citenamefont {Geyer},
  \citenamefont {Howard}, \citenamefont {Schmidt},\ and\ \citenamefont
  {MacKintosh}}]{battle2016broken}%
  \BibitemOpen
  \bibfield  {author} {\bibinfo {author} {\bibfnamefont {C.}~\bibnamefont
  {Battle}}, \bibinfo {author} {\bibfnamefont {C.~P.}\ \bibnamefont
  {Broedersz}}, \bibinfo {author} {\bibfnamefont {N.}~\bibnamefont {Fakhri}},
  \bibinfo {author} {\bibfnamefont {V.~F.}\ \bibnamefont {Geyer}}, \bibinfo
  {author} {\bibfnamefont {J.}~\bibnamefont {Howard}}, \bibinfo {author}
  {\bibfnamefont {C.~F.}\ \bibnamefont {Schmidt}}, \ and\ \bibinfo {author}
  {\bibfnamefont {F.~C.}\ \bibnamefont {MacKintosh}},\ }\href@noop {}
  {\bibfield  {journal} {\bibinfo  {journal} {Science}\ }\textbf {\bibinfo
  {volume} {352}},\ \bibinfo {pages} {604} (\bibinfo {year}
  {2016})}\BibitemShut {NoStop}%
\bibitem [{\citenamefont {Seifert}(2012)}]{seifert2012stochastic}%
  \BibitemOpen
  \bibfield  {author} {\bibinfo {author} {\bibfnamefont {U.}~\bibnamefont
  {Seifert}},\ }\href@noop {} {\bibfield  {journal} {\bibinfo  {journal}
  {Reports on Progress in Physics}\ }\textbf {\bibinfo {volume} {75}},\
  \bibinfo {pages} {126001} (\bibinfo {year} {2012})}\BibitemShut {NoStop}%
\bibitem [{\citenamefont {Ryter}(1981)}]{ryter1981brownian}%
  \BibitemOpen
  \bibfield  {author} {\bibinfo {author} {\bibfnamefont {D.}~\bibnamefont
  {Ryter}},\ }\href@noop {} {\bibfield  {journal} {\bibinfo  {journal}
  {Zeitschrift f{\"u}r Physik B Condensed Matter}\ }\textbf {\bibinfo {volume}
  {41}},\ \bibinfo {pages} {39} (\bibinfo {year} {1981})}\BibitemShut {NoStop}%
\bibitem [{\citenamefont {Eyink}\ \emph {et~al.}(1996)\citenamefont {Eyink},
  \citenamefont {Lebowitz},\ and\ \citenamefont
  {Spohn}}]{eyink1996hydrodynamics}%
  \BibitemOpen
  \bibfield  {author} {\bibinfo {author} {\bibfnamefont {G.~L.}\ \bibnamefont
  {Eyink}}, \bibinfo {author} {\bibfnamefont {J.~L.}\ \bibnamefont {Lebowitz}},
  \ and\ \bibinfo {author} {\bibfnamefont {H.}~\bibnamefont {Spohn}},\
  }\href@noop {} {\bibfield  {journal} {\bibinfo  {journal} {Journal of
  Statistical physics}\ }\textbf {\bibinfo {volume} {83}},\ \bibinfo {pages}
  {385} (\bibinfo {year} {1996})}\BibitemShut {NoStop}%
\bibitem [{\citenamefont {Graham}(1977)}]{graham1977covariant}%
  \BibitemOpen
  \bibfield  {author} {\bibinfo {author} {\bibfnamefont {R.}~\bibnamefont
  {Graham}},\ }\href@noop {} {\bibfield  {journal} {\bibinfo  {journal}
  {Zeitschrift f{\"u}r Physik B Condensed Matter}\ }\textbf {\bibinfo {volume}
  {26}},\ \bibinfo {pages} {397} (\bibinfo {year} {1977})}\BibitemShut
  {NoStop}%
\bibitem [{\citenamefont {Prost}\ \emph {et~al.}(2009)\citenamefont {Prost},
  \citenamefont {Joanny},\ and\ \citenamefont
  {Parrondo}}]{prost2009generalized}%
  \BibitemOpen
  \bibfield  {author} {\bibinfo {author} {\bibfnamefont {J.}~\bibnamefont
  {Prost}}, \bibinfo {author} {\bibfnamefont {J.-F.}\ \bibnamefont {Joanny}}, \
  and\ \bibinfo {author} {\bibfnamefont {J.}~\bibnamefont {Parrondo}},\
  }\href@noop {} {\bibfield  {journal} {\bibinfo  {journal} {Physical Review
  Letters}\ }\textbf {\bibinfo {volume} {103}},\ \bibinfo {pages} {090601}
  (\bibinfo {year} {2009})}\BibitemShut {NoStop}%
\end{thebibliography}%
\end{document}